\documentclass[prb, aps, twocolumn, amsmath, amssymb, graphicx, longbibliography, nofootinbib]{revtex4-2}

\usepackage{bm}
\usepackage{graphicx}
\usepackage{braket}
\usepackage[normalem]{ulem}

\usepackage{color}
\usepackage[usenames, dvipsnames]{xcolor}
\usepackage[colorlinks=true]{hyperref}
\hypersetup{linktoc=page, colorlinks=true, linkcolor=MidnightBlue, citecolor=MidnightBlue, urlcolor=OliveGreen}

\usepackage[]{cleveref}
\Crefname{section}{Sec.}{Secs.}
\Crefname{equation}{Eq.}{Eqs.}
\Crefname{figure}{Fig.}{Figs.}
\Crefname{tabular}{Tab.}{Tabs.}

\usepackage{siunitx}
\usepackage{soul}
\usepackage{url}

\usepackage{tabularx}
\usepackage{array}
\newcolumntype{I}{>{\hsize=0.36\hsize}X}
\newcolumntype{C}{>{\hsize=0.3\hsize\centering\arraybackslash}X}
\newcolumntype{U}{>{\hsize=0.35\hsize\centering\arraybackslash}X}
\newcolumntype{V}{>{\hsize=0.4\hsize\centering\arraybackslash}X}
\newcolumntype{W}{>{\hsize=0.44\hsize\centering\arraybackslash}X}
\newcolumntype{Y}{>{\hsize=0.49\hsize\centering\arraybackslash}X}

\usepackage{multirow}
\usepackage{makecell}

\usepackage{comment}

\usepackage[dvipsnames]{xcolor}

\newcommand{\muB}{\mu_\mathrm{\scriptscriptstyle B}}
\newcommand{\kB}{k_\mathrm{\scriptscriptstyle B}}
\newcommand{\kF}{k_\mathrm{\scriptscriptstyle F}}
\newcommand{\EF}{E_\mathrm{\scriptscriptstyle F}}

\renewcommand{\L}{\mathrm{\scriptscriptstyle L}}
\newcommand{\R}{\mathrm{\scriptscriptstyle R}}
\newcommand{\LL}{\mathrm{\scriptscriptstyle LL}}
\newcommand{\RR}{\mathrm{\scriptscriptstyle RR}}
\newcommand{\LR}{\mathrm{\scriptscriptstyle LR}}
\newcommand{\RL}{\mathrm{\scriptscriptstyle RL}}
\newcommand{\D}{\mathrm{\scriptscriptstyle D}}
\newcommand{\G}{\mathrm{\scriptscriptstyle G}}
\newcommand{\GLL}{G_\LL}
\newcommand{\GRR}{G_\RR}
\newcommand{\GLR}{G_\LR}
\newcommand{\GRL}{G_\RL}
\newcommand{\Gag}{G_\mathrm{\scriptscriptstyle N}}
\newcommand{\Vp}{V_\mathrm{p}}
\newcommand{\Vb}{V_\mathrm{b}}

\newcommand{\Vg}{V_\mathrm{g}}
\newcommand{\Vlc}{V_\mathrm{lc}}
\newcommand{\Vrc}{V_\mathrm{rc}}
\newcommand{\Vdep}{V_\mathrm{dep}}
\newcommand{\DeltaT}{\Delta_\mathrm{\scriptscriptstyle T}}
\newcommand{\xiT}{\xi_\mathrm{\scriptscriptstyle T}}
\newcommand{\ellLoc}{\ell_\mathrm{loc}}

\newcommand{\SM}{\mathrm{\scriptscriptstyle SM}}
\newcommand{\SC}{\mathrm{\scriptscriptstyle SC}}
\newcommand{\DeltaMax}{\Delta_\mathrm{topo}^\mathrm{max}}
\newcommand{\meanDeltaMax}{{\bar\Delta}_\mathrm{topo}^\mathrm{max}}
\newcommand{\DeltaInd}{\Delta_\mathrm{ind}}
\newcommand{\DeltaAl}{\Delta_\mathrm{Al}}
\newcommand{\xiAl}{\xi_\mathrm{Al}}

\newcommand{\NL}{\mathrm{\scriptscriptstyle NL}}

\renewcommand{\sc}{n_{\mathrm{{\scriptscriptstyle 2D}, int}}}
\newcommand{\scu}{10^{12}/\text{cm}^2}

\newcommand{\volSOI}{\mathcal{V}_\mathrm{\scriptscriptstyle SOI2}}
\newcommand{\BavSOI}{B_\mathrm{\scriptscriptstyle SOI2}}
\newcommand{\volSOIbar}{\bar{\mathcal{V}}_\mathrm{\scriptscriptstyle SOI2}}
\newcommand{\BavSOIbar}{\bar{B}_\mathrm{\scriptscriptstyle SOI2}}

\newcommand{\Iinput}{dI^0}
\newcommand{\Vinput}{dV^0}
\newcommand{\Isample}{dI}
\newcommand{\Vsample}{dV}

\newcommand{\Vv}{\textbf{V}}
\newcommand{\Iv}{\textbf{I}}

\newcommand{\Mm}{\textbf{M}}

\renewcommand{\thesection}{\arabic{section}}
\renewcommand\thesubsection{\thesection.\arabic{subsection}}

\makeatletter
\renewcommand*{\p@subsection}{}
\renewcommand*{\p@subsubsection}{}
\makeatother

\begin{document}

\title{InAs-Al Hybrid Devices Passing the Topological Gap Protocol}

\author{Microsoft Quantum$^\dagger$}

\date{\today{}}

\begin{abstract}
We present measurements and simulations of semiconductor-superconductor heterostructure devices that are consistent with the observation of topological superconductivity and Majorana zero modes.
The devices are fabricated from high-mobility two-dimensional electron gases in which quasi-one-dimensional wires are defined by electrostatic gates.
These devices enable measurements of local and non-local transport properties and have been optimized via extensive simulations to ensure robustness against non-uniformity and disorder.
Our main result is that several devices, fabricated according to the design's engineering specifications, have passed the topological gap protocol defined in Pikulin \textit{et al.} [arXiv:2103.12217].
This protocol is a stringent test composed of a sequence of three-terminal local and non-local transport measurements performed while varying the magnetic field, semiconductor electron density, and junction transparencies.
Passing the protocol indicates a high probability of detection of a topological phase hosting Majorana zero modes as determined by large-scale disorder simulations.
Our experimental results are consistent with a quantum phase transition into a topological superconducting phase that extends over several hundred millitesla in magnetic field and several millivolts in gate voltage, corresponding to approximately one hundred micro-electron-volts in Zeeman energy and chemical potential in the semiconducting wire.
These regions feature a closing and re-opening of the bulk gap, with simultaneous zero-bias conductance peaks at \textit{both} ends of the devices that withstand changes in the junction transparencies.
The extracted maximum topological gaps in our devices are 20-\SI{60}{\micro\eV}.
This demonstration is a prerequisite for experiments involving fusion and braiding of Majorana zero modes.
\end{abstract}

\maketitle

\tableofcontents

\section{Introduction}
\label{sec:introduction}

Topological quantum computation offers the promise of a high degree of intrinsic hardware-level fault-tolerance~\cite{Kitaev97, Freedman98, Nayak08, Alicea12a, DasSarma15, Aasen16}, potentially enabling a single-module quantum computing system that is capable of solving critical problems sufficiently rapidly to have societal impact \cite{vonBurg21}.
This approach hinges on (a) reliably producing a stable topological phase of matter that supports non-Abelian quasiparticles or defects and (b) processing quantum information through protected operations, such as braiding.
The former is challenging due to the material parameter and disorder requirements for topological phases of matter.
In this paper, we report on three-terminal semiconductor-superconductor nanowire
devices that pass the stringent topological gap protocol~\cite{Pikulin21} and therefore satisfy these requirements.
We further extract the gap associated with the topological superconducting phase in our devices~\cite{Kitaev01, Sau10a, Lutchyn10, Oreg10}.

Topological phases are a form of matter in which the ground state has long-range quantum entanglement and there is a gap to excited states~\cite{Kitaev06a}.
Unlike phases of matter that can be distinguished completely by local measurements, topological phases are identified by the transformations of their low-energy states that result from fusing and braiding their quasiparticles and defects.
Directly measuring these properties in experiments is rather subtle~\cite{Bonderson21}, hindering efforts to fully determine the topological order of candidate materials.
In the fractional quantum Hall regime, for example, a quantized Hall conductance reveals the presence of a non-trivial topological phase, but many different topological phases can have the same Hall conductance.
Consequently, different measurements are necessary to determine which topological phase is present in a given device \cite{Chamon97, Fradkin98, Bonderson06a, Stern06, Willett10, Grenier11, Banerjee18, Nakamura20}.

In the case of quasi-one-dimensional superconducting wires without any symmetries enforced, there are only two phases~--- one trivial and one topological.
The latter supports Majorana zero modes (MZMs) localized at the ends of the nanowire~\cite{Kitaev01, Lutchyn10, Oreg10}.
While MZMs can be directly detected through fusion and braiding, one of their auxiliary signatures are zero-bias peaks (ZBPs) in the differential tunneling conductance at the nanowire's ends~\cite{Law09, Sau10b, Flensberg10, Wimmer11, Stanescu11, Fidkowski12}.
Indeed, most of the earlier experimental studies of candidate topological superconductors focused on ZBPs~\cite{Mourik12, Das12, Deng12, Finck12, Churchill13, Deng16, Nichele17, Vaitiekenas18, deMoor18, Lutchyn18, Anselmetti19, Zhang21, Vaitieknas20, Banerjee22a}.
However, ZBPs can also be caused by disorder \cite{Bagrets12, Pikulin12, Pan22}, smooth potential variations near the tunnel junction~\cite{Prada12, Kells12, Tewari14, Liu17, Vuik19, Pan21b}, unintentional quantum dots~\cite{Lee12, Reeg18b}, or a supercurrent~\cite{Yu20}.
These trivial ZBPs can persist over a fairly large range of system parameters~\cite{Lee13, Suominen17, Pan20}.

A ZBP associated with an MZM must have a partner at the other end of the wire and should be stable to variations in the electric and magnetic fields in the device.
The stability of MZMs with respect to such variations is determined by the bulk gap.
However, if a device has a sufficiently large number of control parameters, it is likely that it can be tuned into a configuration in which it has trivial ZBPs at both ends.
Meanwhile, the predicted range of stability of a topological phase depends strongly on device geometry, the full stack of materials, and disorder, rendering it difficult to distinguish ``stable'' ZBPs from ``accidental'' ones purely empirically.
Analyzing the detailed shapes of tunneling conductance spectra leads to some loose qualitative patterns, but there is no sharp binary distinction between the local tunneling conductance spectra associated with MZMs and trivial ZBPs at non-zero temperature.
In short, neither more extensive data sets of ZBPs nor more beautiful ZBPs can distinguish the topological and trivial phases.
Therefore, it is crucial to develop a practical, reliable protocol that enables the detection of the topological superconducting phase of a nanowire, and it is clear that additional measurements beyond the tunneling conductance are necessary for such a protocol.

This challenge is addressed by the topological gap protocol (TGP) \cite{Pikulin21}, which is designed to reliably identify the topological phase through a series of stringent experimental tests.
At the heart of this protocol is the fact that there is necessarily a quantum phase transition between the trivial and topological phases~\cite{Read00}.
The protocol detects a bulk phase transition between low-magnetic-field and high-magnetic-field phases via a bulk gap closing.
It establishes that the high-field phase is topological through the stability of its ZBPs, in a manner that we specify below.
The TGP requires three-terminal device geometries, which overcome the limitations of many earlier two-terminal devices.
They allow ZBPs to be simultaneously observed at both ends and also allow for a measurement of the bulk transport gap through the \textit{non-local conductance}.
The protocol is passed when (a) ZBPs are observed in the local conductances measured at tunnel junctions at both ends of a wire, and they are stable to changes in the junction transparency; (b) these stable ZBPs persist over a range of magnetic fields and electron densities in the wire; (c) a closing and re-opening of the bulk transport gap is detected in the non-local conductances; (d) there is a region in the bulk phase diagram whose boundary is gapless and whose interior is gapped and has stable ZBPs; (e) the observed bulk transport gap throughout this region~--- the \textit{topological gap}~--- exceeds the resolution of the measurement.

The hallmarks of most topological phases, including the one discussed here, are rather subtle: there is no signature as immediate as a quantized conductance or Meissner effect since there is no transport coefficient or thermodynamic observable that is a topological invariant of 1D superconductors.
Instead, the existence of a topological phase is imprinted on the measurable properties of the system in a manner that can only be identified through an elaborate measurement and analysis procedure such as the TGP or the even more elaborate procedures necessary for fusion and braiding.
Thus, it is of paramount importance that the TGP has been validated by applying it to simulated transport data, especially since the tunneling spectroscopy and transport measurements comprising the TGP do not measure a topological invariant directly.

In simulated devices, we know whether there is a topological phase since we can compute a topological invariant.
Hence, we tested the TGP on transport data from simulated devices by comparing its output to this topological invariant.
We emphasize that we have \textit{not} attempted to establish qualitative similarities between simulated and measured conductance plots and this is not the purpose of these simulations.
The goal is to see if the TGP correctly distinguishes between regions with trivial and non-trivial topological invariant.

We simulated hundreds of devices with different disorder levels and concluded that if a device passes the TGP, then the probability that the candidate region in the phase diagram is not topological is $< 8\%$ at the 95\% confidence level.
The TGP thereby distinguishes MZMs from trivial Andreev bound states and determines whether topological superconductivity is present in the parameter range scanned in a data set.
Having thus confirmed the reliability of the TGP on simulated data, we formulate the central question of this paper: \textit{can we fabricate and measure devices that pass the TGP?}

We answer this question in the affirmative by presenting data from four devices, named A, B, C, and D, that have passed this protocol with respective maximum topological gaps ranging between 20-\SI{60}{\micro\eV}.
As we explain in more detail in \Cref{sec:device_design}, our devices are based on heterostructures combining indium arsenide (InAs) and aluminum (Al).
The superconducting component is an Al strip, epitaxially-grown on the semiconductor so that it induces superconductivity via the proximity effect.
The semiconducting portion is a shallow InAs quantum well hosting a two-dimensional electron gas (2DEG) that has been depleted by electrostatic gates, except for a narrow conducting wire that remains underneath the aluminum strip.
Within this suite of components, we have used simulations to optimize the material stack and the device geometry with respect to the topological gap.

Disorder is the principal obstacle to realizing a topological phase supporting MZMs.
We have used simulations to predict the (design-dependent) disorder level that the topological phase can tolerate.
These simulations incorporate self-consistent electrostatics, the orbital effect of the magnetic field, and realistic semiconductor-superconductor coupling; see Refs.~\onlinecite{Antipov18, Winkler19, Vaitieknas20} for more details.
Consequently, they show both qualitatively and quantitatively how device design can impact the effective disorder strength.
Many of the resulting specifications are quite demanding, including: (1) higher mobility ($> 60,000\,$cm$^2/$V$\cdot$s) than previously achieved in shallow InAs quantum wells and (2) gate-defined wires that are sufficiently narrow ($< 120\,$nm) as to enable tuning into the single sub-band regime.

Our simulations indicate that mesoscopic fluctuations are important in our \SI{3}{\micro\meter}-long ``topological gap devices'' based on InAs-Al heterostructures, see \Cref{fig:device_SLG}.
Thus, even devices with the same average disorder level can have different TGP outcomes: some disorder realizations will pass while others fail.
The disorder strength determines an expected yield for passing the TGP which is between 0\% and 100\% over a range of disorder levels.
As expected from these simulation results, we have also measured devices that were similar to devices A-D but did not pass the TGP, and we report on data from two of them, which have been named devices E and F.

In summary, each of devices A-D has a high probability of being in the topological phase.
To the best of our knowledge, these devices are the first to have passed as stringent a set of requirements as those encompassed by the TGP, namely (a) concurrent ZBPs that are stable both with respect to changes of the junction parameters and also with respect to changes of the bulk parameters that are larger (in appropriate units) than the bulk gap; and (b) a bulk gap closing and re-opening in response to an increasing magnetic field that is visible in the non-local conductance, indicating a quantum phase transition into a phase with correlated ZBPs.

\section{Topological gap device design and requirements}
\label{sec:device_design}

\subsection{Proximitized semiconductor nanowire model and its topological phase diagram}
\label{sec:proximitized_nanowire_model}

\begin{figure}
\includegraphics[width=8.7cm]{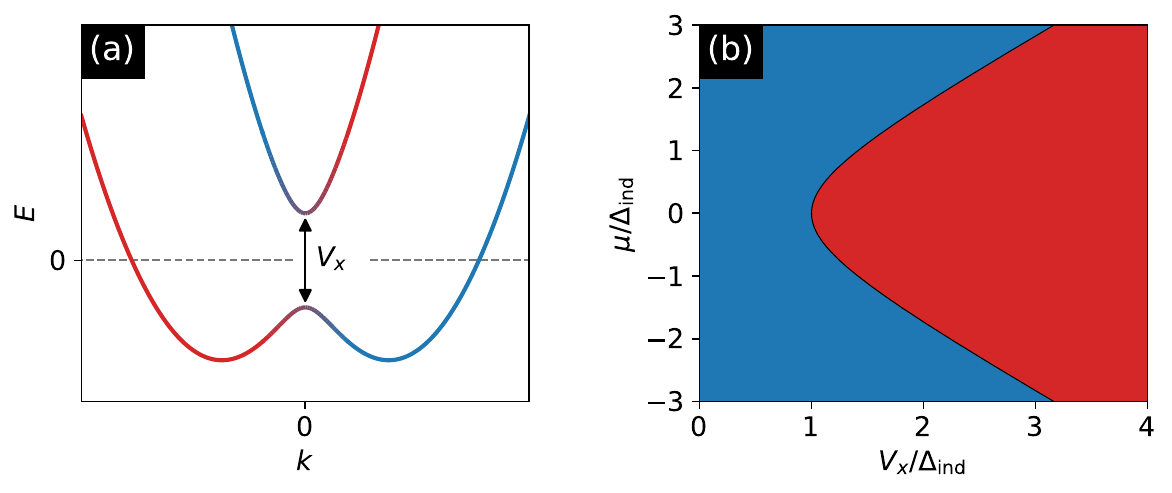}
\vskip -3mm
\caption{
(a)~Energy spectrum as a function of momentum $k$ along the nanowire.
Rashba spin-orbit coupling splits the parabola at the bottom of the band into two parabolas displaced in $k$.
The Zeeman energy $V_x$ is proportional to the external magnetic field perpendicular to the direction of the Rashba spin-orbit coupling.
A non-zero Zeeman energy $V_x$ opens a gap at $k=0$.
When the chemical potential $\mu$ is within the gap, the nanowire Hamiltonian \Cref{eq:H0} has only one pair of Fermi points with spin and momentum locked.
The color gradient represents the change of spin orientation with momentum.
(b)~The topological phase diagram as a function of Zeeman energy $V_x$ and chemical potential $\mu$.
$\DeltaInd$ is the superconducting gap induced in the semiconductor nanowire.
The low-field (blue) phase is a trivial superconductor while the high-field (red) phase is a topological superconductor that supports MZMs at the opposite ends of the nanowire.
}
\label{fig:energy_spectrum_topophase}
\end{figure}

In this section we briefly review the proximitized nanowire model~\cite{Lutchyn10, Oreg10} which supports topological superconductivity over a range of densities and magnetic fields.
The minimal model is comprised of a semiconductor nanowire with Rashba spin-orbit interaction coupled to a conventional ($s$-wave) superconductor.
The effective Hamiltonian for such a system is:
\begin{align}
    & H = H_\SM + \DeltaInd O_\SC, \label{eq:H0} \\
    & H_\SM \!=\! \int\limits_0^L \!dx \, 
    \psi_{\sigma}^\dag(x) \Bigl(
        \!-\frac{\partial_x^2}{2m^*}\!
        -\!\mu\!
        +\!i\alpha \hat{\sigma}_y \partial_x\!
        +\!V_x \hat{\sigma}_x\!
    \Bigr)_{\!\sigma\sigma'}
    \!\psi_{\sigma'}(x), \nonumber \\
    & O_\SC \!=\! \int\limits_0^L \!dx \, \bigl(
        \psi_\uparrow^\dag(x) \psi^\dag_\downarrow(x) 
        + \mathrm{h.c.} 
    \bigr).
    \nonumber
\end{align}
Here, ``SM'' and ``SC'' are abbreviations for, respectively, semiconductor and superconductor, $m^*$, $\mu$ and $\alpha$ are the effective mass, chemical potential, and Rashba spin-orbit coupling, respectively.
$V_x$ is the Zeeman splitting due to the applied magnetic field $B$ along the nanowire: $V_x\!=\!g_\SM \muB B/2$, where $g_\SM$ and $\muB$ are, respectively, the Land\'e $g$-factor and Bohr magneton.
The proximity to the $s$-wave superconductor is effectively described by the pairing operator $O_\SC$, while $\DeltaInd$ is the induced pairing potential.

The zero-temperature phase diagram of the proximitized nanowire Hamiltonian of \Cref{eq:H0} consists of a trivial ($s$-wave-like) phase and a topological phase, as shown in \Cref{fig:energy_spectrum_topophase}.
The latter supports MZMs at the opposite ends of the nanowire and is in the same phase as a spinless $p$-wave superconductor~\cite{Kitaev01}.
The trivial and topological phases are separated by a quantum phase transition at $V_x = \sqrt{\mu^2 + |\DeltaInd|^2}$ which is necessarily accompanied by the closing of the bulk gap.
The stability of a topological phase is characterized by its bulk transport gap or, equivalently, the gap to extended excited states, which we call the topological gap $\DeltaT$.
In the idealized case of \Cref{eq:H0}, this is simply the bulk gap.
This phase has been proposed to occur in quasi-one-dimensional systems composed of chains of magnetic atoms on the surface of a superconductor~\cite{Nadj-Perge13, Klinovaja13, Braunecker13, Pientka13, Nadj-Perge14}; in nanowires that are completely encircled by a superconducting shell in which the order parameter winds around the wire due to the orbital effect of the magnetic field~\cite{Cook11, Hosur11, Vaitieknas20}; and in the vortex cores of three-dimensional superconductors
\cite{Wang18, Kong19}.
The corresponding two-dimensional topological superconducting state can occur in $p + ip$ superconductors~\cite{Read00}, at the surface of a topological insulator~\cite{Fu08, Fu09, Hasan10}, in ferromagnetic insulator-semiconductor-superconductor heterostructures~\cite{Sau10a, Alicea10, Sau10b, Chung11, Duckheim11, Potter12, Lutchyn18}, and in $s$-wave superfluids of ultra-cold fermionic atoms~\cite{Sato09b, Zhang08}.

The model discussed so far neglects many of the ingredients of actual devices, such as additional sub-bands and the orbital effect of the magnetic field.
To address this, we have developed realistic 3D simulations that take these effects into account.
These simulations include self-consistent electrostatics, orbital magnetic field contributions, and renormalization effects due to coupling to the superconductor~\cite{Vuik16, Mikkelsen18, Winkler17, Winkler19, Antipov18, Nijholt16}.
We have validated these simulations through comparison with ARPES~\cite{Schuwalow21}, THz spectroscopy~\cite{Chauhan22}, the Hall bar measurements reported in \Cref{sec:electrostatic_calibration}, and transport through multiple types of previous devices involving proximitized semiconductor nanowires~\cite{Hart19, Vaitieknas20, Kringhoj20, Shen21}.
We also take into account multiple disorder mechanisms such as charged disorder and variations of geometry and composition along the wire length, as discussed in \Cref{sec:disorder_uniformity}.
The superconductor's degrees of freedom are integrated out, yielding a formulation in which it is encapsulated by self-energy boundary conditions \cite{Vaitieknas20, Kringhoj20, Shen21}.
Using this advanced simulation model, we optimized the design for gate-defined devices based on high-quality 2DEG heterostructures in order to minimize the effects of disorder, additional sub-bands, and the orbital effect of the magnetic field.
This design is presented in the next subsection.
We extract the parameters of a minimal model projected to the lowest sub-band (neglecting couplings to higher subbands which are suppressed by large sub-band level spacing) in \Cref{sec:projected_model}.
Our minimal model is similar to \Cref{eq:H0}.
The parameters that define this effective single sub-band model are listed in \Cref{tab:effective_parameters}.
This projected model and the full 3D model show good agreement for bulk quantities in the field and density ranges of interest.
In order to simulate transport properties, we add a realistic description of the junctions (junction design is described in the following section).
We perform these simulations by projecting the full 3D model of our device to the low-energy subspace.
The corresponding results are presented in \Cref{sec:TGP}.
After we have discussed the device design, we describe the general effects of disorder in mesoscopic topological wires, then quantify the effective disorder potential in our devices.

\begin{figure*}
\includegraphics[width=18.0cm]{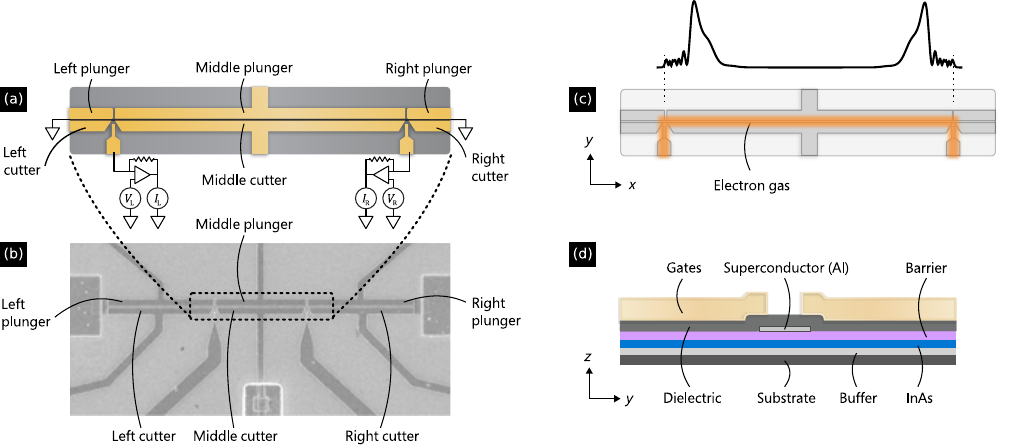}
\vskip -2mm
\caption{
(a)~A schematic of the top view of the active area of a single-layer gate (SLG) topological gap device.
All of the labelled gates serve to deplete the 2DEG in the InAs quantum well to define a high-quality one-dimensional conducting channel.
The left, middle, and right plungers also tune the density in the corresponding sections of the device, while the left and right cutters also open and close the junctions.
The two unlabelled gates are the ``helper gates'' which are used to control the electron density in the junctions and leads, the latter of which are connected to a measurement circuit as shown in this panel.
(b)~An SEM image of a topological gap device.
The dashed line indicates the active region depicted in (a).
(c)~Region of non-zero electron density (orange) in the InAs quantum well when the device is tuned to the operating regime: the middle section (underneath the middle cutter/plunger) is tuned to the topological regime while the outer sections (underneath the left/right cutter/plunger gates) are tuned to the trivial phase using the plunger gates.
The black curve shows the local density of states in the wire near zero energy, computed in the ideal disorder-free limit.
(d)~A schematic of the cross-section of an SLG topological gap device, in which the Al strip induces proximity superconductivity in the one-dimensional InAs nanowire that is defined by the gates shown in panel (a) and extends perpendicular to this cross-sectional view.
The $x$-, $y$-, and $z$-directions are indicated in panels (c,d).
}
\label{fig:device_SLG}
\end{figure*}

\begin{figure}
\includegraphics[width=8.65cm]{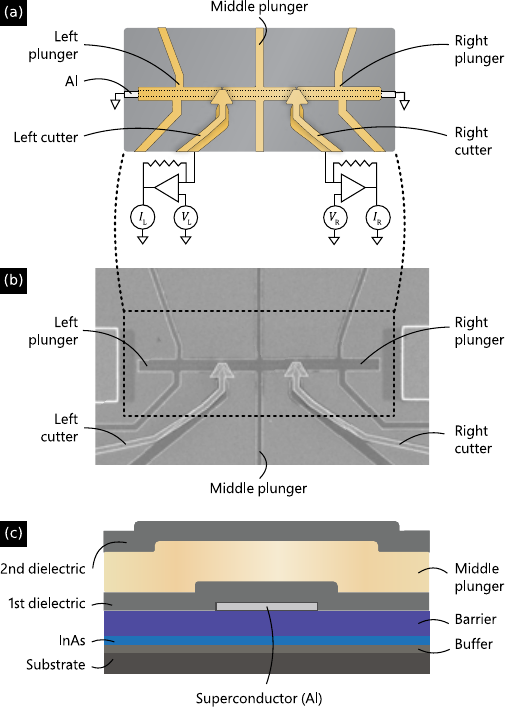}
\vskip -2mm
\caption{
(a)~A schematic of the top view of the active area of a dual-layer gate (DLG) topological gap device.
The plunger gates serve to deplete the 2DEG in the InAs quantum well to define a high-quality one-dimensional conducting channel and to tune the density in the corresponding sections of the device.
The left and right cutters open and close the junctions.
The two unlabelled gates are the ``helper gates'' which are used to increase the electron density in the junctions and leads, the latter of which are connected to a measurement circuit as shown in this panel.
(b)~An SEM image of a DLG topological gap device.
The dashed line indicates the active region depicted in (a).
(c)~A schematic of the cross-section of the middle section of a DLG topological gap device, in which the Al strip induces proximity superconductivity in the one-dimensional InAs nanowire that is defined by the gates shown in panel (a) and extends perpendicular to this cross-sectional view.
The bulk of the wire doesn't have a second gate layer, but the junctions have cutter gates in a second gate layer, and the $2^\text{nd}$ dielectric layer separates them from the plunger gates in the first gate layer.
}
\label{fig:device_DLG}
\end{figure}

\subsection{Gate-defined proximitized nanowire}
\label{sec:layout}

Our devices are defined by an Al strip separated from an InAs quantum well by a barrier layer.
There are two designs which are conceptually similar but have some practical differences.
One has a single-layer gate (SLG) design while the other has a dual-layer gate (DLG) design, shown in \Cref{fig:device_SLG} and \Cref{fig:device_DLG}, respectively.
We will refer to both designs as ``topological gap devices.'' 
A cross-section of an SLG device is shown in \Cref{fig:device_SLG}(d), where the Al strip is light grey.
The strip's dimensions have been optimized using the simulations described above: length \SI{10}{\micro\meter}, width $< 120\,$nm, thickness $< 10\,$nm.

The length is in the direction perpendicular to the cross-section in \Cref{fig:device_SLG}(d).
The Al strip is covered by a several nm thick top oxide formed by controlled oxidation [not shown in \Cref{fig:device_SLG}(d) or \Cref{fig:device_DLG}(c)].
The Al strip features larger Al pads at each end of its \SI{10}{\micro\meter} length, which can be seen at the right and left edges of the scanning electron micrograph (SEM) images in \Cref{fig:device_SLG}(b) and \Cref{fig:device_DLG}(b).
The pads are contacted with Ti/Au or Ti/Al Ohmic leads, by which the Al strip is grounded.
(Both types of contacts are normal in the typical operating regime.)
We will denote the direction perpendicular to the surface of the quantum well as the $z$-direction, while the directions along and perpendicular to the Al strip are the $x$- and $y$-direction, respectively, as shown in \Cref{fig:device_SLG}(c,d).

There is a dielectric layer that separates the superconductor-semiconductor heterostructure from the electrostatic gates that are at the top of the cross-section in \Cref{fig:device_SLG}(d) and \Cref{fig:device_DLG}(c).
The gates deplete the 2DEG except underneath the Al, which partially screens their electric fields, thereby creating a high-quality nanowire.
The top view in \Cref{fig:device_SLG}(a) and the SEM image in \Cref{fig:device_SLG}(b) show that the split-gate structure of the SLG design
is divided into three sections: three plunger gates and three cutter gates.
The three plunger gates serve to deplete the 2DEG on their side of the Al strip while the three cutter gates deplete the 2DEG on the other side.
Once the 2DEG has been depleted, operating the plunger gates at even more negative voltages tunes the density underneath the Al via the fringe electric fields that remain after screening by the Al.
The densities in the left, middle, and right sections can be controlled independently by the three plunger gates.
We operate in the low-density limit in which only the lowest $z$-direction sub-band is occupied so, here and henceforth, will use the term ``sub-band'' for $y$-direction sub-bands.
The left and right plungers control the densities underneath the corresponding sections of the Al, which are normally set for full depletion (no occupied sub-bands) underneath the Al.
The width of the Al strip was chosen to enable this for moderate gate voltages $\Vdep > -3\,$V and also to minimize the orbital effects of a magnetic field in the $x$-direction.

There are two side tunnel junctions at the boundaries between the middle cutter gate and the left/right cutter gates, enabling the $3$-terminal measurements \cite{Rosdahl18, Menard20, Puglia21, Poeschl22, Banerjee22b} of the conductance matrix that are necessary for the TGP, as we discuss in \Cref{sec:TGP}.
In addition to depleting the 2DEG on the opposite side of the Al strip from the plungers, the left and right cutter gate voltages $\Vlc$ and $\Vrc$ are also used to vary, respectively, the transparency of the left and right tunnel junctions.
The split-gate geometry with plunger-cutter pairs ensures independent tuning of density and junction transparency for each section of the gate-defined nanowire.
The two junctions are typically tuned into the tunneling regime in which the above-gap low-temperature differential tunneling conductance $\Gag$ is $\lesssim e^2/h$, while the Al strip is grounded.
The junctions are connected to Ohmic contacts via conducting paths in the 2DEG.
There are two ``helper'' gates, which are the unlabelled gates at the bottom of \Cref{fig:device_SLG}(a); they extend from the junctions to the bottom edge of the SEM in \Cref{fig:device_SLG}(b).
The helper gates define these conducting paths by accumulating carrier density in the 2DEG underneath them and keeping it conducting.
The orange region in \Cref{fig:device_SLG}(c) shows where the electron density is non-zero in the 2DEG in the device's normal operating regime: underneath the middle section of the Al strip and underneath the helper gates.

In the DLG design, instead of a split-gate geometry, the plunger gates cover the Al strip completely, as illustrated schematically in \Cref{fig:device_DLG}a and in an SEM image shown in \Cref{fig:device_DLG}(b).
This makes it considerably easier to align the gates with the Al strip.
Moreover, the plunger gates have a single role, which is to control the electron density in the 2DEG~--- to fully deplete it underneath the regions adjacent to the Al strip and to either fully deplete it or to tune it to the lowest sub-band directly underneath the Al strip.
The function of controlling the bulk density is separated from the function of opening and closing the junctions, which is accomplished by cutter gates that are in a second gate layer, separated from the first gate layer by a second dielectric layer.
The cutter gates in the DLG design only cover the junctions, so they do not affect the bulk density in the wire.
Although the above differences between the SLG and DLG designs are practically important, the basic principles and length scales are the same in both.

We will call the semiconductor underneath the middle section ``the wire,'' and the superconducting gap that is induced in the wire at $B=0$ via the proximity effect the ``induced gap'' $\DeltaInd$.
We denote the middle plunger gate voltage by $\Vp$, which tunes the density in the wire.
At the optimal operation point, the wire is tuned to the single-sub-band regime that occurs just before full depletion $\Vp \gtrsim \Vdep$.
We will focus on the phase diagram of the wire as a function of the middle plunger gate voltage $\Vp$ and the magnetic field $B$.

We comment briefly on the length of the wire here and discuss it in greater detail in \Cref{sec:length_scales}.
To operate the device in the optimal regime, the nanowire should be much longer than the coherence length in the topological superconducting state, so that MZMs are well localized at the opposite ends of the nanowire [the situation depicted in \Cref{fig:device_SLG}(c)].
In this case, MZMs would lead to ZBPs that are stable with respect to local perturbations.
When the coherence length is comparable to or larger than the nanowire length, a ZBP at one end of the wire may arise from an Andreev state extending from the opposite end~\cite{Hess21}.
In this case, however, we do not expect ZBPs to be stable with respect to local perturbations.
Our simulations suggest that, for these designs and material stacks, the coherence length in the topological state varies between 100-250\,nm in the absence of disorder.
The wire is designed to be much longer than this.
Disorder in the bulk of the nanowire suppresses the topological gap and increases the coherence length which, as we discuss in \Cref{sec:length_scales}, leads to a non-trivial requirement for the wire length which depends on the stack geometry/composition and disorder level.
On the other hand, the wire cannot be too long since the visibility of gap closings will be strongly suppressed if the length of the wire is more than several times the normal-state localization length~\cite{Rosdahl18}.

Assuming weak to moderate disorder, the optimal wire length in our devices is \SI{3}{\micro\meter}.
This length choice also ensures that when a transport gap closing is observed, there is a non-zero density of states in the bulk at zero energy which has non-vanishing matrix elements to both leads so that non-local conductance is above the noise floor~\cite{Puglia21}.

Finally, the outer sections (underneath the left/right cutter/plunger gates) must be significantly longer than the coherence length of the parent superconductor in order to prevent quasiparticle transport below the parent gap at full depletion.

\subsection{Material stack}
\label{sec:material_stack}

The material stack of the topological gap device is optimized to produce a large topological gap.
To achieve a topological phase, the semiconductor stack needs to produce a large spin-orbit coupling and a large nominal $g$-factor in the confined 2DEG.
In addition, the heterostructure should provide a low disorder environment, typically parameterized by high 2DEG mobility at low temperatures.
Given the lack of suitable insulating and lattice-matched substrates, the active region is grown on an InP substrate employing a graded buffer layer to accommodate lattice mismatch.

The active region consists of the Al superconductor, an upper barrier, the InAs quantum well, and the buffer.
The upper barrier layer plays a critical role in fine-tuning the coupling between the superconductor and the 2DEG residing in the quantum well.
To drive the device into the topological phase, $B$ needs to be increased until the Zeeman energy $\muB |g^{\star}| B / 2$ exceeds the induced gap $\DeltaInd$.
Here, $g^{\star}$ is the renormalized $g$-factor in the superconductor-semiconductor heterostructure, which is given by $g^{\star}=g_\mathrm{SM}\,\Delta/(\Gamma+\Delta)$ if we neglect the $g$-factor of aluminum; a more general form of the renormalization factor is discussed in \Cref{sec:projected_model}.
For strong coupling $\Gamma$ between the wire and the Al strip, $\DeltaInd$ would approach the gap in the Al strip $\Delta_\mathrm{Al}$ and the electronic wavefunction of the single occupied sub-band of the wire would have large weight in the Al strip.
In this case, $|g^{\star}|$ would be renormalized to small values.
In such a case, $\muB |g^{\star}| B / 2$ would not approach $\DeltaInd$ until the magnetic field is very large ($>2.5\,$T), close to the critical in-plane field of the Al strip~\cite{Stanescu11, Doru17, Antipov18, Reeg18a, Cole15}.
Conversely, if the coupling between the superconductor and semiconductor were too weak, the maximum attainable topological gap would be small, since it is bounded above by $\DeltaInd$.
Hence, the material stack must satisfy $\kB T \ll \DeltaInd < \Delta_\mathrm{Al}$.

As we shall see in \Cref{sec:experimental_data}, the parent gap in the Al strip is $\Delta_\mathrm{Al} \approx \SI{300}{\micro\eV}$ (this is strongly dependent on the Al thickness).
For an optimized heterostructure, according to our simulations of the device of \Cref{fig:device_SLG}(a), we expect $\SI{100}{\micro\eV} < \DeltaInd < \SI{200}{\micro\eV}$, corresponding to an induced gap to parent gap ratio of $0.33 < \DeltaInd / \Delta_\mathrm{Al} < 0.67$, and $4 < |g^{\star}| < 7$.

Another function of the upper barrier layer is to separate the quantum well states from disorder on the dielectric-covered surface of the stack, thus enhancing the electron mobility.
The quantum well thickness is chosen to minimize orbital effects from the magnetic field applied in the $x$-direction, to allow electrostatic tuning, and to retain the desirable properties of InAs, including optimally renormalized $g^{\star}$.

Rashba spin-orbit coupling in the wire, characterized by the parameter $\alpha$, enables superconductivity to co-exist with the magnetic field $B$.
Although $\alpha$ does not determine the critical field for the transition into the topological phase, it does contribute to the size of the topological gap and the extent of the topological phase in parameter space.
The spin-orbit coupling in a 2DEG heterostructure covered with the superconductor is difficult to measure directly.
Using weak anti-localization measurements in shallow InAs 2DEGs, see, for example, Ref.~\onlinecite{Shabani16}, and typical values of the electric field (obtained from simulations assuming band offset parameter measured in Ref.~\onlinecite{Schuwalow21}), we estimate that the Rashba spin-orbit coupling is in the range of $5$ to $15\,\mathrm{meV}{\cdot}\mathrm{nm}$.

In this paper, we present the results of measurements and simulations of devices based on four different material stacks satisfying the requirements given in this subsection.
While they all feature an InAs quantum well, there are important differences in the quantum well width, barrier composition and thickness, and dielectric.
In \Cref{tab:effective_parameters}, we give the
effective parameters that encapsulate the effect of these materials changes, such as the the effective mass, $g$-factor, and spin-orbit coupling.
We will call these different materials stacks $\beta$, $\delta$, $\delta'$, and $\varepsilon$.

\begin{figure}
\includegraphics[width=8.7cm]{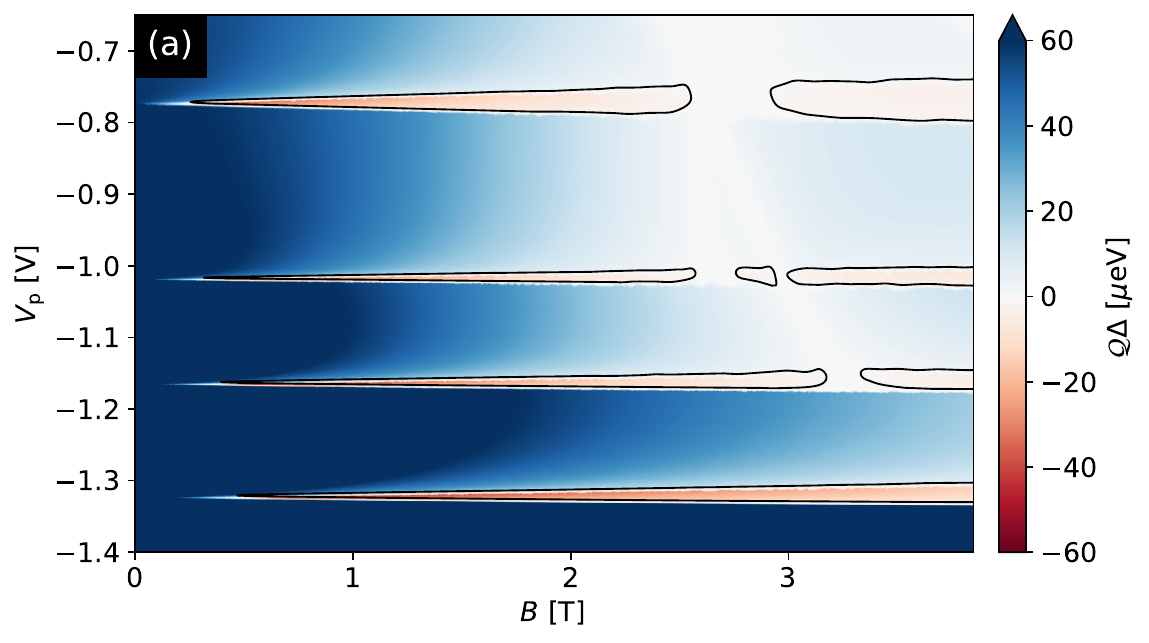}
\includegraphics[width=8.6cm]{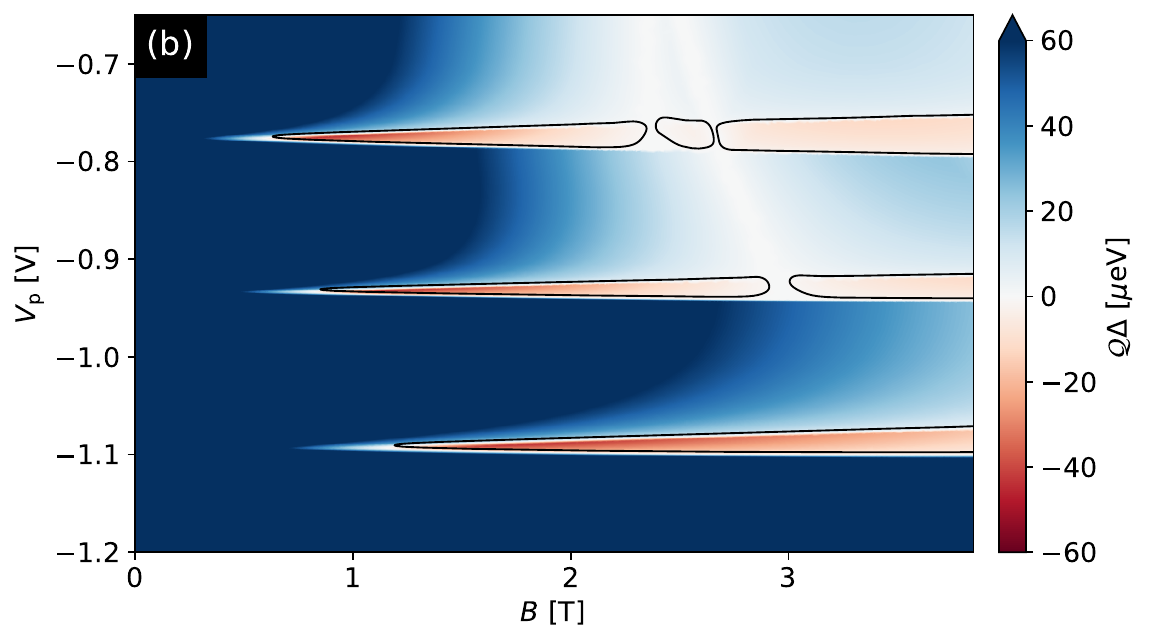}
\vskip -3mm
\caption{
(a)~The simulated phase diagram of the SLG device design shown in \Cref{fig:device_SLG}(d) and the $\beta$ material stack in the ideal disorder-free limit.
(b)~The simulated phase diagram of the DLG device design shown in \Cref{fig:device_DLG}(c) and the $\varepsilon$ material stack in the ideal disorder-free limit.
Here, the Pfaffian invariant $\mathcal{Q} = +1$ in the trivial phase and $\mathcal{Q} = -1$ in the topological phase.
Hence, the color scale indicates the size of the gap in the trivial (blue) and topological (red) phases.
The black curves indicate the phase transition where the topological invariant changes sign.
The axes $B$ and $\Vp$, the magnetic field and plunger gate voltage, respectively, are the actual control parameters of the device.
Most of the phase diagrams in this paper will similarly be in the $(B,\Vp)$ plane.
This phase diagram is for a wire of infinite length.
The maximum topological gap in the lowest sub-band is approximately \SI{50}{\micro\eV}.
The data and scripts required to reproduce this and other simulated figures are available in Ref.~\onlinecite{code_and_data}.
}
\label{fig:simulated_infinite_clean_gap}
\end{figure}

\subsection{Phase diagram of ideal devices}
\label{sec:clean_device}

For the SLG and DLG device designs described in \Cref{sec:layout} and the material stacks described in \Cref{sec:material_stack}, we have computed the phase diagrams in the ideal disorder-free limit as a function of the actual control parameters of the device, $\Vp$ and $B$.
This is to be contrasted with \Cref{eq:H0} and \Cref{fig:energy_spectrum_topophase}, which contain the effective parameters $\mu$ and $V_x$.
The bare spin-orbit coupling in the semiconductor is taken to be $\alpha_0 = 10\,\mathrm{meV}\cdot\mathrm{nm}$ in both the SLG-$\beta$ and DLG-$\varepsilon$ designs.
The color scheme in \Cref{fig:simulated_infinite_clean_gap} is determined by the Pfaffian invariant \cite{Kitaev01, Wimmer12} (see \Cref{sec:topological_invariants} for a brief description of this invariant).
Darker red corresponds to larger topological gap and darker blue corresponds to larger trivial superconducting gap, as indicated by the color scale on the right-hand-side of the figure.

The red parabola in \Cref{fig:energy_spectrum_topophase} has now become a sequence of red slivers in the ideal phase diagrams of an SLG device built on the $\beta$ stack in \Cref{fig:simulated_infinite_clean_gap}(a) and a DLG device built on the $\varepsilon$ stack in \Cref{fig:simulated_infinite_clean_gap}(b).
The red slivers are topological phases with different numbers of occupied 1D sub-bands in the wire \cite{Lutchyn11, Stanescu11}.
When we zoom in on any one of these slivers, we see that it has the parabolic lobe-like shape $|g^{\star}| \muB B/2 > \sqrt{\mu^2 + \DeltaInd^2}$ that follows from \Cref{eq:H0}.
Here, the single-sub-band topological phase is at $\Vp \approx -1.35\,$V, and it has a larger topological gap than when there are more occupied sub-bands.
Recall that one of the design criteria was that the single-sub-band regime could be reached for moderate gate voltages; this figure confirms that it is satisfied by this design.
As we increase $\Vp$, thereby increasing the number of occupied sub-bands, the effective cross-sectional area of the gate-defined nanowire increases and, at some point, the orbital effect of the applied magnetic field becomes very important.
In the second sub-band, an orbital-field-induced gap closing is visible at $B \approx 3\,$T and $\Vp \gtrsim -1.2\,$V.
It occurs at $B \approx 2.5\,$T in the third sub-band and at lower fields in higher sub-bands.
In contrast, in the lowest sub-band, an orbital-field-induced gap closing does not occur over the relevant field range.
(At fields higher than $3.5\,$T, the Al parent gap can close, so an orbital-field-induced gap closing would be a sub-leading effect anyway.) Thus, in order to maximize both the accessible volume of the topological phase and its maximum gap, it is necessary to tune the device into the single-sub-band regime.

There is very little difference between the SLG and DLG designs in the bulk of the wire; the principle difference is in the junctions, which have no effect on the ideal bulk phase diagram.
However, the $\varepsilon$ stack has larger $\DeltaInd$ and smaller $g_\SM$ so the topological phase occurs at higher $B$ for this stack.
Hence, the DLG-$\varepsilon$ phase diagram in the clean limit has a lowest sub-band topological phase that is pushed to higher fields, as may be seen in \Cref{fig:simulated_infinite_clean_gap}.

Within the lowest sub-band, the effective mass $m^*$, effective Rashba spin-orbit coupling $\alpha^*$, effective $g$-factor $g^*$, superconductor-semiconductor coupling $\Gamma$, and lever arm $d\mu/d\Vp$ take the values given in \Cref{tab:effective_parameters}.
As a result of the projection to the lowest sub-band, the bare Rashba spin-orbit coupling $\alpha_0$ is replaced by the effective parameter $\alpha$ given in the table.
The precise definition of the effective single-band model governed by these parameters is given in \Cref{sec:projected_model}.

\begin{table}[t!]
\begin{center}
\begin{tabularx}{\columnwidth}{|I|C|W|C|C|W|}
\cline{1-6}
\shortstack[l]{\noalign{\vskip 1.0ex} Design, \\ stack} &
\shortstack[c]{$m^*$ \\ $[m_\mathrm{e}]$} &
\shortstack[c]{$\alpha^*$ \\ $[\mathrm{meV}{\cdot}\mathrm{nm}]$} &
\shortstack[c]{$g^*$ \\ $\phantom{1}$} &
\shortstack[c]{$\Gamma$ \\ $[\mathrm{meV}]$} &
\shortstack[c]{$d\mu/d\Vp$ \\ $[\mathrm{meV/V}]$} \\ [0.5ex]
\cline{1-6}
SLG-$\beta$ & 0.032 & 8.7 & $-11.8$ & 0.13 & 85 \\
\cline{1-6}
DLG-$\delta$ & 0.032 & 8.4 & $-11.5$ & 0.21 & 79 \\
\cline{1-6}
DLG-$\varepsilon$ & 0.032 & 8.3 & $-11.4$ & 0.32 & 78 \\
\cline{1-6}
\end{tabularx}
\end{center}
\vskip -3mm
\caption{
Single-band effective model parameters obtained for various device designs.
The $\delta'$-stack is not simulated.
It has similar effective parameters to the $\delta$-stack, but differs in lever arm.
}
\label{tab:effective_parameters}
\end{table}

\subsection{Disorder and uniformity requirements}
\label{sec:disorder_uniformity}

We now discuss the level of imperfection that our device designs can tolerate and still have a topological phase with coherence length $\xi(0)$ shorter than the wire length $L=\SI{3}{\micro\meter}$.
See \Cref{sec:length_scales} for a discussion of the coherence length $\xi(0)$ and other important length scales.

In our devices, there are many different sources of disorder, including geometric and charged disorder~\cite{Stanescu11, Lutchyn18, Woods21}.
Even small local variations in any of a number of device parameters can cause significant variations in the potential experienced by the electrons along the wire.
As we discuss in \Cref{sec:projected_model}, we can extend the single-sub-band effective model \Cref{eq:projected_nanowire} parameterized by the couplings given in \Cref{tab:effective_parameters} to include disorder, leading to the Hamiltonian \Cref{eq:nanowire_with_disorder}.
When the various disorder mechanisms are projected into this single-sub-band model, most of them can be characterized by the quenched Gaussian disorder model~\cite{Giamarchi04} in which disorder is represented by a random potential $V(x)$ whose probability distribution is approximately described by the second-order cumulant defined in \Cref{eq:cumulant}.
Both the strength of disorder $\delta V$ and its correlation length $\kappa$ depend on each disorder source in a manner that is highly dependent on the specific design and must be calculated in a full three-dimensional model, as we describe below.
The designs in \Cref{fig:device_SLG} and \Cref{fig:device_DLG} have been optimized to be as forgiving as possible by requiring that the design minimize the projected disorder for fixed microscopic disorder.

Even in such an optimized design, the topological phase is impossible if the disorder strength $\delta V$ exceeds a critical value.
For somewhat smaller disorder strengths, there will be a topological phase, but the coherence length $\xi(0)$ will be very long.
We need still smaller $\delta V$ in order to have a topological phase with $\xi(0)<L$.
Hence, it is essential to understand and minimize the sources of disorder that contribute to $\delta V$.

\begin{figure}
\vskip -1.4cm
\includegraphics[width=8.5cm]{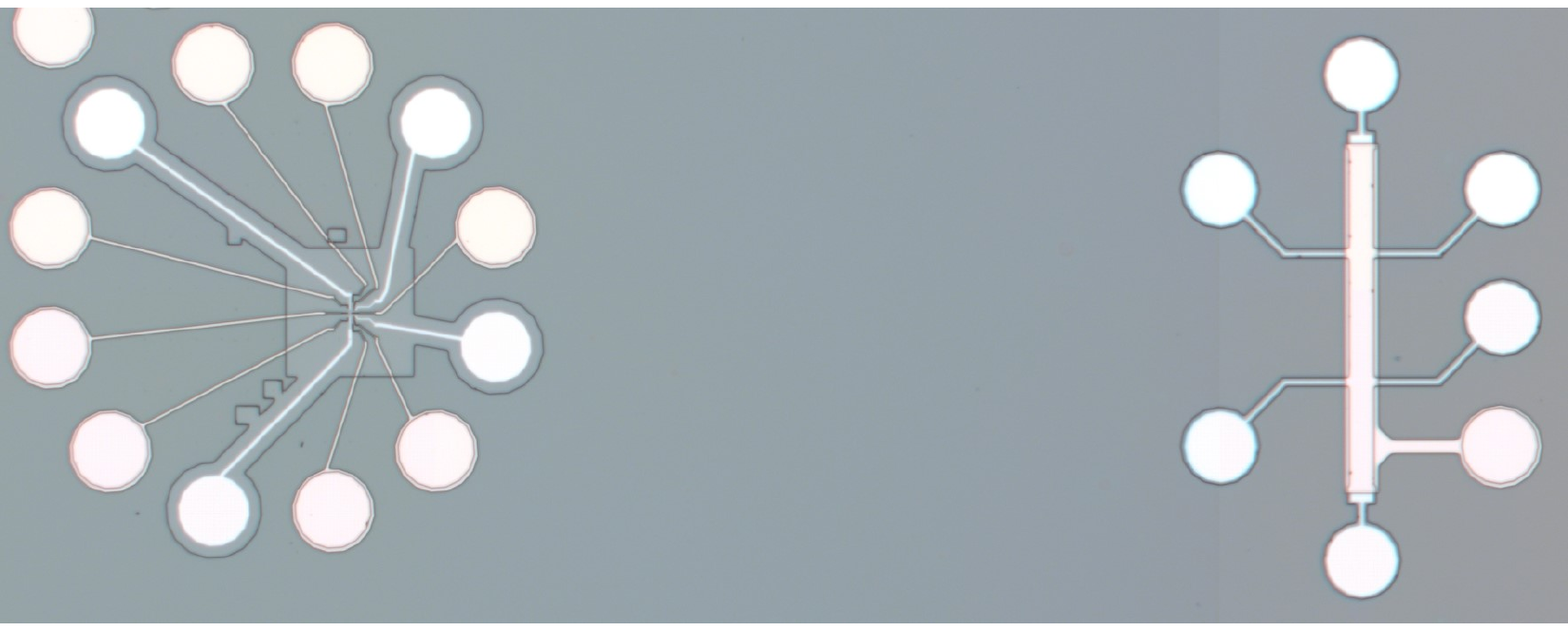}
\vskip -1.6cm
\caption{
There are Hall bars on the same chip as our topological gap devices, visible at the right and left sides, respectively, of this optical image.
Both device types undergo the same processing steps, so the charged defect density at the semiconductor-dielectric interface $\sc$ extracted from Hall mobility measurements is reflective of the semiconductor-dielectric interface in the neighboring topological gap device.
}
\label{fig:hall_bar_chip}
\end{figure}

In the regime of interest~--- the low-density regime with single sub-band occupancy~--- charged disorder dominates~\cite{Boutin22}.
From an analysis of the density-dependence of the mobility of Hall bars, we conclude that charged disorder is located primarily at the interface between the semiconductor surface and the gate dielectric.
Hall bar measurements allow us to extract the average density of charged imperfections at the semiconductor-dielectric interface, denoted by $\sc$, and the lever arm $d\mu/d\Vp$.
This is illustrated in \Cref{sec:electrostatic_calibration}.
Each chip studied in this paper has both topological gap devices and Hall bars, as shown in \Cref{fig:hall_bar_chip}, enabling us to extract the average density of charged imperfections for each chip and to assess the impact on topological gap devices of chip-to-chip changes in the disorder level.
\textit{Any impact that post-growth fabrication has on the semiconductor-dielectric interface in a topological gap device will be present in its partner Hall bar as well since they are processed together on the same chip.}
If any fabrication processes increase the density of charged imperfections in a topological gap device, we will detect this in the corresponding Hall bar.

We have optimized the device geometry with respect to charged imperfections at the semiconductor-dielectric interface by choosing the Al width as wide as possible while still maintaining the ability to tune into the single sub-band regime.
This keeps the active region in the InAs quantum well as far as possible from charged disorder at the interface between the semiconductor and the dielectric [see \Cref{fig:device_SLG}(d)].
(As we discussed in \Cref{sec:material_stack}, the barrier layer plays a similar role in separating charged disorder as much as possible from the active region.)
We use self-consistent electrostatics calculations~\cite{Antipov18, Winkler19} to find the disorder potential underneath the Al.
For realistic densities of charge defects $\sc$, we find the variance of the projected disorder potential and correlation length to vary between $\delta V \approx 0.5$-1.5\,meV and $\kappa\approx 75$-125\,nm, respectively.
In \Cref{fig:disorder_strength_vs_n2D}, we show how $\delta V$ depends on $\sc$ for the SLG and DLG designs of, respectively, \Cref{fig:device_SLG} and \Cref{fig:device_DLG} in the $\beta$, $\delta$, or $\varepsilon$ stacks.

\begin{figure*}
\includegraphics[width=18cm]{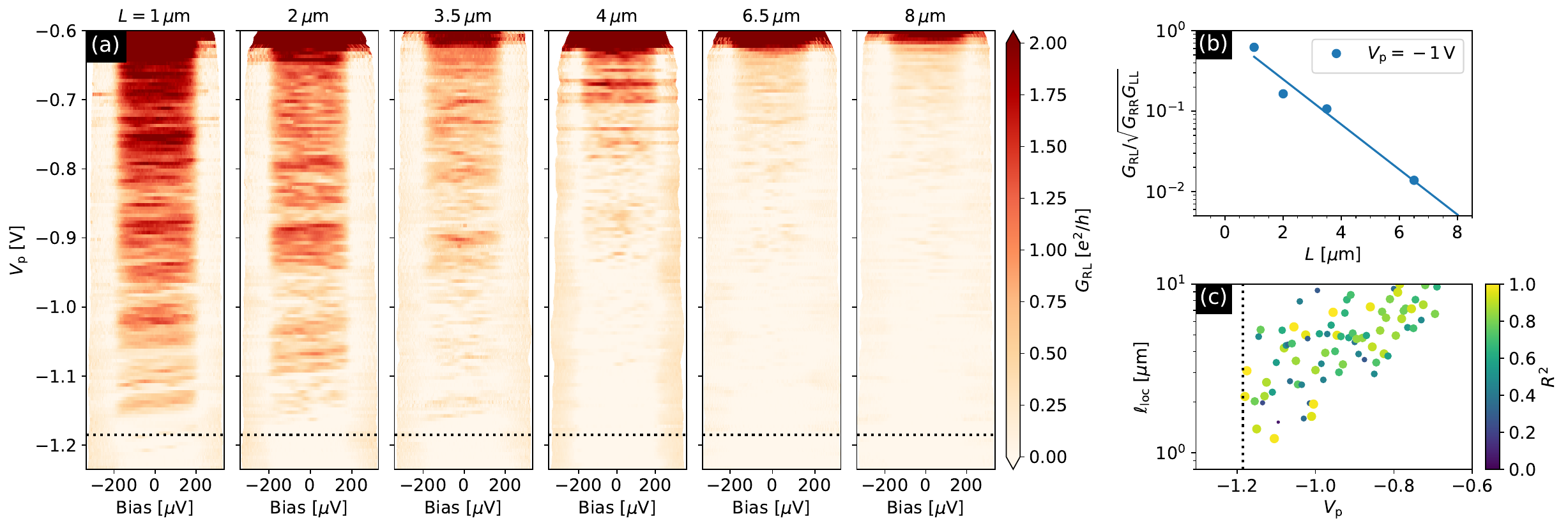}
\vskip -2mm
\caption{
The experimentally-measured non-local conductance $-\GRL$ across sections of lengths $L = 1$, 2, 3.5, 4, 6.5, $\SI{8}{\micro\meter}$ in the same wire (see \Cref{sec:localization_length} for a description of the device).
Panel (a) shows the non-local conductance vs plunger and bias voltage for different length segments.
Increasing the length directly leads to a significantly reduced conductance.
Around $\Vp = -1.185\,$V the conductance of the $\SI{1}{\micro\meter}$ wire becomes larger than 0.05$e^2/h$.
The onset of conductance appearing around this gate voltage for multiple wire lengths suggests this signifies the onset of the first subband (indicated by black dotted lines).
Around $\Vp = -0.6\,$V the conductance becomes very high in all segments because the entire 2DEG becomes conductive.
In panel (b) an example of the localization length extraction is shown.
The conductance is averaged over a small bias window of $\pm \SI{20}{\micro\eV}$ to improve signal quality.
The localization length is then extracted by fitting the data to the expected value of the typical conductance $-\GRL/\sqrt{\GRR \GLL} = A \exp(-2 L/\ellLoc)$~\cite{Beenakker97}.
Fit parameters $\ellLoc$ and $A$ are obtained by the linear fit of $\ln(-\GRL)$ vs $L$ with $R^2$ describing the quality of the fit.
We normalize the non-local conductance $\GRL$ by local conductances $\sqrt{\GRR \GLL}$ at zero bias to minimize the contributions of the local effects.
Panel (c) shows the extracted localization length as a function of plunger.
We note that it is above $\SI{1}{\micro\meter}$ throughout the measured range.
We do not show points above $\SI{10}{\micro\meter}$ because the method cannot reliably determine $\ellLoc$ values greater than $L$.
}
\label{fig:hmp}
\end{figure*}

From a transfer matrix calculation of $\xi(0)$ for the model in \Cref{eq:nanowire_with_disorder}, we can obtain the disorder strength $\delta V$ at which the minimum value of the coherence length $\xi(0)$ begins to exceed our device length.
\Cref{fig:disorder_strength_vs_n2D} enables us to translate that value into a target $\sc$.
In particular we obtain for SLG-$\beta$ parameters that this occurs for $\sc > 3 \cdot \scu$.
The $\delta$ and $\varepsilon$ stacks have slightly different requirements as a result of their stronger coupling to the superconductor, $\Gamma$ (which is still within the required range of ${\DeltaInd}/{\DeltaAl}$).
Hence, an initial target for dielectric quality is $\sc < 3 \cdot \scu$.
In this paper, we show data from devices that are below and above this target.
The topological phase is present in the thermodynamic limit even for relatively high disorder~\cite{Boutin22}, but with large $\xi(0)$, which renders it unusable in an $L = \SI{3}{\micro\meter}$ wire.
The condition that $\xi(0)<L$ is significantly more restrictive.
As we shall see when we consider the case of a single disorder realization in \Cref{sec:pd_single_realization}, the condition that the gap not be too small is also more restrictive.

We estimate that the corresponding bound on the peak mobility (as a function of density) for Hall bar devices fabricated on the same material stack is $\mu_\mathrm{2D} > 60,\!000\,$cm$^2/$V$\cdot$s at electron densities $n_\mathrm{e} \sim 0.6$-$0.8 \cdot \scu$.
The 2DEGs used in this paper have peak mobility in the range 60,000-100,000$\,$cm$^2/$V$\cdot$s in this density range.
Additional details are in \Cref{sec:electrostatic_calibration}.

In a similar fashion, we have optimized the design with respect to other disorder mechanisms including variations of the following parameters along the length of the wire: thickness and dielectric constant of the oxide, barrier thickness and composition, wire width, quantum well thickness, buffer composition and thickness.
We have extracted these disorder parameters from measurements and used them in our simulations of topological gap devices.
We have also taken into account disorder induced by imperfections in the substrate and as well as disorder resulting from inhomogeneous superconductor growth.

We now discuss how we have verified that these design, growth, and fabrication advances have led to superconductor-semiconductor nanowires with long localization length, as required for a topological phase.
We have fabricated a variation on our topo gap device that has multiple junctions defining segments of different lengths, as we explain in more detail in \Cref{sec:localization_length}.
This enables us to measure the non-local conductance for different segment lengths $L$ and, thereby, extract the electron localization length $\ellLoc$ in the semiconductor.
This device, shown in Fig.~\ref{fig:1d_mobility_schematic}, was fabricated according to the same process as the DLG-$\delta$ topo gap device.
We apply an in-plane magnetic field perpendicular to the wire $B\sim 1\,$T to suppress the induced gap in all wire segments.
Consequently, there is a signal in $\GRL$ and $\GLR$ at low bias.
The junctions are operated in the open junction regime, see Fig.~\ref{fig:hmp}.
The typical non-local conductance decays with length as $\sim \exp(-2L/\ellLoc)$.
For fitting our measured conductances to this form, we normalize it by the local conductances to reduce the effect of the junctions.
From this fit we find that our gate-defined nanowires have localization length $\ellLoc \gtrsim \SI{1}{\micro\meter}$ in the single sub-band regime.
Thus, the localization length $\ellLoc$ of electrons in the wire underneath the Al is much longer than the mean-free-path of electrons in the Hall bar devices at a similar density due to screening of charged imperfections by Al in the former device type.
This observation also confirms that the Al-2DEG interface is of high quality ({i.e.}, the deposition of Al does not introduce new significant disorder mechanisms in our topo gap devices) and, thus, corroborates our disorder root-cause analysis discussed above.

Finally, we note, as a point of comparison to the previous works trying to realize topological superconductivity in quasi-one-dimensional nanowires~\cite{Lutchyn18}, that ``bottom-up'' vapor-liquid-solid (VLS) nanowires have been measured with field-effect mobilities of $10^{3}$-$10^{4}\,$cm$^{2}$/V$\cdot$s in InAs~\cite{Chang14, Heedt16} and InSb~\cite{Gul15} nanowires yielding localization lengths of 10-$100\,$nm in the few subband regime.
The origin of the dominant disorder mechanisms in VLS nanowires has not been established but is likely due to surface charged impurities.
Thus, half-shell proximitized VLS nanowires are likely to have a much shorter localization length than the topo gap devices considered here.
Field effect mobilities as high as $44,000\,$cm$^{2}$/V$\cdot$s have been observed in stemless InSb nanowires \cite{Badawy19}.
It would be interesting to extract the corresponding localization length underneath a superconductor for such a nanowire by a measurement similar to that described above and in \Cref{sec:localization_length}.
This can determine if topological superconductivity is possible.

\begin{figure*}
\includegraphics[width=18cm]{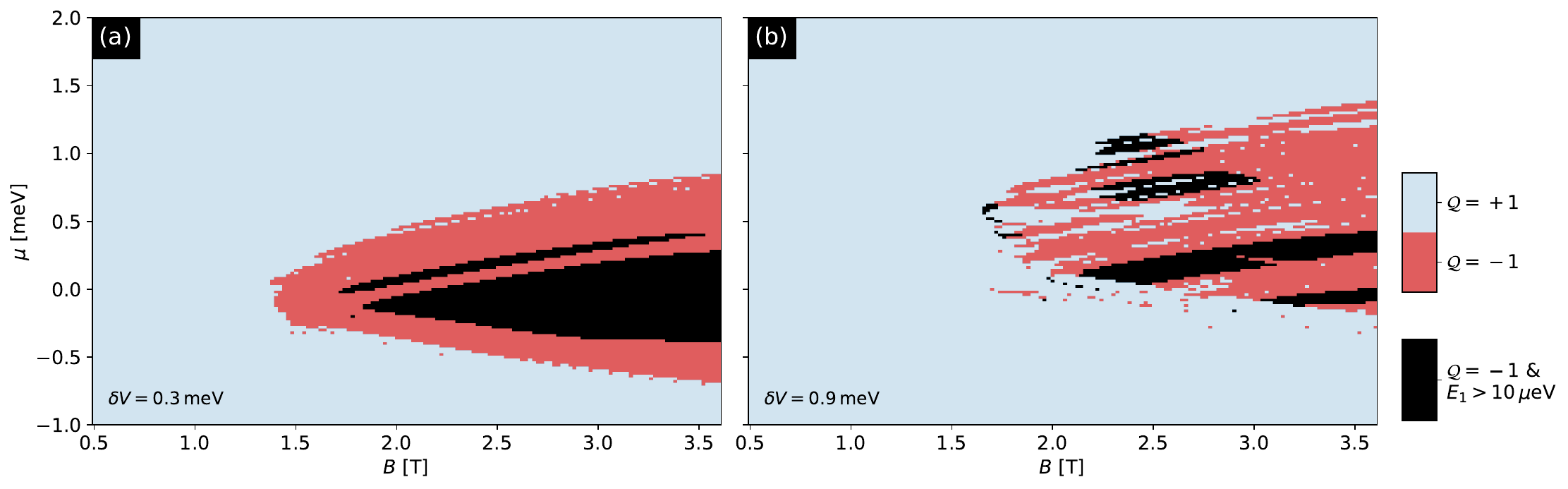}
\vskip -3mm
\caption{
Phase diagrams for a single disorder realization at $\delta V = 0.3\,$meV and $\delta V = 0.9\,$meV, as a function of the magnetic field $B$ and chemical potential $\mu$ for an $L = \SI{3}{\micro\meter}$ wire with the $\varepsilon$-stack single-band model parameters given in \Cref{tab:effective_parameters}.
Blue represents the trivial phase and red is the topological phase, which is identified by the Pfaffian invariant $\mathcal{Q} = \pm 1$.
Compared to the ideal topological phase in \Cref{fig:energy_spectrum_topophase} and relatively weak disorder effect in (a), the lobe in (b) moves to higher values of $\mu$ and splinters as a result of increasing disorder, consistent with \Cref{fig:xi0_epsilon_stack}b.
A subset of the topological phase is highlighted in black, where the second lowest eigenvalue, representative of the gap, also has a high value.
This more restrictive notion of topological region is even more ``splintered" and prone to finite-size effects and mesoscopic fluctuations.
}
\label{fig:pd_pfaffian}
\end{figure*}

\subsection{Topological phase diagram for a single disorder realization}
\label{sec:pd_single_realization}

Even when a device satisfies the requirements explained in the previous subsection and has a topological phase, disorder can cause the shape of the phase diagram to be rather complicated.
To gain a getter understanding, it is helpful to examine the phase diagram for a few representative disorder realizations.
In this section, we diagonalize the Hamiltonian in \Cref{eq:nanowire_with_disorder} and calculate the Pfaffian topological invariant \cite{Kitaev01, Wimmer12} for two independent disorder realizations.
In any finite-sized system, the disorder-driven phase transition between the topological and trivial phases is rounded into a crossover.
Consequently, a topological phase can be found in the phase diagram in some percentage of devices even for average disorder levels that exceed the critical value $\delta V_\mathrm{c}$ obtained in the thermodynamic limit.
Conversely, some percentage of devices will not have a topological region of the phase diagram even for average disorder levels for which there would be a topological phase in the thermodynamic limit.
Although disorder induces low-energy states~--- by creating domain walls between topological and non-topological regions, for instance~--- the density of such states may be low enough that an appreciable fraction of even reasonably long devices may not have any.
Thus we can also characterize the phase diagram by the spectral gap in the $\mathcal{Q} = -1$ region, taken as the second-lowest eigenvalue $E_1$ of $H$ (the lowest corresponds to the Majorana zero mode pair splitting).

In \Cref{fig:pd_pfaffian}, we show the phase diagrams of two different simulated devices.
Both have the DLG-$\varepsilon$ design, but with two different disorder realizations, one with $\delta V = 0.3\,$meV (panel a) and one with $\delta V = 0.9\,$meV (panel b).
For weak disorder, the lobe structure of the topological phase is preserved, and the spectral gap remains high over a large region inside the lobe.
For stronger disorder, mesoscopic fluctuations are important,
as we discuss in \Cref{sec:length_scales}.
The parabolic-shaped lobe of the topological phase of \Cref{fig:energy_spectrum_topophase}~--- as identified by the Pfaffian topological invariant $\mathcal{Q} = -1$~--- is splintered into several disconnected regions of narrow range in $\mu$ and larger extent in $B$.
This effect is even more dramatic if we additionally condition on a large spectral gap (black regions in \Cref{fig:pd_pfaffian}).
We will call these long, narrow regions of topological phase \textit{splinters} of the single-sub-band lobe.

There precise shape of these splinters varies from one disorder realization to the next.
We expect such mesoscopic fluctuations in our devices.
In \Cref{sec:disorder_pd,sec:length_scales}, we will discuss disorder-averaged parameters, such as the localization length $\xi(E)$.

\subsection{Statement on confidential information}
\label{sec:disclaimer}

In summary, the principles behind the design of our devices and material stacks are that they should enable three-terminal transport and:
(1)~be based on a 2DEG residing in a low-defect quantum well;
(2)~have a charged defect density $\sc$ at the semiconductor-dielectric interface that is less than $3 \cdot \scu$, as measured on a Hall bar on the same chip;
(3)~allow tuning to the lowest sub-band and full depletion of the wire; and (4)
have an induced gap to parent gap ratio in the lowest sub-band that satisfies $0.33 < \DeltaInd / \Delta_\mathrm{Al} < 0.67$.
Hall bar measurements can be used to measure progress towards satisfying requirements (1) and (2); while zero-field transport measurements of topological gap devices (described in the next section) can be used to determine when (3) and (4) are satisfied.
We present data from such measurements which directly verifies that the $\beta$, $\delta$, $\delta'$, and $\varepsilon$ material stacks in either SLG or DLG designs fulfill them.

The barrier thickness and composition, quantum well thickness, dielectric composition and deposition method, and Al strip width are critical factors that determine whether a device meets these prerequisites.
The details of these design parameters and fabrication methods are Microsoft intellectual property that we cannot disclose.
However, we have explained the principles by which we determined these parameters and processes in this Section (particularly \Cref{sec:layout,sec:material_stack,sec:disorder_uniformity}).

The following are some of the key ideas.
We grow our superconductor-semiconductor heterostructure by molecular beam epitaxy on an insulating InP substrate.
There is a graded buffer layer that modifies the lattice constant while preventing extended defects from reaching the active region \cite{Hatke17}.
In this regime, charged defects at the interface to the dielectric are the primary source of disorder.
We have engineered the electron wavefunction in order to minimize the effective disorder level while maintaining a near-optimal induced gap in the semiconductor.
In particular, we have varied the thickness of the InAs quantum well, $t_\mathrm{\scriptscriptstyle QW}$, over the range $7\,\mathrm{nm} < t_\mathrm{\scriptscriptstyle QW} < 11\,\text{nm}$, and we have varied the thickness of an InAlAs barrier, $t_\mathrm{\scriptscriptstyle B}$, over the range $4\,\text{nm} < t_\mathrm{\scriptscriptstyle B} < 12\,\text{nm}$.
These parameters have been optimized within these windows to maximize the distance from the active region to the dielectric while simultaneously targeting a gap ratio $\DeltaInd/ \Delta_\mathrm{Al} \approx 0.5$.
We have chosen InAs for the quantum well because: (a) its renormalized $g$ factor and spin-orbit coupling can reach the minimum required values of $4$ and $4\,\text{meV}{\cdot}\text{nm}$, respectively; (b) there are known lattice-matched barriers; and (c) it has a larger temperature window for subsequent processing steps than alternative materials.
We have chosen aluminum for the superconductor because it has demonstrated $2e$-periodic Coulomb blockade peaks, which is essential for the qubits that we discuss in \Cref{sec:looking_ahead}, and it has a superconducting gap that is known to increase with decreasing thickness.
The aluminum strip was chosen to be as wide as possible (in order to keep defects in the dielectric as far as possible from the active region in the quantum well) while still allowing full depletion of the wire at plunger gate voltages $\Vp>-3\,$V.
Meanwhile, we have varied dielectric deposition conditions in order to find a process point at which $\sc< 3\times\scu$.
Our devices can be reproduced through similar optimization steps, combining simulation and experimentation.

All of the key material and design parameters feed into the effective parameters given in \Cref{tab:effective_parameters}, together with $\delta V$ and $\kappa$.
They define the projected single-sub-band model in \Cref{eq:nanowire_with_disorder} from which our simulations of bulk properties of our devices can be reproduced.
Any device that replicates our design and material stack will have similar effective parameters.

\section{Topological gap protocol}
\label{sec:TGP}

\begin{figure}
\includegraphics[width=8.5cm]{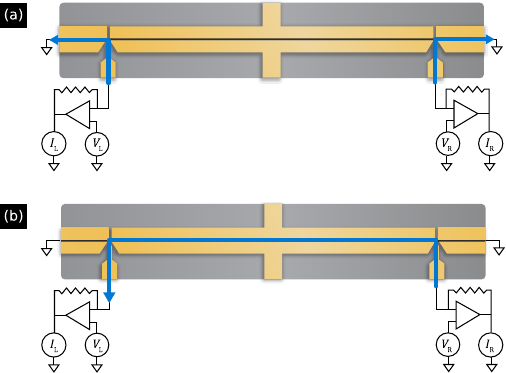}
\vskip -2mm
\caption{
Schematic illustrations of the current paths (blue arrows) that contribute to (a) $\GRR$, $\GLL$ and (b) the non-local conductance $\GLR$ discussed in \Cref{sec:TGP}.
}
\label{fig:local_nonlocal_conductance}
\end{figure}

The goal of the TGP is to identify whether there are regions in the experimental parameter space that show signatures consistent with a topological phase.
The full source code of the TGP and raw data sets are available in Ref.~\onlinecite{code_and_data}.
The device's outer sections are kept in the trivial superconducting phase by tuning their densities with the right and left plunger gates.
In the topological phase of the wire, MZMs are localized at the boundaries between the topological and trivial sections, see \Cref{fig:device_SLG}(c).
Provided that $L$ is smaller than or, at least, not too much larger than the localization length $\xi(\DeltaT)$, see ~\Cref{sec:length_scales}, there will also be an observed non-zero bulk transport gap.
When this condition is satisfied, a non-zero above-gap non-local conductance is observable, enabling an identification of the gap, as we discuss further in \Cref{sec:length_scales}.
In the TGP \cite{Pikulin21}, the presence of MZMs and a bulk transport gap is detected by measuring the differential conductances
\begin{equation}
\begin{pmatrix}
    \GLL & \GLR \\
    \GRL & \GRR
\end{pmatrix}
= \begin{pmatrix}
    dI_\L / dV_\L & dI_\L / dV_\R \\
    dI_\R / dV_\L & dI_\R / dV_\R
\end{pmatrix}
\label{eq:G_matrix}
\end{equation}
as a function of $\Vp$ and $B$ as well as the voltages $\Vrc, \Vlc$ controlling the tunnel junction transparencies, and the bias voltages $\Vb=V_\R, V_\L$, which can be increased in order to tunnel current into states of higher energies.
The currents and voltages $I_\R, I_\L, V_\R, V_\L$ are illustrated in \Cref{fig:local_nonlocal_conductance}.
We use the cutter gates to open and close the junctions; when $\Vrc$ is more negative, the junction is more closed, and similarly with $\Vlc$.
We discard all devices in which one of the junctions cannot be completely closed at a pinch-off voltage $>-3\,$V.
Even among devices that pass this basic health check, there is considerable device-to-device variation in the pinch-off voltages and, more generally, in the relation between $\Vrc$, $\Vlc$ and the conductances through the junctions.
This is, presumably, due to the different disorder configurations in the different junctions; these differences have a large effect because the junctions are depleted, leaving charged impurities unscreened, unlike in the bulk of the wire where the Al strip can suppress the effects of charged impurities via screening.

We want to vary the cutter gate voltages so that the local electrostatic environments at the two junctions change by enough to change the energy of bound states that are accidentally at zero energy for one cutter gate configuration.
But since the cutter gate voltage change required to open or close a junction varies significantly from one junction to another as a result of disorder, we cannot simply choose the same sequence of $\Vrc$, $\Vlc$ values for each device.
Instead, we use the above-gap conductance $\Gag$ at $B=0$ and a bias voltage of \SI{500}{\micro\volt} as a measure of the junction transparencies.
In each device, we find sequences of cutter gate voltages $\Vrc$, $\Vlc$ for which $\Gag$ at both junctions take values between $\approx 0.1 e^2/h$ and $\approx e^2/h$.
They are slightly different in each device, but they always cover a substantial fraction of this range.
When we say, as a shorthand, that we are varying the junction transparencies, we mean that we vary $\Vrc$, $\Vlc$ in this manner.

In the tunneling regime (i.e.~for $\Gag < e^2/h$), the current paths contributing to $\GRR$, $\GLL$ are illustrated in \Cref{fig:local_nonlocal_conductance}(a).
In this regime, $\GRR$ and $\GLL$ directly measure the local density of states in the wire at the boundaries between the middle and, respectively, the right and left sections.
ZBPs are determined by the condition that $- d^3 I_\R/dV_\R^3$ exceeds the noise level and similarly for the left junction, i.e. the second derivative of the $dI/dV$ curve is more negative than the noise level.
ZBPs in $\GRR$ and $\GLL$ in the tunneling regime indicate the presence of zero-energy states in the wire with sufficient tunneling matrix elements to the leads, consistent with MZMs but also with trivial zero-energy Andreev bound states.
A zero-energy state (either MZM or trivial ABS) at the right junction will be manifested as a ZBP in $\GRR$ and similarly for a zero-energy state at the left junction and a ZBP in $\GLL$.
Trivial ABS are not generically stable with respect to local perturbations whereas well-separated MZMs are.
Therefore, the ZBP stability criterion, discussed below, allows one to better identify the region of interest.

\begin{figure*}
\includegraphics[width=18cm]{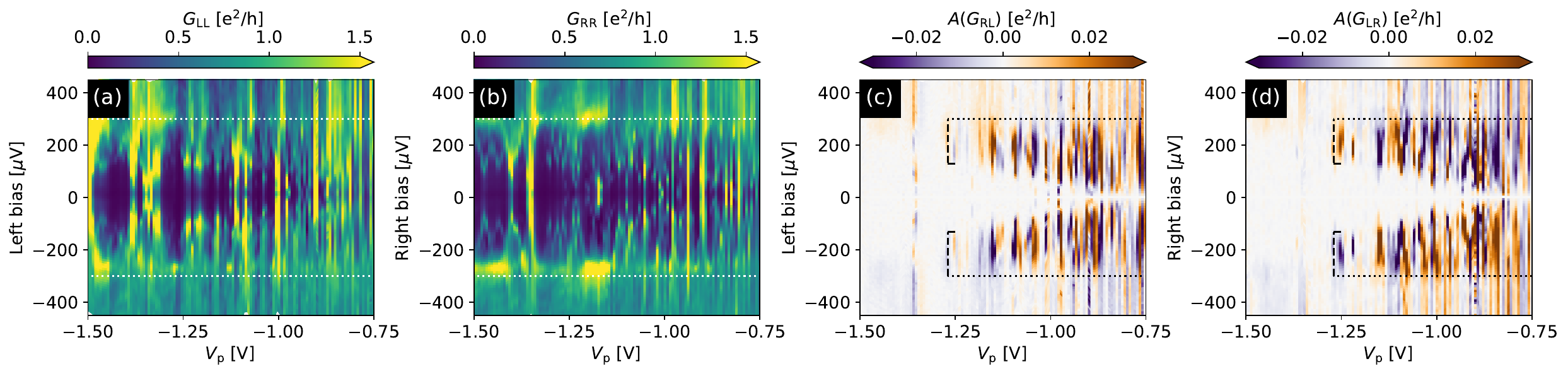}
\vskip -3mm
\caption{
The experimental local and anti-symmetrized non-local conductances for device A at zero magnetic field as a function of plunger gate voltage $\Vp$ and bias voltage.
The anti-symmetrized non-local conductances in panels (c) and (d) are suppressed above the parent Al gap, which is indicated by horizontal dotted lines.
They are non-vanishing down to small bias for $\Vp$ more positive than $\approx -0.9\,$V, which indicates that there is conduction through the region of the 2DEG that is not underneath the Al strip.
This region of the 2DEG is depleted for $\Vp$ more negative than approximately $-0.9\,$V.
The anti-symmetrized non-local conductances vanish for plunger gate voltages below $\Vp \approx -1.27\,$V, which we identify as the bottom of the lowest sub-band.
For $-0.9\,$V$ \lesssim \Vp \lesssim -1.27\,$V, the induced gap is indicated by a dashed curve which terminates in dashed vertical lines at the depletion voltage.
The parent gap is visible in the local conductances in panels (a) and (b), but it is more challenging to identify the induced gap due to sub-gap states in the junctions.
}
\label{fig:deviceA_conductance}
\end{figure*}

The current path contributing to $\GLR$ is illustrated in \Cref{fig:local_nonlocal_conductance}(b); $\GRL$ is determined by the reverse path.
For an intuitive understanding of $\GLR$ and $\GRL$, we first note that in the thermodynamic limit of the wire, the clean limit, and the tunneling limit of both junctions, a current injected at bias voltage above the Al parent gap will flow through the Al strip to ground via the contacts at the ends of the device unless it relaxes to energies between the induced gap and the parent gap.
Hence, at bias voltages above the Al parent gap, $\GRL$ and $\GLR$ are strongly suppressed and are non-zero only as a result of these weak relaxation processes \cite{Rosdahl18, Puglia21, Wang21}.
At zero-temperature, in the thermodynamic limit of the wire, the clean limit, and the tunneling limit of the junctions, current cannot be injected into the wire at bias voltages below the induced gap, except by Andreev processes, which inject supercurrent that also flows to the grounded contacts at the ends of the device.
Now consider a finite-length disordered wire.
At bias voltages at which the localization length $\xi(eV)$ is less than the length of the wire, $\GRL$ and $\GLR$ are strongly suppressed and are non-zero only as a result of non-zero temperature and finite ratio $L/\xi(eV)$.
(In an infinite wire, $\GRL$ and $\GLR$ would vanish at all bias voltages because all states are localized, except precisely at the transition.
For a further discussion, see \Cref{sec:length_scales}.)
Consequently, the highest bias voltage below which $\GRL$ and $\GLR$ are nearly vanishing (in a sense that we make more precise below) can be interpreted it as the transport gap $\Delta_\mathrm{tr}$ that we define in \Cref{sec:length_scales}.
As we discuss in \Cref{sec:TGP_measurements}, we perform this gap extraction with the parts of the non-local conductances that are antisymmetric in bias voltage, $A(\GRL)$, $A(\GLR)$:
\begin{equation}
    A[(\GRL(\Vb)] \equiv \left[\GRL(\Vb) - \GRL(-\Vb)\right]/2
\end{equation}
and similarly for $\GLR$.

The high-dimensional nature of the parameter space that is explored by the TGP makes it prudent to narrow the measured parameter range.
We explained above how the range of junction transparencies is limited.
Meanwhile, the parameter range of $\Vp$ is chosen to be close to the bottom of the first sub-band.
When the chemical potential is below the first sub-band, the wire is fully depleted.
The depletion point is identified by scanning the non-local conductance as a function of bias and $\Vp$.
This can be done at $B = 0$ or at non-zero $B$, with $B$ below the critical field of the superconductor, where the signal is generally larger.
Recall that, as noted above, the non-local conductances are essentially zero outside the range of bias voltages between the induced and parent gaps, except for finite-size effects, thermal activation, and relaxation effects.
Hence, full depletion of the wire causes the non-local conductance at bias voltages below the Al gap to drop below the noise floor.
We use this depletion point to identify the single-sub-band regime.

In \Cref{fig:deviceA_conductance}, we show the four elements of the experimentally-measured conductance matrix as a function of bias voltage $\Vb$ and plunger gate voltage $\Vp$ at zero magnetic field in one of our devices, which we label device A, to illustrate how the depletion point is identified.
As may be seen from \Cref{fig:deviceA_conductance}(c,d), the anti-symmetrized non-local conductances are small above the parent gap $\Delta_\mathrm{Al} = 295 \pm \SI{8}{\micro\eV}$, which is indicated by horizontal dotted lines in \Cref{fig:deviceA_conductance}(c,d).
The anti-symmetrized non-local conductances are non-vanishing down to small bias for $\Vp \gtrsim -0.9\,$V, which indicates that there is conduction through 2DEG regions not contacted by the Al for these plunger gate voltages.
For $\Vp$ more negative than $\approx -0.9\,$V, these 2DEG regions are depleted, and the induced gap opens up.
As discussed previously, the anti-symmetrized non-local conductances are large between the induced and parent gaps, are suppressed above the parent Al gap, and are very strongly suppressed below the induced gap.
As $\Vp$ is decreased further, the induced gap increases, eventually reaching its maximum measured value of $\DeltaInd = 129 \pm \SI{12}{\micro\eV}$.
At $\Vp \approx -1.25\,$V, the anti-symmetrized non-local signal drops sharply while local conductances remain large.
For more negative $\Vp$, the anti-symmetrized non-local signal is very small, and there is no longer a visible bias range between the induced and parent gaps.
This is interpreted as full depletion of the semiconductor below the Al strip.
The single-sub-band regime occurs just before wire depletion.

In summary, the TGP makes the parameter space of our devices manageable by focusing on the most favorable region: $\Vp$ near the bottom of the lowest sub-band; $B$ from zero up to $2.5\,$T; and a range of junctions transparencies $\Gag$ between $\approx 0.1 e^2/h$ and $\approx e^2/h$.

The steps of the TGP are divided into two stages.
\textit{Stage 1:} (1)~From an analysis of $\GRR$ and $\GLL$, identify ZBPs at each end of the wire that are stable to variations of the junction transparencies and variations in local junction potential (which are controlled by $\Vrc$, $\Vlc$ in the manner discussed above).
(2)~Find clusters of points in the $B$-$\Vp$ plane where there are stable ZBPs at both ends of the wire.
These clusters and their surrounding neighborhoods define the regions of interest ROI$_1$ that are the focus of Stage 2.
If there are no such clusters, the device fails Stage 1.

\textit{Stage 2:} (3)~Focusing on smaller $(B, \Vp)$ ranges containing ROI$_1$s and restricting to cutter gate voltage pairs for which the junction transparency is approximately the same at both ends, confirm the existence of stable zero bias peaks in $\GRR$ and $\GLL$ and recover the clusters of points in the $B$-$\Vp$ plane where there are stable ZBPs at both ends of the wire.
This step is important when there is a drift in $\Vp$ between Stages 1 and 2.
The cutter voltages can either be set such that the junction transparencies are set on average to target conductance values, or compensated as a function of $\Vp$ such that the transparencies are stabilized to the target values.
(4)~Use $A(\GRL)$ and $A(\GLR)$ to determine the bulk energy gap as a function of $(B, \Vp)$ for each pair of cutter gate settings.
(5)~For each pair of cutter gate settings, find ZBP clusters identified in step 3 whose interiors are gapped and whose boundaries are gapless.
We will denote them by $\mathcal{C}^A_i$ where $i$ is a index for the pair of cutter gate settings and $A$ is an index that distinguishes different gapped ZBP clusters with gapless boundaries that might occur for the same pair of cutter gate settings.
(6)~Find the sets of clusters $T$ in the $B$-$\Vp$ plane consisting of $\mathcal{C}^A_i$ that overlap for different cutter gate settings.
To be more precise, we define $T \equiv \{\mathcal{C}^A_i \,|\,\mathcal{C}^A_i \cap \mathcal{C}^B_j \neq \emptyset \mbox{ for some } j\neq i \mbox{ and } \bigcup \mathcal{C}^A_i \mbox{ connected}\}$.

The device passes the TGP if there is a $T$ such that there is a $\mathcal{C}^A_i \in T$ for a number of cutter gate settings $i$ that exceeds some threshold, as we make more precise in \Cref{sec:TGP_parameters}.
In this case we define the region of interest $\mathrm{ROI}_2 = \bigcup_{\mathcal{C}^A_i\in T} \mathcal{C}^A_i$.
Note that for a given device, there can be several $T$'s and $\mathrm{ROI}_2$s.
We will call the clusters $\mathcal{C}^A_i \in T$ ``subregions of interest SOI$_2$ belonging to a region of interest ROI$_2$.''

Note that Stage 2 of the TGP typically uses $5$ or fewer cutter gate values, chosen so that the above-gap conductance at each junction varies by $\sim {e^2}/h$ between the most closed and most open configuration.
Stability of ZBPs to variation over a denser set of cutter gate values is neither necessary nor sufficient for passing the TGP.
Further details are discussed in \Cref{sec:TGP_details}.

There are a number of important measurement complexities that we discuss in \Cref{sec:TGP_measurements}.
The TGP is formulated with several thresholds which we explain in \Cref{sec:TGP_parameters}: the minimum percentage of cutter gate settings for which a ZBP must be present in order to be considered stable, denoted by $(\mathrm{ZBP}\%)_\mathrm{th}$; the minimum percentage of the boundary of a ZBP cluster that must be gapless in order for the whole boundary to be considered gapless, denoted by $(\mathrm{GB}\%)_\mathrm{th}$; the conductance value below which we consider it to be effectively zero up to finite-size effects, denoted by $G_\mathrm{th}$; and the minimum percentage of cutter gate settings for which an overlapping SOI$_2$ must be present in order to form an ROI$_2$, denoted by $(\mathcal{C}_i \%)_{\rm th}$.

The TGP captures the key physics of topological superconductivity because it requires a device to show stable ZBPs at both ends and also a bulk gap closing and re-opening.
However, we can make a much stronger quantitative statement about its reliability by testing it on simulated devices.
We simulated 349 devices of different designs, material stacks, and disorder levels and applied the TGP to transport data from these devices.
To test its reliability, we compared the ROI$_2$s located by the TGP with the ``scattering invariant''~\cite{Fulga11}, a topological index that is defined for open systems (see \Cref{sec:topological_invariants} for a brief description of this invariant).
When the topological index is $-1$ in some region of the phase diagram, the region is topological; when it is $+1$, the region is trivial.
However, trivial regions of the phase diagram can exhibit relatively stable ZBPs in their transport data, and the TGP was designed to avoid misidentifying such regions as topological.

We classify ROI$_2$s as true positives (TP) if they contain any region with non-trivial topological index and as false positives (FP) otherwise.
The false discovery rate (FDR) is the probability that an ROI$_2$ is trivial:
\begin{align}
    \textrm{FDR} 
    & \equiv P(\textrm{ROI$_2$ is trivial}) \nonumber\\
    & = \lim_{N\to\infty}{\mathrm{FP}}/(\mathrm{FP} + \mathrm{TP}),
    \label{eq:FDR_preview}
\end{align}
where $N$ is the total number of devices.
In essence, the FDR is the probability that if a device passes the TGP then the ROI$_2$ that it identifies has a completely trivial explanation, such as a trivial ABS.
We estimate the FDR from the TP and FP numbers obtained from a large~--- but finite~--- number of simulated devices.
As $N\rightarrow \infty$, the ratio ${\mathrm{FP}}/(\mathrm{FP} + \mathrm{TP})$ approaches the FDR.
For finite $N$, the best that we can do is estimate upper and lower bounds on the FDR.
We use the Clopper-Pearson confidence interval at the 95\% confidence level to estimate these bounds.

Our results are shown in \Cref{tab:TGP_FDR}.
Since we found no false positives, the confidence interval for the FDR is between zero and the upper bound that we list in the rightmost column.
We find that if a device passes the TGP, there is a $< 8\%$ probability that the ROI$_2$ that it finds does {\rm not} contain a topological phase, provided that the simulated data is drawn from the same probability distribution as the data produced by real devices.
For the DLG-$\varepsilon$ design, the probability is $< 6\%$.
We simulated several different disorder levels to investigate whether the TGP is more likely to give false positives when disorder is higher.
Our results indicate that the TGP is reliable over the entire range $\sc = 0.1$-$ 4 \cdot \scu$, which is the range of charged disorder levels in the measured devices discussed in \Cref{sec:experimental_data}.%
\footnote{Note that the threshold $G_\mathrm{th}$ depends on the level of disorder in the system and is taken differently at $0.1 \cdot \scu$ charged disorder compared to the other cases.
See Appendices \ref{sec:thresholds} and \ref{sec:testing_TGP} for details.}
Similarly, the differences between the SLG-$\beta$ and DLG-$\varepsilon$ stacks and designs have no effect on the accuracy of the TGP.
The small dependence of our FDR estimates on disorder level and design that may be seen in \Cref{tab:TGP_FDR} are entirely a consequence of the different numbers of ROI$_2$s that were found at different disorder levels.
Further details are given in \Cref{sec:TGP_calibration_testing}.
As we discuss in \Cref{sec:discussion}, the statistical properties of the ROI$_2$s that we find in our simulations agree with the corresponding experimental values, thereby further validating the simulation model used estimate the FDR.
This analysis addresses open questions regarding the reliability of the TGP~\cite{Akhmerov22}.

\begin{table}
\begin{center}
\begin{tabularx}{\columnwidth}{|I|W|*{3}{C|}}
\cline{1-5}
\shortstack[l]{\noalign{\vskip 1.0ex} Design, \\ stack} & 
\shortstack[c]{$\sc$ \\ $[\scu]$} & 
\shortstack[c]{TP \\ $\phantom{1}$} & 
\shortstack[c]{FP \\ $\phantom{1}$} & 
\shortstack[c]{FDR \\ $\phantom{1}$}
\\ [0.5ex]
\cline{1-5}
\multirow{3}{*}{SLG-$\beta$}
& 1.0 & 244 & 0 & $< 1.5$ \\ 
\cline{2-5}
& 2.7 & 46 & 0 & $< 7.7$ \\ 
\cline{2-5}
& 4.0 & 45 & 0 & $< 7.9$ \\
\cline{1-5}
\multirow{3}{*}{DLG-$\varepsilon$}
& 0.1 & 125 & 0 & $< 2.9$ \\ 
\cline{2-5}
& 1.0 & 97 & 0 & $< 3.7$ \\ 
\cline{2-5}
& 2.7 & 67 & 0 & $< 5.4$ \\ 
\cline{2-5}
& 4.0 & 66 & 0 & $< 5.4$ \\
\cline{1-5}
\end{tabularx}
\end{center}
\vskip -3mm
\caption{
Statistics of TGP results for simulated transport data from SLG-$\beta$ and DLG-$\varepsilon$ devices with $50$ different disorder realizations for each of the average disorder strengths given in the first column.
Devices with $\sc \leqslant \scu$ typically have multiple distinct ROI$_2$s, so we have more total ROI$_2$s for low disorder.
The listed range of FDR values is the confidence interval at the 95\% confidence level.
}
\label{tab:TGP_FDR}
\end{table}

\section{Experimental data}
\label{sec:experimental_data}

\subsection{Measurements of device A}
\label{sec:deviceA1}

In the remainder of this paper, we focus on measurements of devices such as the one shown in \Cref{fig:device_SLG}.
In this section, we focus on data from device A, which is a \SI{3}{\micro\meter} long SLG device built on a $\beta$-stack.
We discuss three experimental measurements from this device.
The raw data is available in Ref.~\onlinecite{code_and_data}.
Measurement A1 was taken in one dilution refrigerator while measurements A2-A3 were taken in a different cooldown of device A in a different dilution refrigerator.
The measured zero-field superconducting gap in the Al strip is $\Delta_\mathrm{Al} = 295 \pm \SI{8}{\micro\eV}$ and the maximum induced gap at zero $B$-field is $\DeltaInd = 129 \pm \SI{12}{\micro\eV}$, which indicates that the induced gap to parent gap ratio $\DeltaInd/\Delta_\mathrm{Al} = 0.44$ is well within the desired range.%
\footnote{For the extraction of the zero-field induced gap, see \Cref{fig:deviceA_characterization}.}
The effective charged impurity density at the interface with the dielectric is $\sc = 2.7 \cdot \scu$, as is discussed in \Cref{sec:electrostatic_calibration}.
This value satisfies the specification explained in \Cref{sec:disorder_uniformity}, which is based on the assumption that the average charged impurity density at the dielectric-semiconductor interface in the Hall bar is the same as at the dielectric-semiconductor interface in a topological gap device [the boundary between light blue and grey on either side of the Al strip in \Cref{fig:device_SLG}(d)] on the same chip.
The critical field, $B_\mathrm{c}$, for the thin Al strip is $> 4.5\,$T for magnetic fields in the direction of the strip.
The single sub-band regime, as determined from the non-local conductance in the same manner as in \Cref{fig:deviceA_conductance}, is reached at $\Vp$ between $-1.2$ and $-1.4\,$V (depending on the cooldown).
This is consistent with our simulations for device A; see \Cref{fig:simulated_infinite_clean_gap}.
The base temperature in our measurements is $\sim 20$~mK and, using NIS thermometry \cite{Feschenko15}, we measured an electron temperature $T_\mathrm{e} < 40\,$mK.

\subsubsection{TGP Stage 1}
\label{sec:deviceA1_stage1}

\begin{figure*}
\includegraphics[width=18cm]{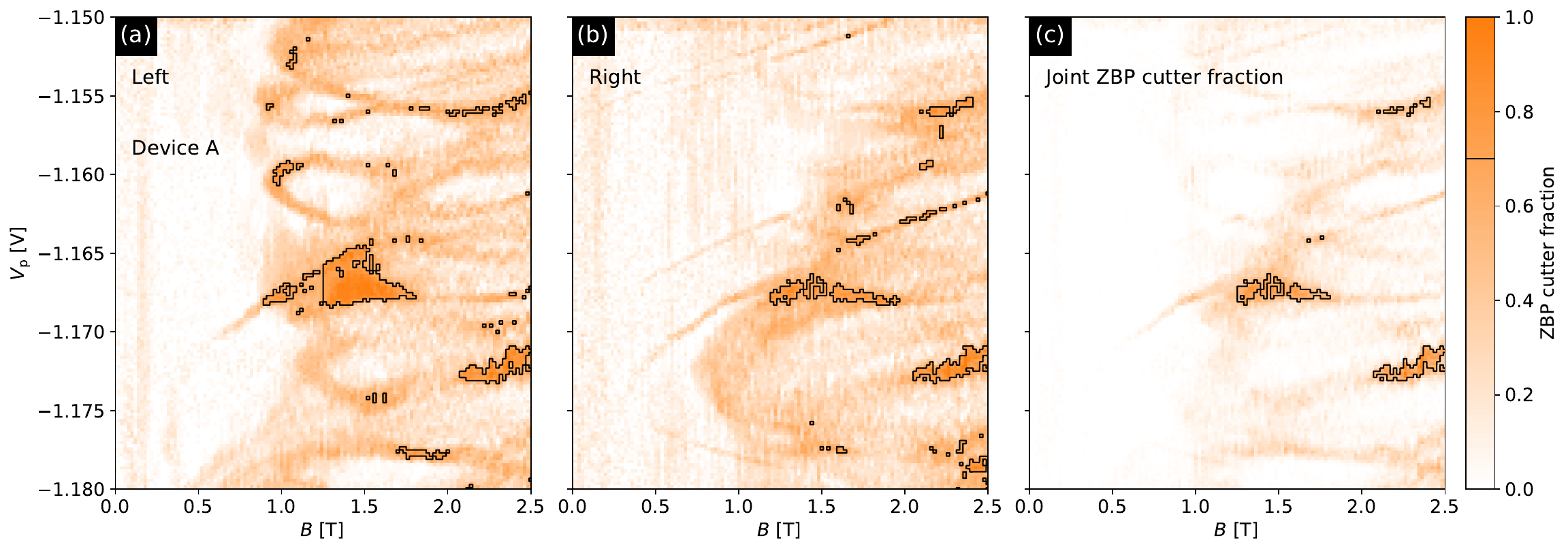}
\vskip -3mm
\caption{
Experimental ZBPs for device A, measurement A1, Stage 1 identifying ROI$_1$ for \Cref{fig:deviceA1_stage2}.
(a)~The cutter gate fraction for which there is a ZBP at the left junction as a function of $B$ and $\Vp$.
(The color scale is at the far right.) 
(b)~The cutter gate fraction for ZBPs at the right junction as a function of $B$ and $\Vp$.
(c)~The cutter gate fraction for ZBPs at both junctions.
In all three panels, regions for which the cutter gate fraction for ZBPs is at least $70\%$ are surrounded by black lines.
The cutter gate fractions for ZBP are defined in the third paragraph of \Cref{sec:deviceA1_stage1}.
ZBPs that occur at only one junction or for a small cutter gate fraction are likely to be due to trivial Andreev bound states.
The data and scripts needed to reproduce this and other experimental figures are available in Ref.~\onlinecite{code_and_data}.
}
\label{fig:deviceA_stage1}
\end{figure*}

We begin by finding the single-sub-band regime, following the method discussed in \Cref{sec:TGP_measurements}.
The non-local signal below the Al parent gap vanishes for $\Vp < -1.18\,$V, which we interpret as the point at which the wire is fully depleted.
We focus our Stage 1 scans on a $\Vp$ range of $30\,$mV above this value.

In \Cref{fig:deviceA_stage1}(a,b), we show the cutter gate fraction for ZBPs at, respectively, the left and right junctions as a function of $B$ and $\Vp$.
The black lines in \Cref{fig:deviceA_stage1}(a,b) encloses the regions in which the cutter gate fraction for ZBPs at the left or right junction is greater than $> 70\%$.
Finally, in \Cref{fig:deviceA_stage1}(c), we show the fraction of junction transparencies at which there are ZBPs at both junctions, plotted as a function of $B$ and $\Vp$.
The black line indicates the part of the phase diagram where the cutter gate fraction for ZBPs at both junctions is $> 70\%$.

Stage 1 data was taken for 21 different cutter gate voltages at each junction, chosen such that $\Gag$ at each junction is in the range 0.01 to 0.85$e^2/h$.
These 21 cutter voltages were found by the following procedure including a calibration measurement prior to the TGP Stage 1 (TGP1) measurement.
21 targets for $\Gag$ were chosen as equidistant points in the range 0.01 to 0.85$e^2/h$.
The change of local potential corresponding to this conductance change is estimated to be a few meVs.
In the calibration measurement, the cutter voltage for each junction yielding each of the 21 targets (within $\pm$ 0.02$e^2/h$) was recorded at 61 equidistant points along the plunger voltage range $V_p$ to be used in TGP1.
For each $\Gag$ target, the median cutter voltage along the plunger voltage axis is chosen as the cutter voltage for that $\Gag$ target.
This procedure returned the 21 cutter voltages for each junction.
Each of the 21 cutter voltage pairs used in TGP1 is a pair in which each of the voltages is drawn with no replacement from this list.

In Stage 1, we find $(B, \Vp)$ values at which there are ZBPs at both junctions for more than $15$ out of $21$ $(\Vlc, \Vrc)$ pairs or, in other words, for which the cutter gate fraction for ZBPs at both junctions is $> 70\%$.
Clusters of such points are the candidate regions of topological phase yielded by Stage 1 of the TGP, dubbed ROI$_1$ in Ref.~\onlinecite{Pikulin21}.

There are several key features in \Cref{fig:deviceA_stage1} worth emphasizing.
First, we expect that the topological phase in proximitized nanowires should have a lobe-like shape $|g^{\star}| \muB B/2 > \sqrt{\mu^2 + \DeltaInd^2}$ in the absence of disorder.
As a result of disorder, we expect the lobe to be splintered, as shown in the simulations in \Cref{fig:pd_pfaffian}.
In Stage 1 data from simulated device R1, this manifested as splintered regions in which there are stable ZBPs at both ends of the device, as may be seen in \Cref{fig:simulated_SLG_beta_R1_stage1}.
The $30\,$mV field of view in \Cref{fig:deviceA_stage1} corresponds to a single lobe,which we identify as the lowest sub-band according to the method discussed in \Cref{sec:TGP_measurements}.The structure that is visible in the phase space locations of stable ZBPs at the left and right junctions and, especially, in ROI$_1$ resembles the splintering of the lobe.

We have observed very similar ROI$_1$s in several devices (such as devices B, C, D, and E).
In more disordered devices (such as device F, which is discussed in \Cref{sec:other_devices}), ZBPs are scattered throughout phase space, and there is no structure, which suggests a non-topological phase of matter.

The data is reproducible between successive measurement runs on the same device, as we show in \Cref{sec:measurementsA2A3}.
The system is very stable, provided that $\Vp$ is varied by $30\,$mV or less.
If the voltage is varied by more than $100\,$mV, features shift in $\Vp$ but we can recover the same ROI$_1$.
If a device idles for approximately a week near an ROI$_1$, we find that voltages drift by at most a few mV, as we will see when we compare measurements A2 and A3.

We emphasize that the main goal of Stage 1 is to identify promising regions in parameter space for measurements of both the local and non-local conductances over a range of bias voltages, which are the focus of Stage 2.

\subsubsection{TGP Stage 2: Measurement A1} 
\label{sec:deviceA1_stage2}

\begin{figure*}
\includegraphics[width=18cm]{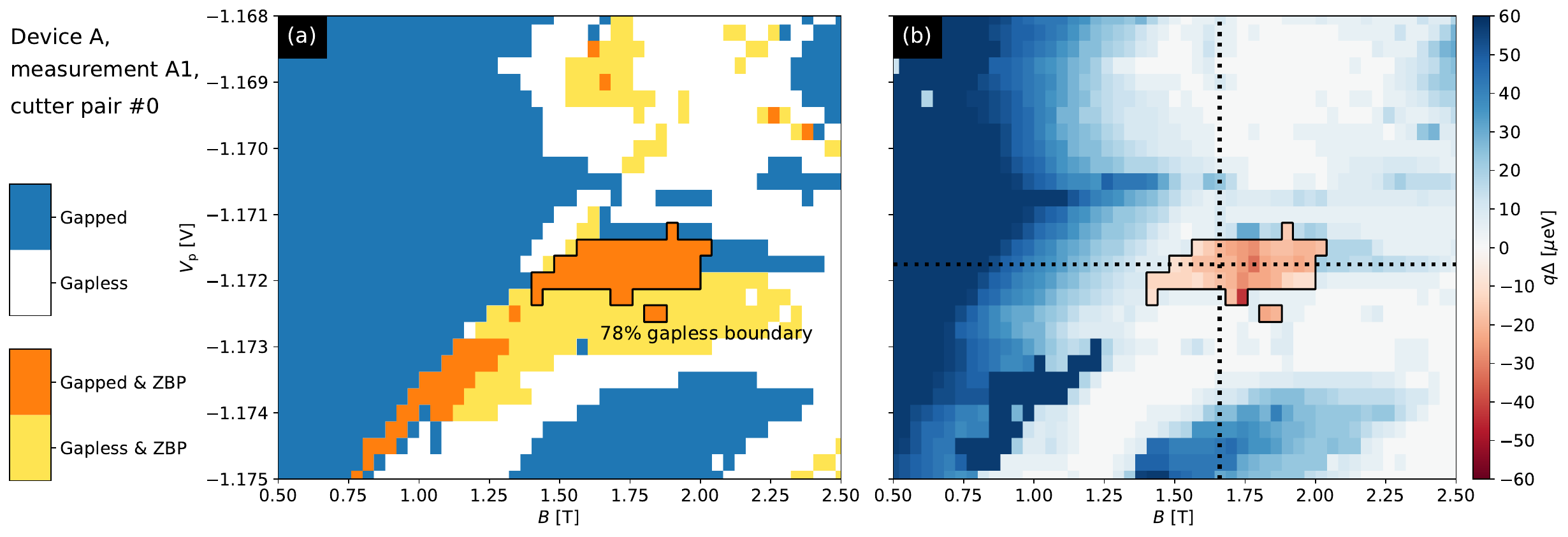}
\includegraphics[width=18cm]{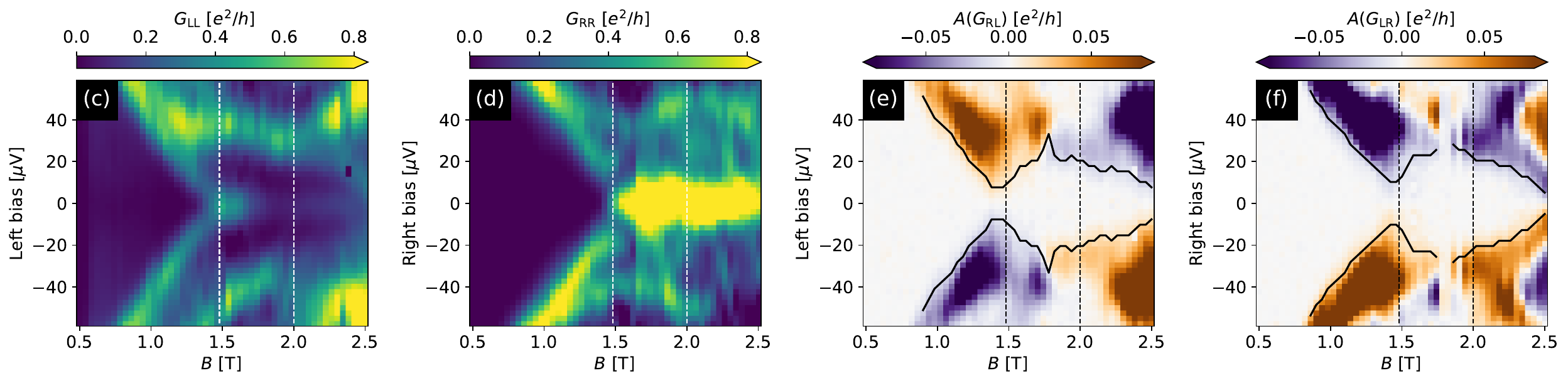}
\includegraphics[width=17.9cm]{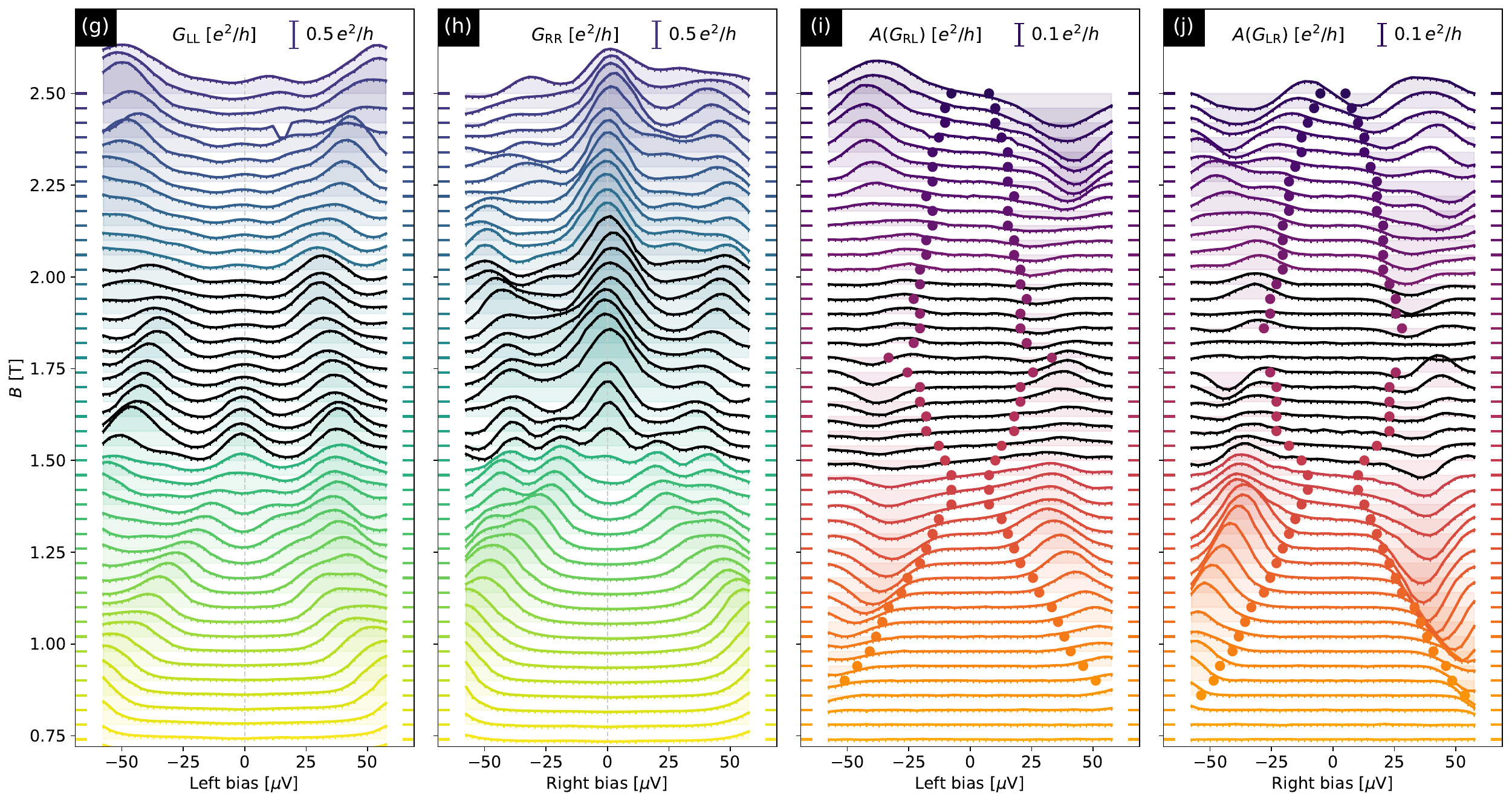}
\vskip -3mm
\caption{
(a)~The experimental phase diagram of device A (measurement A1) in the color scheme shown at the left.
The stability of ZBPs is determined by varying the cutter gates so that for both junctions $\Gag$ takes the values $0.3$, $0.5$, and $0.7e^2/h$ (cutter pairs \#0, \#1, and \#2, correspondingly).
The boundary of the SOI$_2$ is interpreted as a phase transition line, consistent with a visible gap closure along 78\% of it.
(b)~The experimental phase diagram, showing trivial/topological phases, which the TGP identifies with the exterior/interior ($q = \pm 1$) of the SOI$_2$.
The color scale shows the size of the trivial (positive sign) or topological (negative sign) gap.
The protocol assigns a maximum topological gap $\DeltaMax = \SI{23}{\micro\eV}$.
Measured local and anti-symmetrized non-local conductances along the horizontal line in panel b at $\Vp = -1.17175\,$V: (c)~$\GLL$, (d)~$\GRR$, (e)~$A(\GRL)$, (f)~$A(\GLR)$.
The SOI$_2$ lies between the vertical lines.
Panels (g)-(j) are ``waterfall'' plots representing the same measured data.
The data shown in (c)-(j) was obtained for $\Gag \approx 0.3 e^2/h$ for both sides (we call it cutter pair \#0).
The black curves in panels (e) and (f) and the dots in panels (i) and (j) are \textit{not} guides to the eye;
they indicate where the non-local signal drops below a threshold value, as described in the text.
The analysis of cutter pairs \#1 and \#2 is shown in \Cref{fig:deviceA1_stage2_cutter1,fig:deviceA1_stage2_cutter2}.
}
\label{fig:deviceA1_stage2}
\end{figure*}

In Stage 2, we focus on the regions of the $B$-$\Vp$ plane where there are clusters of points with stable ZBPs at both junctions.
We map out the full conductance matrix \Cref{eq:G_matrix} as a function of $B$, $\Vp$, $\Vlc$, $\Vrc$, and, in addition, $\Vb$.
Since we are now exploring a higher-dimensional parameter space, we restrict the $\Vp$ sweep to the vicinity of ROI$_1$ identified in Stage 1, which is typically $\delta\Vp \approx 5$-$15\,$mV.
We further restrict the parameter space by taking scans for 3-5 pairs of cutter gate settings (rather than the $>$ 20 pairs of Stage 1).
For each pair of cutter settings, the cutters are compensated as a function of $\Vp$ to achieve the target $\Gag$ values at each side.
In the measurement of device A displayed in \Cref{fig:deviceA1_stage2}, there were 3 cutter gate pairs $(\Vlc(\Vp), \Vrc(\Vp))$.
These cutter gate settings correspond to $\Gag$ targets of $0.3$, $0.5$, and $0.7e^2/h$ at both junctions.
In \Cref{fig:deviceA1_stage2}, we show data for the representative cutter gate settings for which $\Gag = 0.3e^2/h$ for both junctions and the discussion below focuses on this data.
Qualitatively similar observations hold for the other two settings.

Since, as was previously mentioned, there is typically a small voltage drift between Stages 1 and 2, we start the analysis of the Stage 2 data by determining the regions with stable zero bias peaks anew.
We call the ZBPs stable if they are present for at least 2 out of 3 cutter gate settings.
In \Cref{fig:deviceA1_stage2}(c,d), we illustrate ZBPs for our representative cutter gate setting by showing $\GLL$ and $\GRR$ for $\Vp = -1.17175\,$V.
In \Cref{fig:deviceA1_stage2}(b), we see that the corresponding horizontal line passes through a region with stable ZBPs, indicating that these ZBPs are present at least one other cutter gate setting as well.
The $\GLL$ and $\GRR$ data shown in \Cref{fig:deviceA1_stage2}(c,d) is displayed as ``waterfall'' plots in \Cref{fig:deviceA1_stage2}(g,h), which is an alternate but equivalent method of representing the same data.
We re-emphasize that the conductances $\GRR$, $\GLL$ are not topological invariants and are not expected to have quantized values at non-zero temperature and non-zero junction transparency.
So a ZBP, no matter how stable or well-quantized, cannot prove that the system is in a topological phase.
Conversely, the existence of a topological phase in a device is not disproven by a ZBP that has a small magnitude, such as the ZBPs at the left junction in \Cref{fig:deviceA1_stage2}(c,g).
In the TGP, we classify ZBPs by their stability to parameter changes.
In particular, ZBPs that are stable with respect to changes of the cutter gate voltages is a mandatory requirement of the TGP.
We give an example from a Stage 2 measurement in \Cref{fig:deviceA3_stable_ZBPs}.

Next, we use bias scans of the non-local conductances $A(\GRL)$, $A(\GLR)$ to determine the bulk transport gap at each point in the phase diagram.
We illustrate this in \Cref{fig:deviceA1_stage2}(e,f), where we show $A(\GRL)$ and $A(\GLR)$ as a function of $B$ at the $\Vp = -1.17175\,$V horizontal line in \Cref{fig:deviceA1_stage2}(b).
The black curves in \Cref{fig:deviceA1_stage2}(e,f) show the transport gap extracted from, respectively, $\GRL$ or $\GLR$ as a function of $B$ for this $\Vp$ value.
The black curves are determined according to the procedure explained in \Cref{sec:TGP_measurements}.
The transport gap is obtained by taking the minimum of the values extracted from $A(\GRL)$ and $A(\GLR)$.
There is a clear bulk transport gap closing and re-opening visible in $A(\GRL)$ at $B \approx 1.5\,$T.
The gap remains open from $B \approx 1.5\,$T to $B \approx 2.5\,$T.

\begin{figure*}
\includegraphics[width=17.9cm]{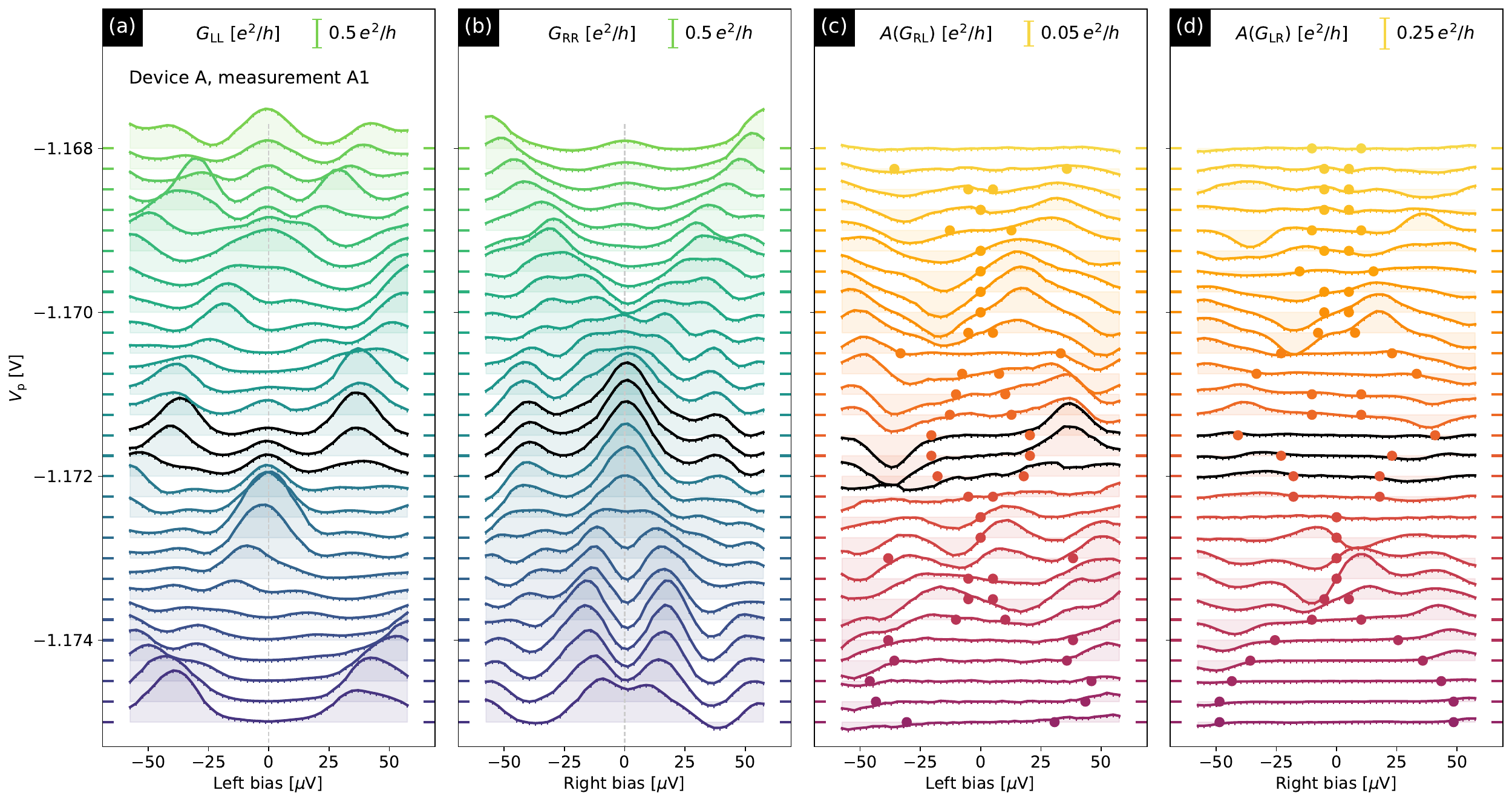}
\vskip -3mm
\caption{
``Waterfall'' conductance plots of (a)~$\GLL$, (b)~$\GRR$, (c)~$A(\GRL)$, (d)~$A(\GLR)$ as a function of the corresponding bias and plunger gate voltage $\Vp$ for device A (measurement A1).
The data is the $B = 1.66\,$T vertical line in \Cref{fig:deviceA1_stage2}(b).
ZBPs and extracted gap points corresponding to SOI$_2$ are shown in black.
}
\label{fig:deviceA1_stage2_conductance_waterfall}
\end{figure*}

We further illustrate the behavior of $A(\GRL)$, $A(\GLR)$ by taking a vertical cut through the $B$-$\Vp$ plane.
In \Cref{fig:deviceA1_stage2_conductance_waterfall}, we show waterfall plots of local and non-local conductances as a function of $\Vp$ at fixed $B = 1.66\,$T.
The black dots in \Cref{fig:deviceA1_stage2_conductance_waterfall}(c,d) indicate the bulk transport gaps extracted from $\GRL$ and $\GLR$.
A gap closing and re-opening is clearly visible in these plots.
We emphasize that these cuts through the phase diagram are a very small sample of the data comprising Stage 2 of measurement A1.

While these illustrative cuts are highly enlightening, they are not the primary goal of Stage 2 of the TGP, which is to derive an experimental phase diagram from the measured conductance matrix as a function of $B$, $\Vp$, $\Vlc$, $\Vrc$, and $\Vb$.
The TGP yields the experimental phase diagrams in \Cref{fig:deviceA1_stage2}(a,b) for our representative cutter gate setting.
All three cutter gate settings yield similar phase diagrams.
The color scheme in \Cref{fig:deviceA1_stage2}(a,b) is the same as in the simulated phase diagrams in \Cref{fig:simulated_SLG_beta_R1_stage2}(a,b).
The most salient feature of \Cref{fig:simulated_SLG_beta_R1_stage2}(a,b) is the presence of an SOI$_2$.
In this region, there are stable ZBPs at both ends of the device, and there is a non-zero bulk transport gap; 78\% of the boundary of this region in the $B$-$\Vp$ plane is gapless.
Device A passed the TGP.

\begin{figure*}
\includegraphics[width=18cm]{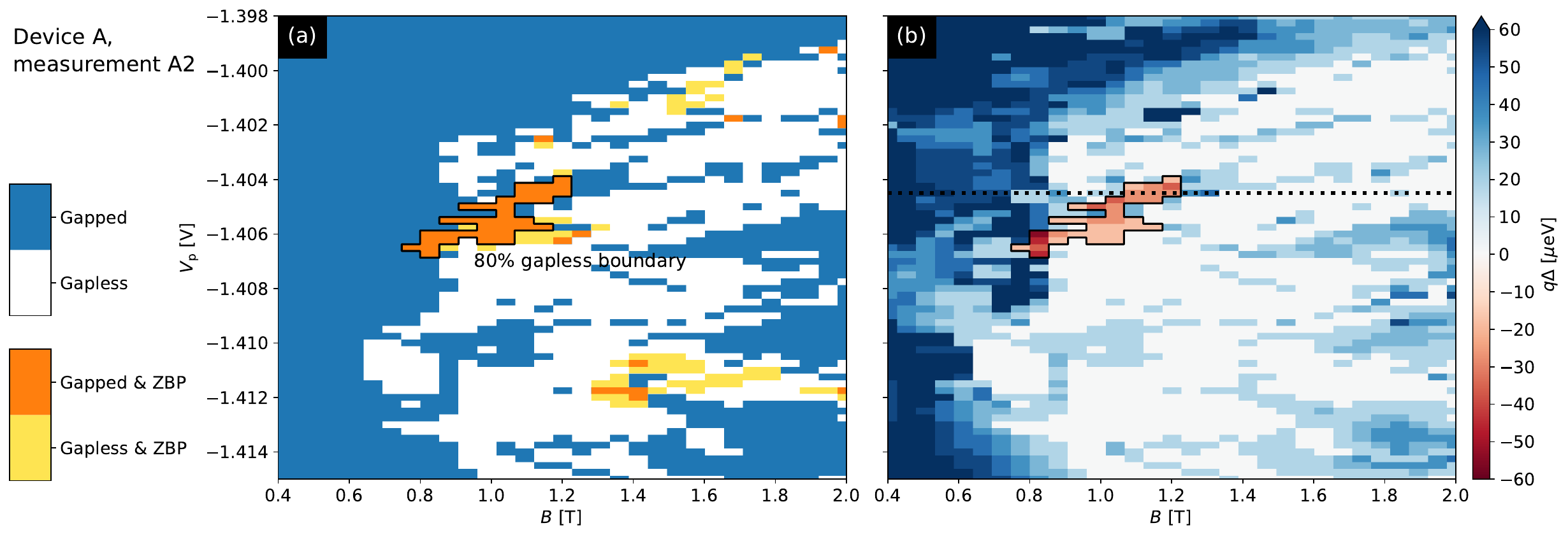}
\includegraphics[width=18cm]{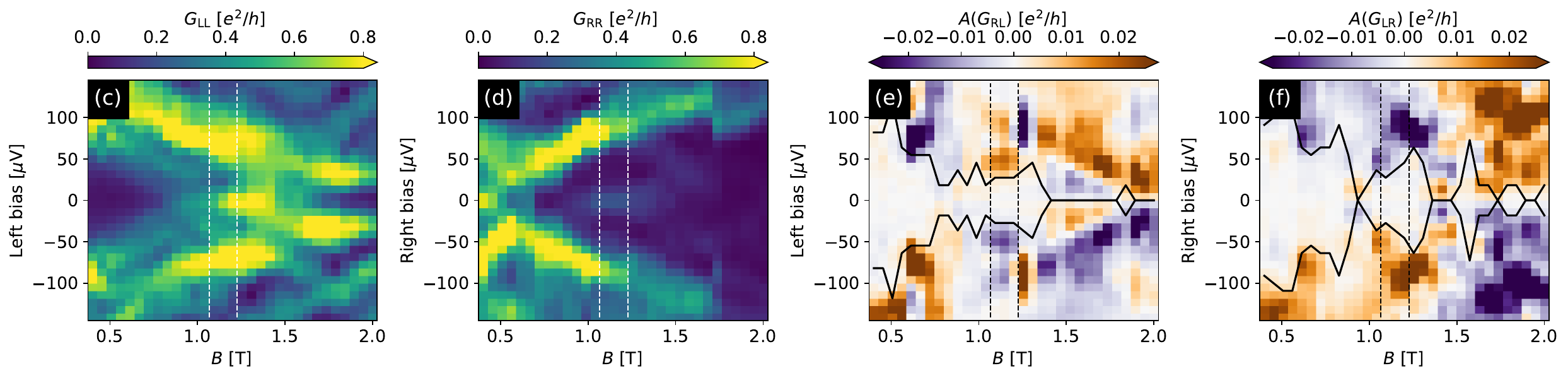}
\includegraphics[width=17.9cm]{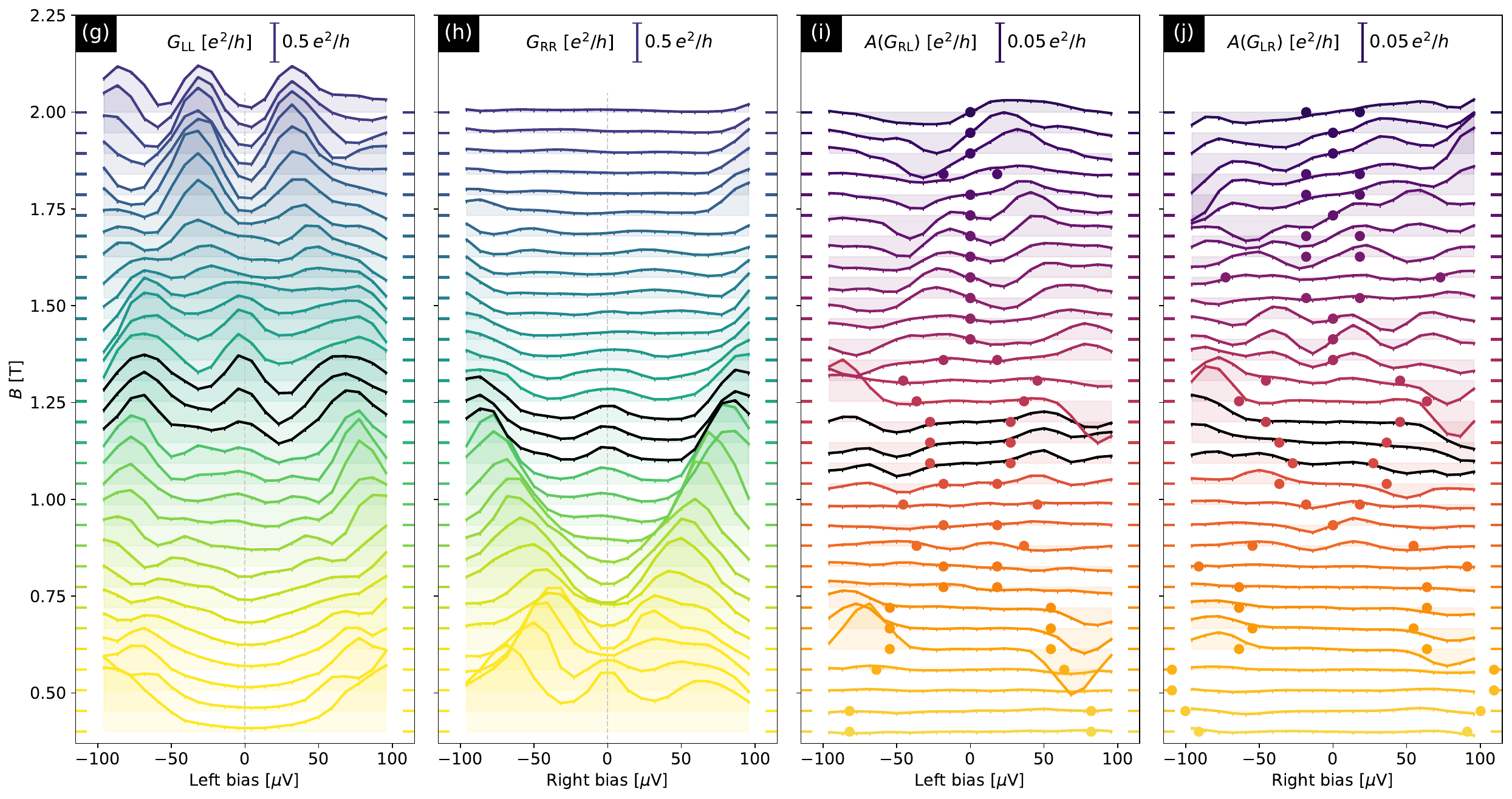}
\vskip -3mm
\caption{
(a)~The experimental phase diagram of device A (measurement A2) that results from combining the clusters of stable ZBPs at both junctions with the map of the locus of zero/non-zero gap.
The stability of ZBPs is determined by varying the cutter gate settings so that for both junctions $\Gag$ take the values 0.3, 0.5, 0.7, and $0.9e^2/h$.
The boundary of the SOI$_2$ is interpreted as a phase transition line, consistent with a visible gap closure along 80\% of it.
(b)~The experimental phase diagram, showing trivial/topological phases, which the TGP identifies with the exterior/interior ($q = \pm 1$) of the SOI$_2$.
The color scale shows the size of the trivial (positive sign) or topological (negative sign) gap.
The protocol assigns a maximum topological gap $\DeltaMax = \SI{29}{\micro\eV}$.
Measured local and anti-symmetrized non-local conductances along the horizontal line in panel b at $\Vp = -1.4045\,$V: (c)~$\GLL$, (d)~$\GRR$, (e)~$A(\GRL)$, (f)~$A(\GLR)$.
The SOI$_2$ lies between the vertical lines.
Panels (g-j) are ``waterfall'' plots representing the same measured data.
The data shown in (c-j) was obtained for left (right) $\Gag \approx 0.5 e^2/h$ ($\approx 0.8 e^2/h$).
The black curves in panels (e,f) and the dots in panels (i,j) indicate where the non-local signal drops below a threshold value, as described in the text.
}
\label{fig:deviceA2_stage2}
\end{figure*}

\begin{figure*}
\includegraphics[width=18cm]{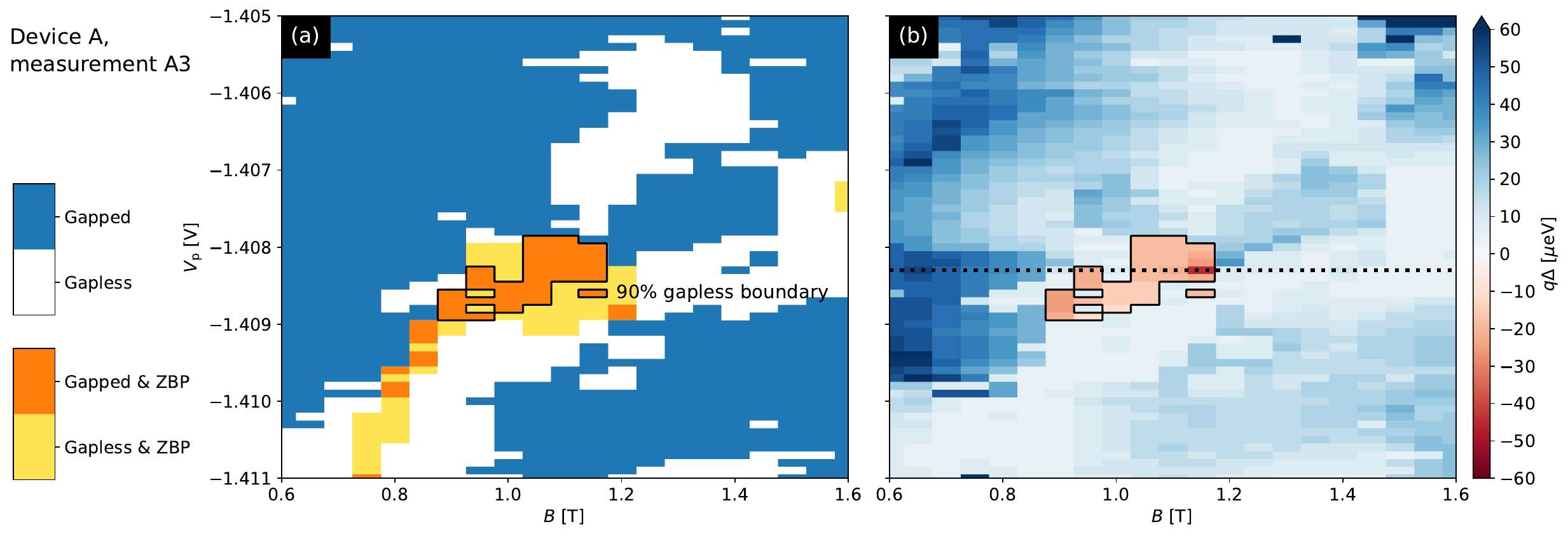}
\includegraphics[width=18cm]{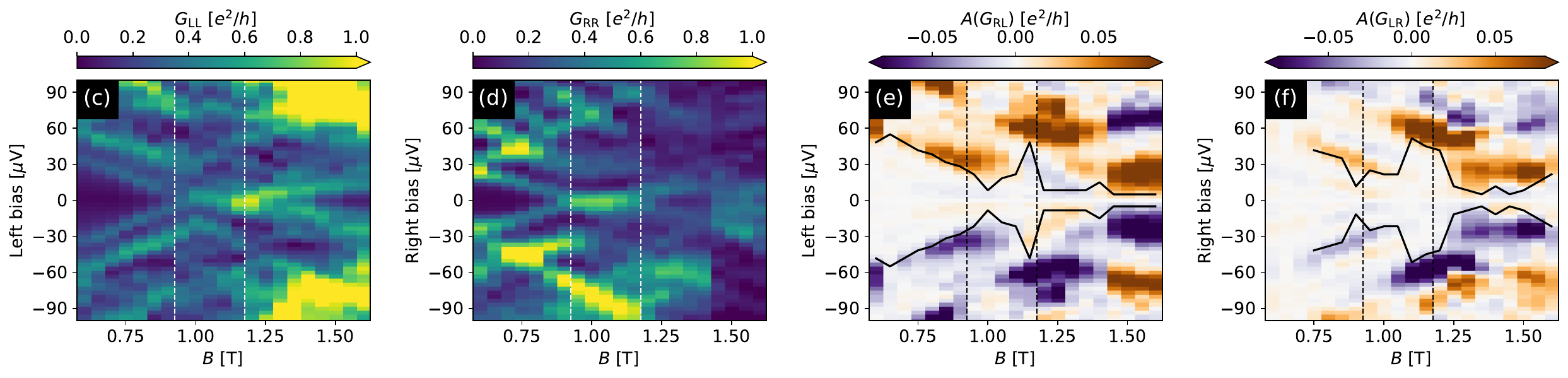}
\includegraphics[width=17.9cm]{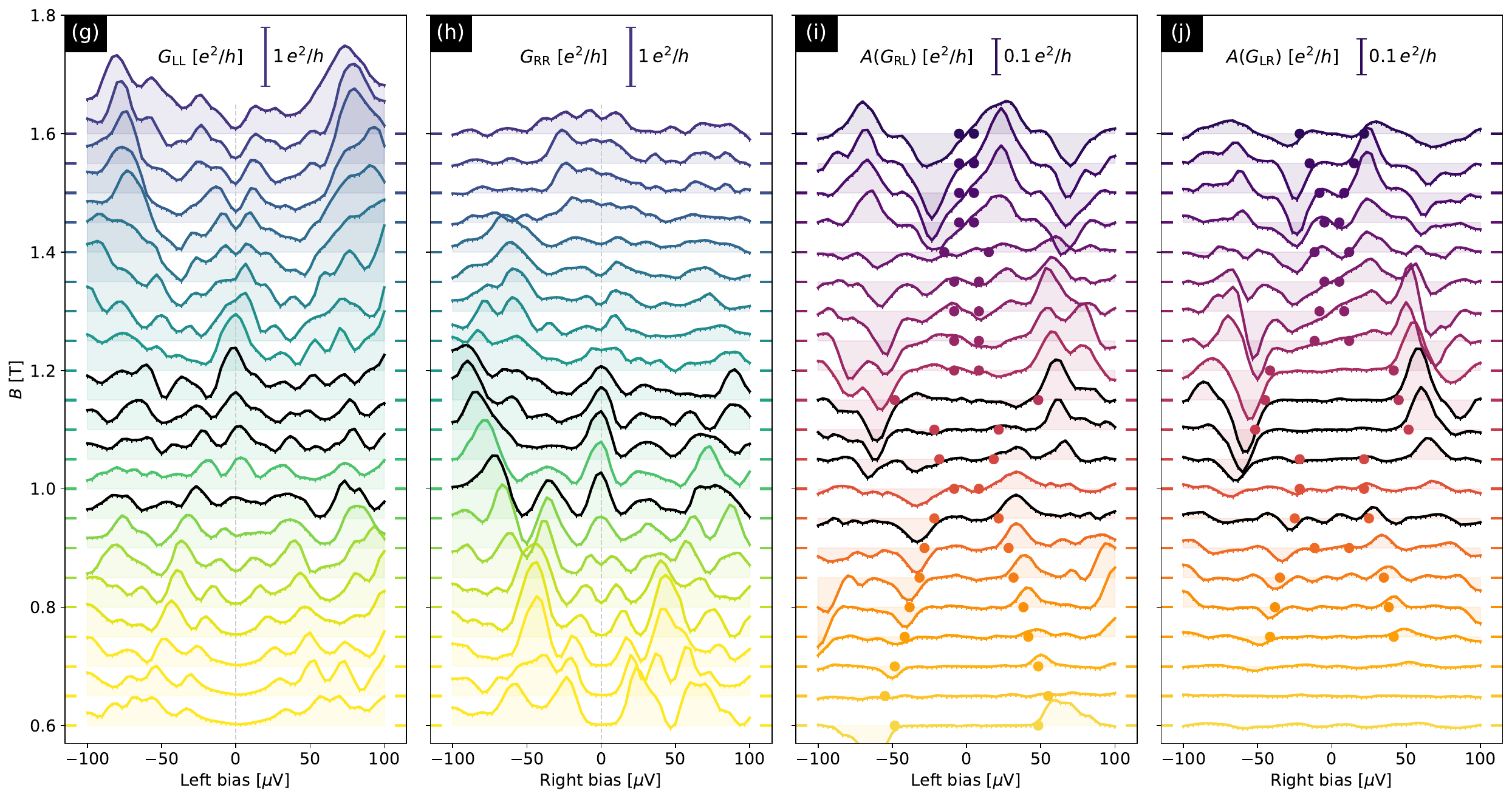}
\vskip -3mm
\caption{
(a)~The experimental phase diagram of device A (measurement A3) that results from combining the clusters of stable ZBPs at both junctions with the map of the locus of zero/non-zero gap.
The stability of ZBPs is determined by varying the cutter gate settings so that for both junctions $\Gag$ take four values between $0.3$ and $0.8e^2/h$ at $B = 0$.
The boundary of the SOI$_2$ is interpreted as a phase transition line, consistent with a visible gap closure along 90\% of it.
(b)~The experimental phase diagram, showing trivial/topological phases, which the TGP identifies with the exterior/interior ($q = \pm 1$) of the SOI$_2$.
The color scale shows the size of the trivial (positive sign) or topological (negative sign) gap.
The protocol assigns a maximum topological gap of $\DeltaMax = \SI{22}{\micro\eV}$.
Measured local and anti-symmetrized non-local conductances along the horizontal line in panel b at $\Vp = -1.4083\,$V: (c)~$\GLL$, (d)~$\GRR$, (e)~$A(\GRL)$, (f)~$A(\GLR)$.
The SOI$_2$ lies between the vertical lines.
Panels (g)-(j) are ``waterfall'' plots representing the same measured data.
The data shown in (c)-(j) was obtained for left (right) $\Gag\approx 0.6 e^2/h$ ($\approx 0.4 e^2/h$).
The black curves in panels (e) and (f) and the dots in panels (i) and (j) indicate where the non-local signal drops below a threshold value, as described in the text.
}
\label{fig:deviceA3_stage2}
\end{figure*}

The crucial point of the TGP is to not rely on a single feature to identify a topological phase, but instead to rely on the totality of the data to provide evidence for the observation of a topological phase.
Indeed, each pixel in \Cref{fig:deviceA1_stage2}(b) is determined by conductance data in a neighborhood of points around that pixel and for a range of cutter gate settings.

From $A(\GRL)$, $A(\GLR)$, we infer a bulk gap closing and re-opening, which is a signature of a second-order phase transition.
It is important to distinguish such behavior in the non-local conductances $A(\GRL)$, $A(\GLR)$ from apparent gap closings/re-openings in the local conductances $\GRR$, $\GLL$, which could easily be the motion of a local state towards zero energy, rather than a bulk phenomenon.

This phase transition line separates the high-field gapped phase from the gapped trivial superconducting phase that is present at low fields.
It does not quite surround the high-field gapped phase: 78\% of the boundary shows a gap closing in $A(\GRL)$, $A(\GLR)$.
This surpasses $(\mathrm{GB}\%)_\mathrm{th}$; it is similar to the percentage of the boundary of the SOI$_2$ that is gapless in the simulated data of \Cref{fig:simulated_SLG_beta_R1_stage2} and is typical for simulations of this device design and disorder level.
Consequently, we believe that the second order phase transition line that surrounds 78\% of our putative topological phase is, in fact, part of an unbroken transition line surrounding the entire phase.
As noted in our discussion of $(\mathrm{GB}\%)_\mathrm{th}$ in \Cref{sec:TGP_parameters}, one possibility is that the gap closing is not visible along 22\% of the boundary of the SOI$_2$ due to a suppression of the signal by disorder/non-uniformity while another is that the topological region is larger than the SOI$_2$.
Indeed, for the cutter gate setting shown in \Cref{fig:deviceA1_stage2}, there are ZBPs at both junctions up to $B = 2.5\,$T.%
\footnote{At the left junction, this ZBP is small, but above the measurement resolution, as may be seen in \Cref{fig:deviceA1_stage2}(c,g).}
However, at the other cutter gate settings, there is no visible ZBP at the left junction.

The high-field gapped phase is characterized by stable ZBPs at both ends of the wire, which is consistent with the topological phase.
For some $\Vp$ values, the ZBPs appear before the gap re-opens, including at the $\Vp = -1.17175\,$V horizontal line in \Cref{fig:deviceA1_stage2}(b).
This is consistent with a scenario in which quasi-MZMs~\cite{Prada12, Kells12, Tewari14, Liu17, Vuik19, Pan21b} are precursors to the transition into the topological phase, which is frequently seen in simulations.

The maximum topological gap is $\DeltaMax = \SI{23}{\micro\eV}$ for this cutter gate setting.%
\footnote{The extracted gap can depend on the cutter gate setting.
Stability of the gap extraction with respect to cutter gate setting is not a requirement of the TGP.}
Over the SOI$_2$, which has an extent of $\delta B \approx 500\,$mT, $\delta\Vp \approx 1.5\,$mV, the extracted topological gap increases from zero to $\DeltaMax$ in such a way that its median value over the region within the black line in \Cref{fig:deviceA1_stage2}(a,b) is \SI{20}{\micro\eV}.
From the phase diagram in \Cref{fig:deviceA1_stage2}(a,b), we see that the lowest field at which the gap closes near the SOI$_2$ is $\approx 0.8\,$T, which implies an effective $g$-factor of at least $|g_\mathrm{eff}| \approx 5.6$.
Here, we define $|g_\mathrm{eff}| \equiv 2\DeltaInd / \muB B_\mathrm{min}$, where $B_\mathrm{min}$ is the lowest field at which the gap closes.%
\footnote{Here we define the effective $g$-factor as the average slope of the extracted induced gap vs $B$-field.
This is different from the conventional definition of the spin $g$-factor in terms of $dE/dB$ at $k=0$.
The former depends on spin-orbit coupling and orbital physics whereas the latter does not.
However, in the single sub-band regime, where the lowest energy state has momentum close to $k=0$, both the orbital effect of the $B$ field and spin-orbit coupling effects are small.
In this case, $|g_\mathrm{eff}|$ is a good proxy for $|g^{\star}|$.}
This value of $|g_\mathrm{eff}|$ is close to the optimal value for this device design and material stack.

The induced gap (and all structure associated with its closing/re-opening) decreases rapidly when the magnetic field is rotated away from the direction of the wire, as expected.
The transition to the topological phase should become more smeared as the temperature is increased, but it is difficult to study this systematically due to voltage drifts.

Comparing the experimental data in \Cref{fig:deviceA1_stage2} to the simulated data in \Cref{fig:simulated_SLG_beta_R1_stage2}, we note both the qualitative and quantitative similarity between the phase diagrams.
In both simulated and measured data, there are gap closings at similar $\Vp$-dependent $B$-field values, and the extent of both the gapless regions and the SOI$_2$s are of similar size in the $B$-$\Vp$ plane.
However, we emphasize again that the main role of simulated data such as that shown in \Cref{fig:simulated_SLG_beta_R1_stage2} is to test the TGP on (simulated) devices for which we know the phase diagram and not to reproduce the experimental phase diagram.

\subsubsection{Reproducibility of the data: Measurements A2 and A3}
\label{sec:measurementsA2A3}

We now present experimental data from a different cooldown in which measurements A2 and A3 were performed one week apart.
These measurements produced similar data sets, both passing the TGP, indicating the reproducibility of our data and the device's stability from one measurement run to another.
Both of these data sets are consistent with measurement A1 shown in \Cref{sec:deviceA1}.

\begin{figure*}
\includegraphics[height=3.95cm]{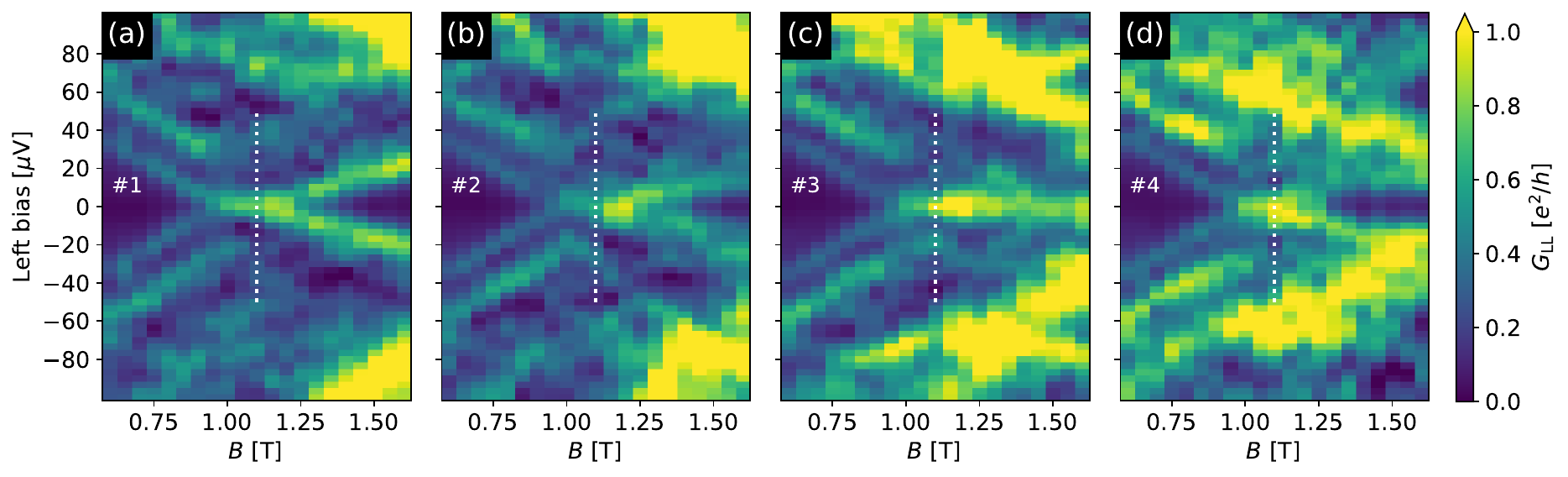}
\hspace{3mm}\includegraphics[height=3.95cm]{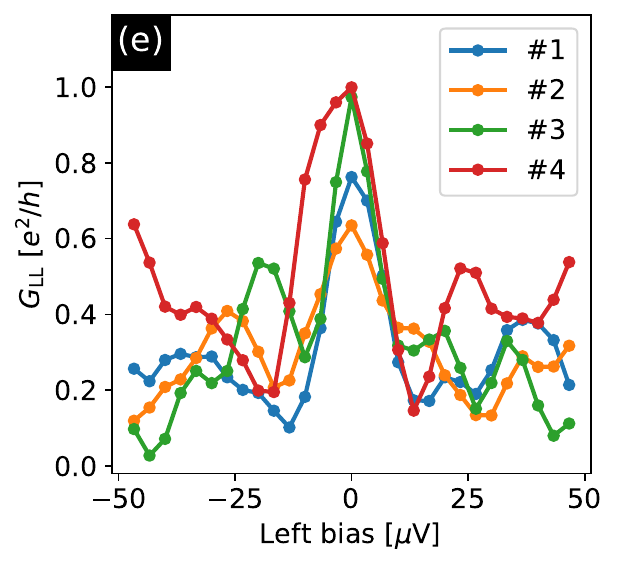}
\vskip -2mm
\caption{(a)-(d) Bias-field cuts of measurement A3 at $\Vp = -1.4083\,$V.
$\GLL$ is shown
for 4 different left cutter gate settings \#1-4 corresponding, respectively, to
$\Gag$ of approximately $0.5$, $0.6$, $0.7$,
and $0.8 e^2/h$.
Panel (b) for cutter \#2 shows the same data as \Cref{fig:deviceA3_stage2}(c).
(e) Line-cut at $B = 1.1\,$T for the above cutter gate settings.
The chosen values of $B$ and $\Vp$ lie within the SOI$_2$ shown in \Cref{fig:deviceA3_stage2}(b) for left and right cutters corresponding to $\Gag$ $0.6$ and $0.4 e^2/h$, respectively.
Note that the ZBP height is of the order of $e^2/h$ for all left cutter gate settings.}
\label{fig:deviceA3_stable_ZBPs}
\end{figure*}

In our simulations, we saw that devices can pass the
TGP for some disorder configurations but not others.
Each cooldown typically leads to a somewhat different disorder configuration, resulting, for example, in a shift of the gate voltages at which we see the depletion of the lowest sub-band.
As mentioned previously, when the device idles for a week, the disorder configuration can also drift slightly.
Hence, we expect that the same device will pass the TGP in some measurements but not in other measurements occurring a week or more apart or in different cooldowns.
This was the case with device A.
It regularly passed the TGP, but also failed sometimes.
In this subsection, we focus on measurements A2 and A3, in which device A passed the TGP with a topological phase that shifted in parameter space.
For measurement A2, device A was warmed-up, removed from the dilution refrigerator in which A1 was performed, cooled down in a different dilution refrigerator, and re-measured.
In accordance with the TGP, we performed Stage 1 measurements and identified an ROI$_1$ with stable ZBPs near $\Vp = -1.4\,$V.
The results of the subsequent Stage 2 measurement are shown in \Cref{fig:deviceA2_stage2}.

The phase diagrams in \Cref{fig:deviceA2_stage2}(a,b) have the same basic features as those in \Cref{fig:deviceA1_stage2}(a,b).
The primary differences are as follows.
The lowest gap closing point in A2 is slightly lower in field than in A1, leading to an effective $g$-factor of $|g_\mathrm{eff}| \approx 6.4$.
The topological phase starts at lower fields, close to 0.8\,T, which is closer to the lowest fields at which the gap closes than in A1, and it extends over a larger range of plunger gate voltages, $\delta\Vp = 2.5\,$mV but a similar magnetic field range $\delta B \approx 500\,$mT.
The maximum topological gap in measurement A2 is $\DeltaMax = \SI{29}{\micro\eV}$, see \Cref{fig:deviceA2_stage2}(b), which is the largest observed for device A, and the percentage of the boundary that is gapless is 80\%.
The local conductances are shown in \Cref{fig:deviceA2_stage2}(c,d).
In addition to ZBPs, there is also a strong local resonance at the right junction which is evident in \Cref{fig:deviceA2_stage2}(d).
However, this resonance moves away from zero bias as $B$ is increased.
Furthermore, this resonance disappears in the subsequent measurement shown in \Cref{fig:deviceA3_stage2}(d), indicating that this is an accidental feature due to an impurity that moved between measurements A2 and A3.
This trivial resonance partially obscures the stable ZBP, which has a smaller amplitude.

In \Cref{fig:deviceA3_stage2}, we show measurement A3 which was performed in the same cooldown as measurement A2 but one week later.
The electrostatic environment of the system drifted by 2-3\,mV during the week between the two measurement runs, as is typically the case in our devices.
However, the main qualitative features are reproduced from one run to the next: there is a topological phase with a comparable critical field and similar overall shape.
The size of the maximum topological gap has decreased from its value in A2 to $\DeltaMax = \SI{22}{\micro\eV}$, which is close to value found in measurement A1.

Finally, we discuss the stability of ZBPs with respect to local perturbations.
In \Cref{fig:deviceA3_stage2}(c,d) one can see stable ZBPs at both junctions over a magnetic field range $\delta B \approx 0.5\,$T.
They are similarly stable with respect to changes in $\Vp$, as may be seen from the vertical extent of the orange region in \Cref{fig:deviceA3_stage2}(a).
These peaks are also stable with respect to cutter changes modulating the transparency of the junctions, as we show in \Cref{fig:deviceA3_stable_ZBPs}.
There are ZBPs with height $O(e^2/h)$ that are present for 4 cutter gate settings.
These changes in left cutter gate settings tune $\Gag$ on both sides to vary over the range $0.1$ and $1.0 e^2/h$.
Thus, while these cutter changes significantly modify the junction transparencies, the ZBPs remain stable with respect to these perturbations.
In SOI$_2$, there are ZBPs exhibiting this type of stability at both junctions.

To conclude this subsection, we believe that the totality of the data from device A~--- passing the TGP in multiple cooldowns and re-measurements, qualitative and quantitative consistency with simulations, and the stability of the SOI$_2$ with respect to various perturbations~--- provides strong evidence for the observation of a stable topological superconducting phase supporting MZMs in this device.
We now turn to the reproducibility of these results in other devices.

\begin{figure*}
\includegraphics[width=18cm]{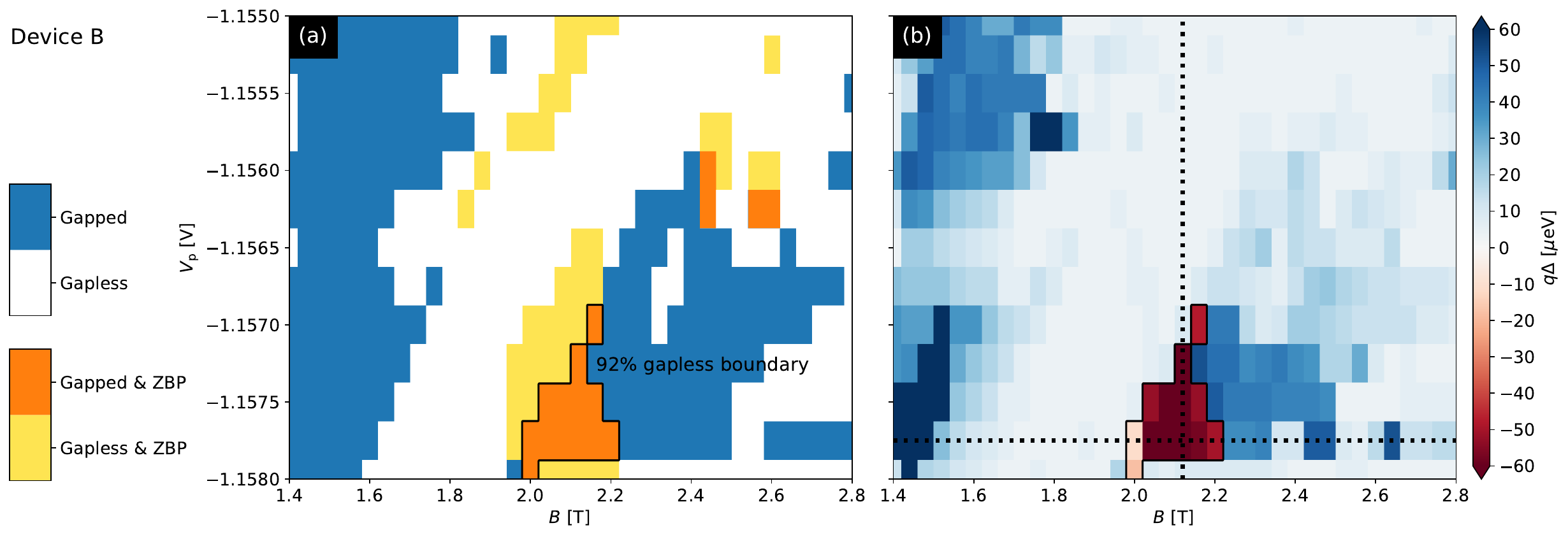}
\includegraphics[width=18cm]{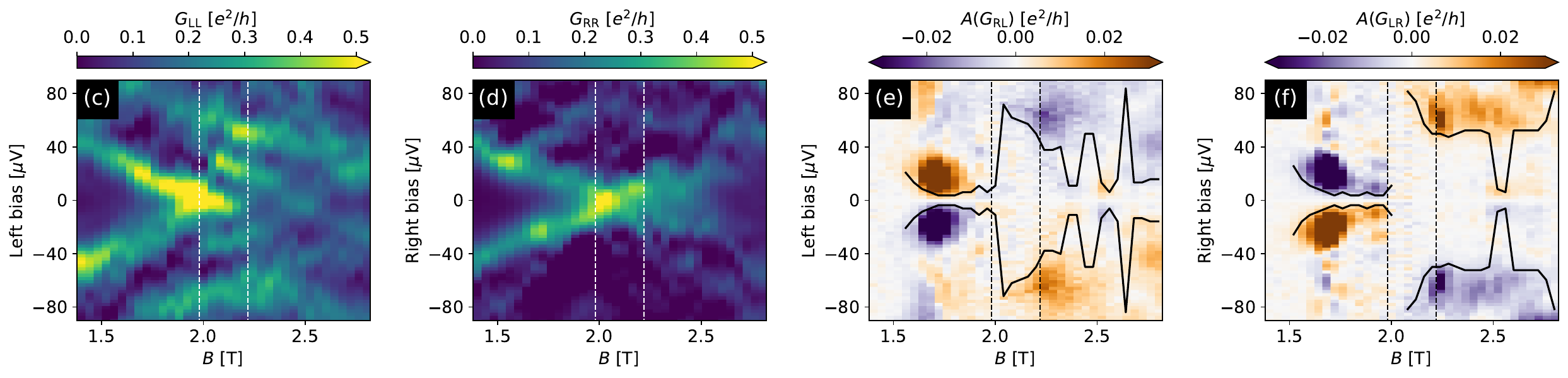}
\includegraphics[width=17.9cm]{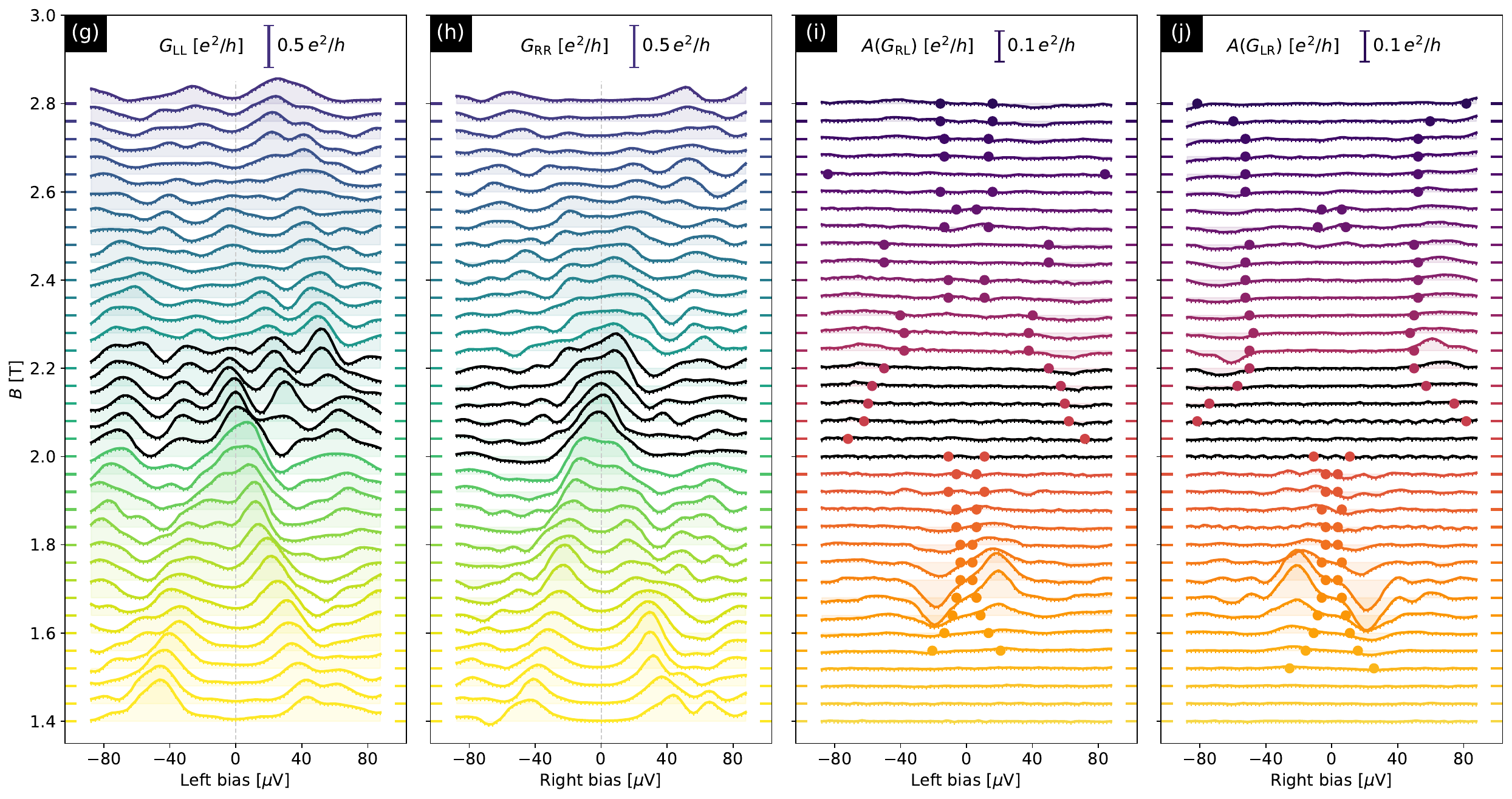}
\vskip -3mm
\caption{
(a)~The experimental phase diagram of device B that results from combining the clusters of stable ZBPs at both junctions with the map of the locus of zero/non-zero gap.
The stability of ZBPs is determined by varying the cutter gate settings so that for both junctions $\Gag$ take the values 0.5, 0.41, 0.33, 0.24, and $0.15 e^2/h$ at $B = 1.4\,$T.
The boundary of the SOI$_2$ is interpreted as a phase transition line, consistent with a visible gap closure along 92\% of it.
(b)~The experimental phase diagram, showing trivial/topological phases, which the TGP identifies with the exterior/interior ($q = \pm 1$) of the SOI$_2$.
The color scale shows the size of the trivial (positive sign) or topological (negative sign) gap.
The protocol assigns a maximum topological gap of $\DeltaMax = \SI{61}{\micro\eV}$.
Measured local and anti-symmetrized non-local conductances along the horizontal line in panel (b) at $\Vp = -1.15775\,$V: (c)~$\GLL$, (d)~$\GRR$, (e)~$A(\GRL)$, (f)~$A(\GLR)$.
The SOI$_2$ lies between the vertical lines.
Panels (g)-(j) are ``waterfall'' plots representing the same measured data.
The data shown in (c)-(j) was obtained for $\Gag\approx 0.41 e^2/h$ on both sides.
The black curves in panels (e) and (f) and the dots in panels (i) and (j) indicate where the non-local signal drops below a threshold value, as described in the text.
}
\label{fig:deviceB_stage2}
\end{figure*}

\begin{figure*}
\includegraphics[width=18cm]{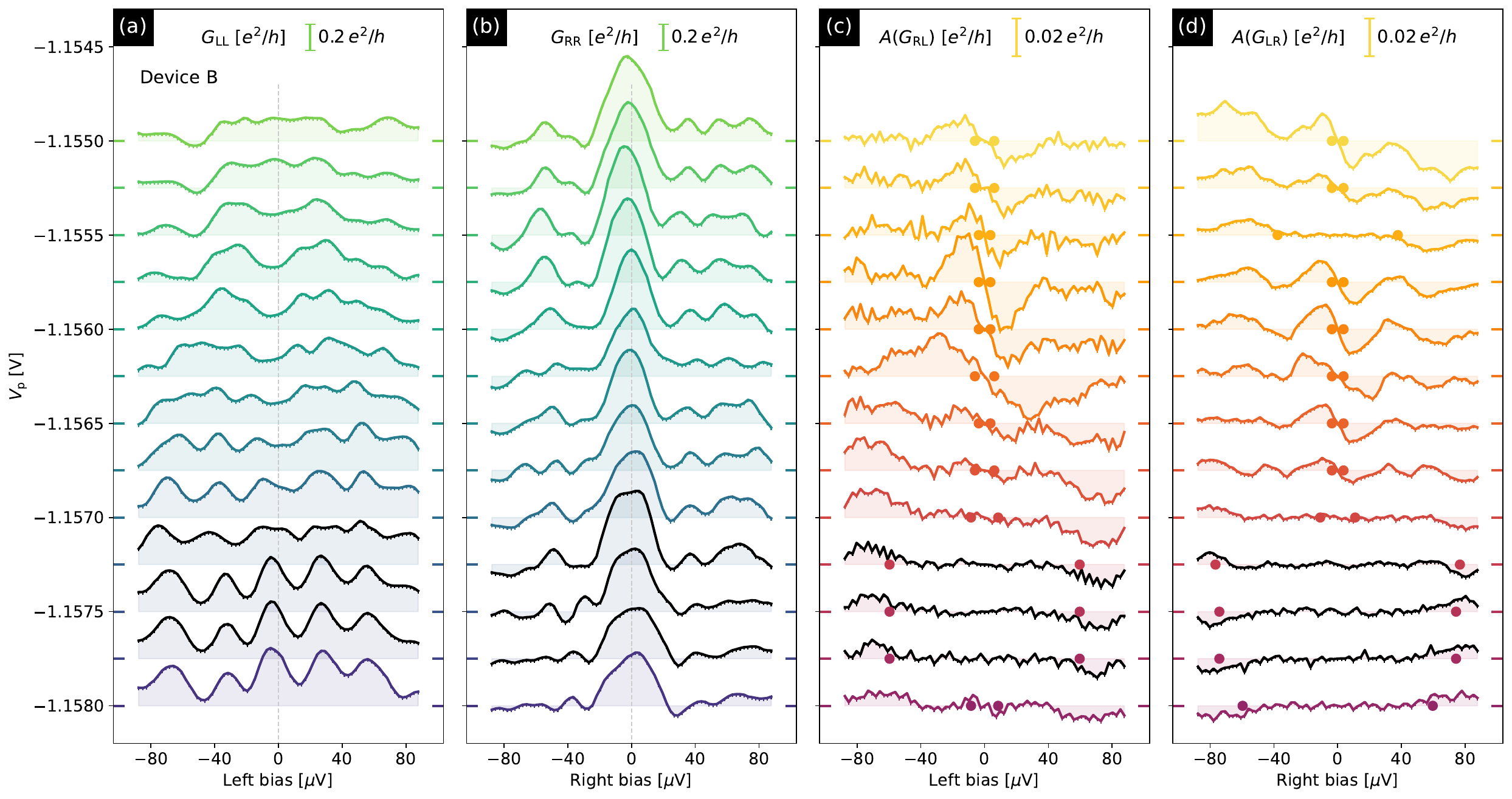}
\vskip -3mm
\caption{
``Waterfall'' conductance plots of (a)~$\GLL$, (b)~$\GRR$, (c)~$A(\GRL)$, (d)~$A(\GLR)$ as a function of the corresponding bias and plunger gate voltage $\Vp$ for device B.
The data is the $B = 2.12\,$T vertical line in \Cref{fig:deviceB_stage2}(b).
ZBPs and extracted gap points corresponding to SOI$_2$ are shown in black.
}
\label{fig:deviceB_stage2_waterfall_plunger}
\end{figure*}

\begin{figure*}
\includegraphics[width=18cm]{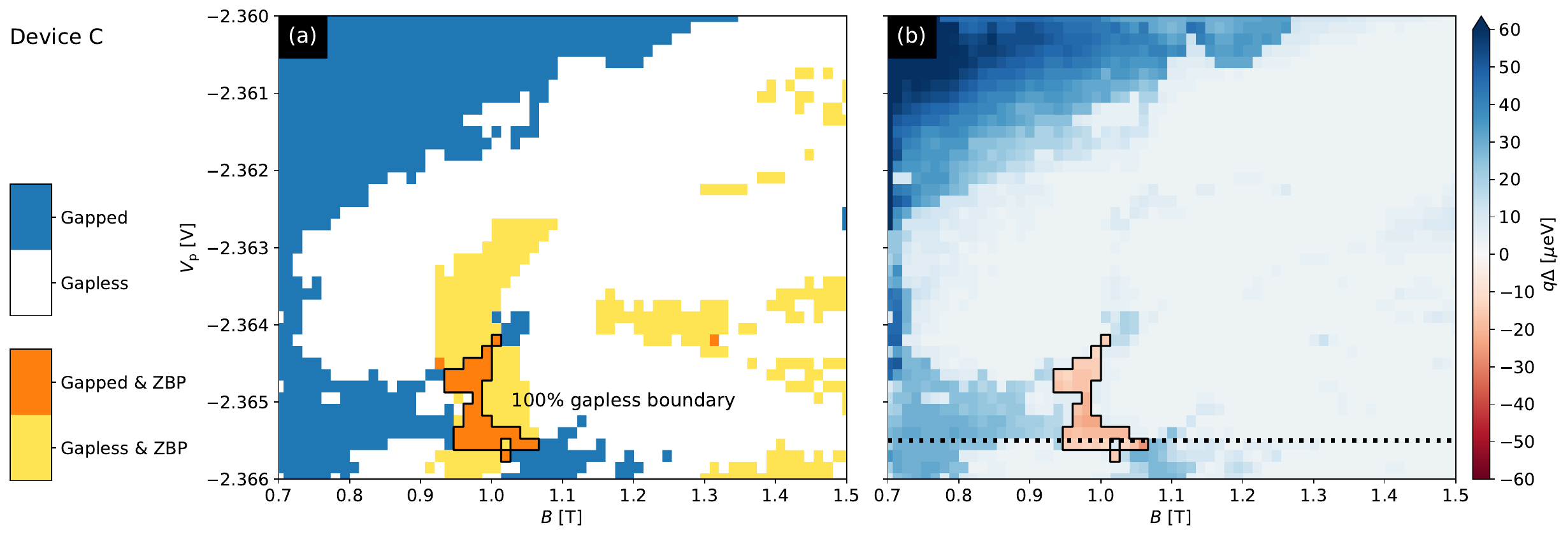}
\includegraphics[width=18cm]{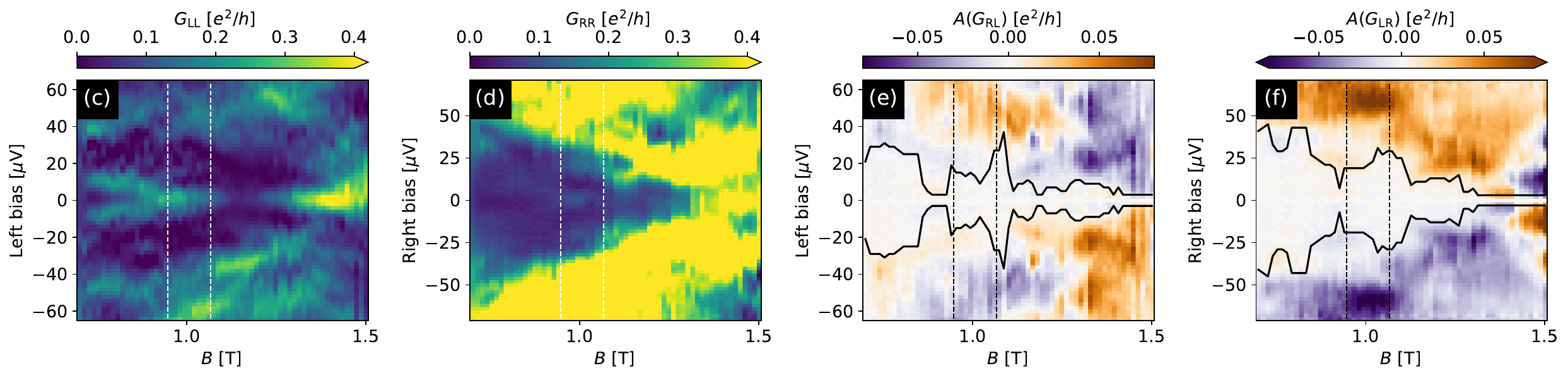}
\includegraphics[width=17.9cm]{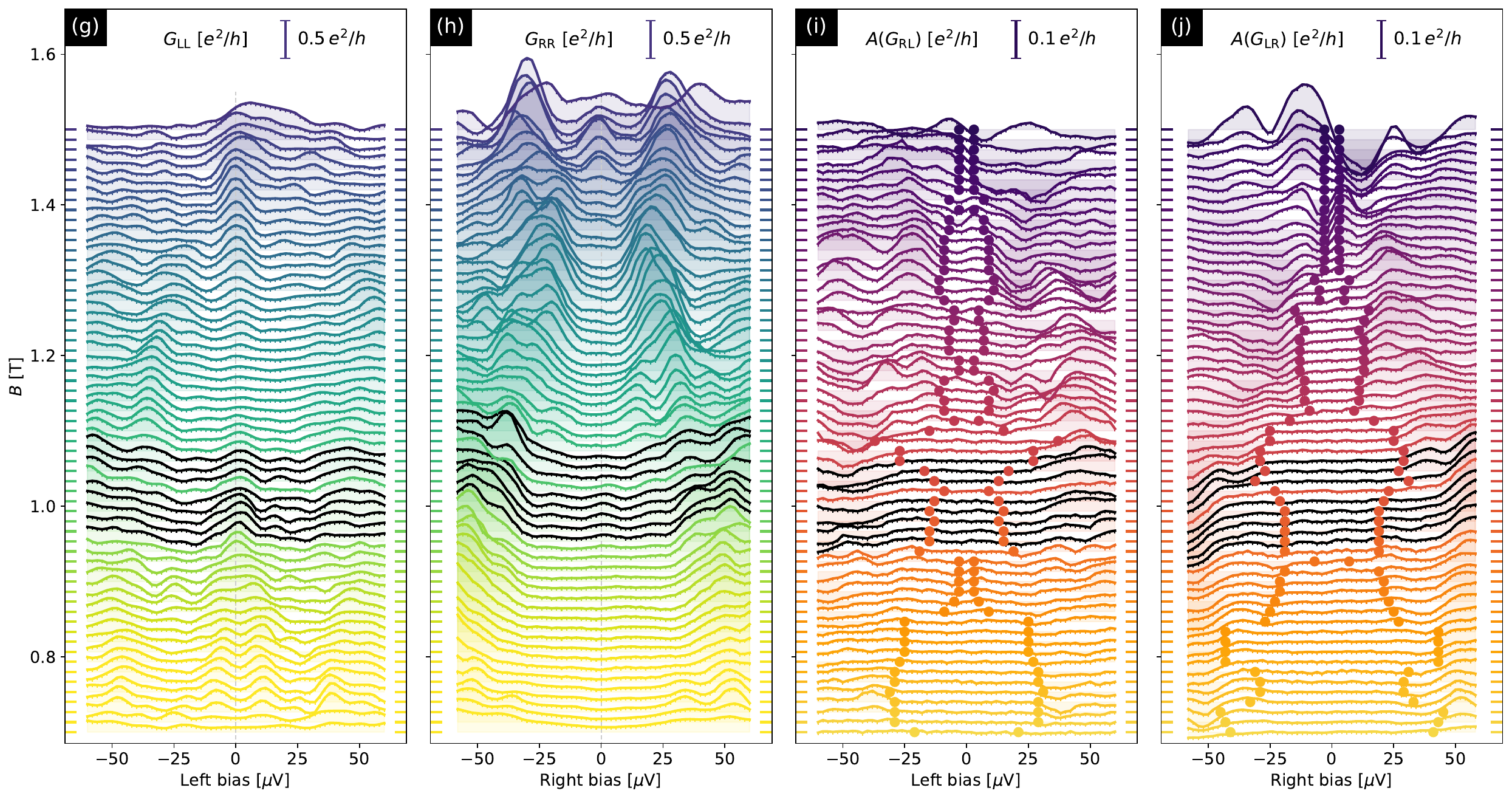}
\vskip -3mm
\caption{
(a)~The experimental phase diagram of device C that results from combining the
clusters of stable ZBPs at both junctions with the map of the locus of zero/non-zero gap.
The stability of ZBPs is determined by varying the cutter gate settings so that $\Gag$ range from $0.6$ and $1.0 e^2/h$ on the left and from $0.2$ and $0.4 e^2/h$ on the right.
The boundary of the SOI$_2$ is interpreted as a phase transition line, consistent with a visible gap closure along 100\% of it.
(b)~The experimental phase diagram, showing the trivial/topological phases, which the TGP identifies with the exterior/interior ($q = \pm 1$) of the SOI$_2$.
The color scale shows the size of the trivial (blue) or topological (red) gap.
The protocol assigns maximum topological gap of \SI{19}{\micro\eV}.
Measured local and anti-symmetrized non-local conductances along the horizontal line in panel (b) at $\Vp = -2.3655\,$V: (c)~$\GLL$, (d)~$\GRR$, (e)~$A(\GRL)$, (f)~$A(\GLR)$.
The SOI$_2$ lies between the vertical lines.
Panels (g)-(j) are ``waterfall'' plots representing the same measured data.
The data shown in (c)-(j) was obtained for left (right) $\Gag$ of approximately $0.6 e^2/h$ ($0.2 e^2/h$).
The black curves in panels (e) and (f) and the dots in panels (i) and (j) indicate where the non-local signal drops below a threshold value, as described in the text.
}
\label{fig:deviceC_stage2}
\end{figure*}

\begin{figure*}
\includegraphics[width=18cm]{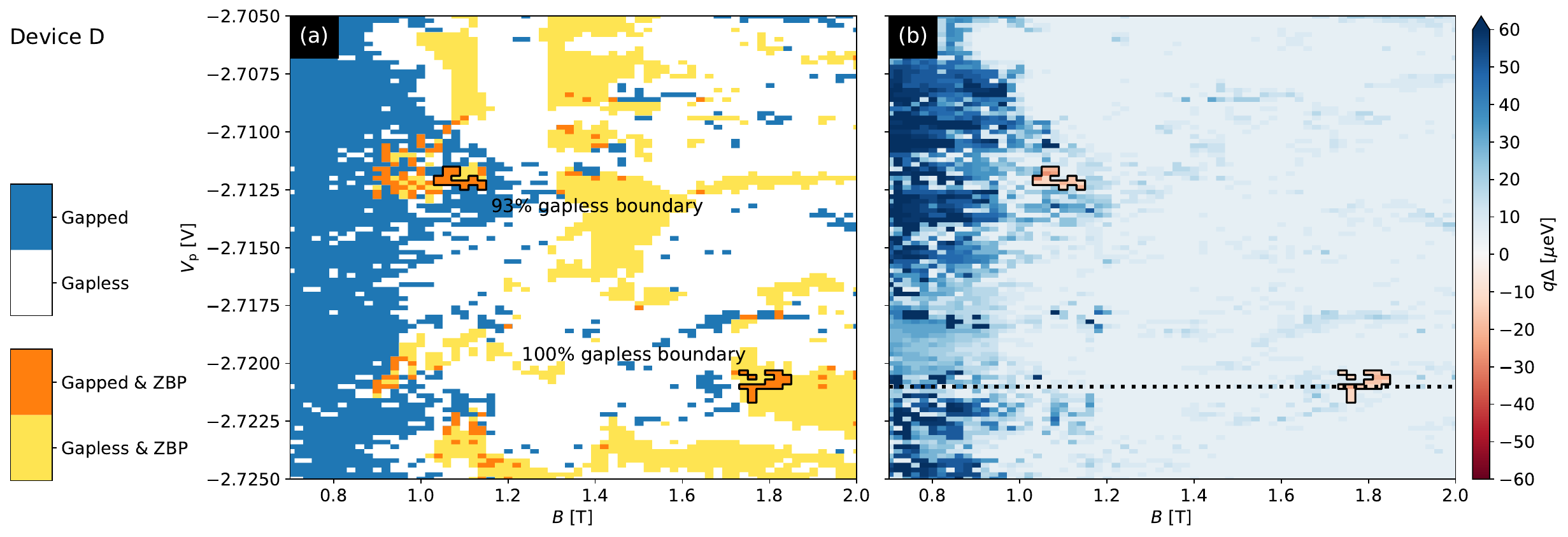}
\includegraphics[width=18cm]{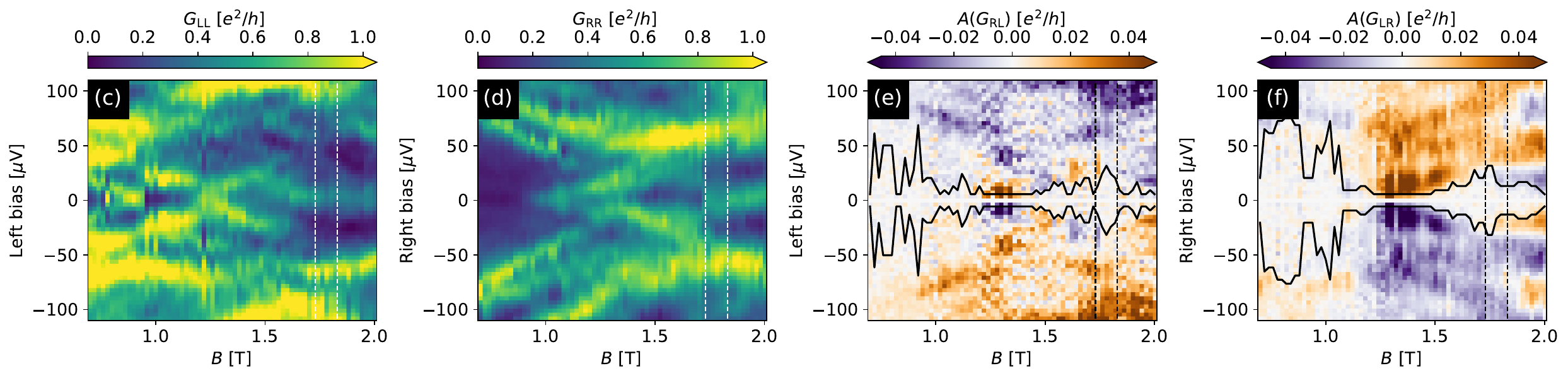}
\includegraphics[width=17.9cm]{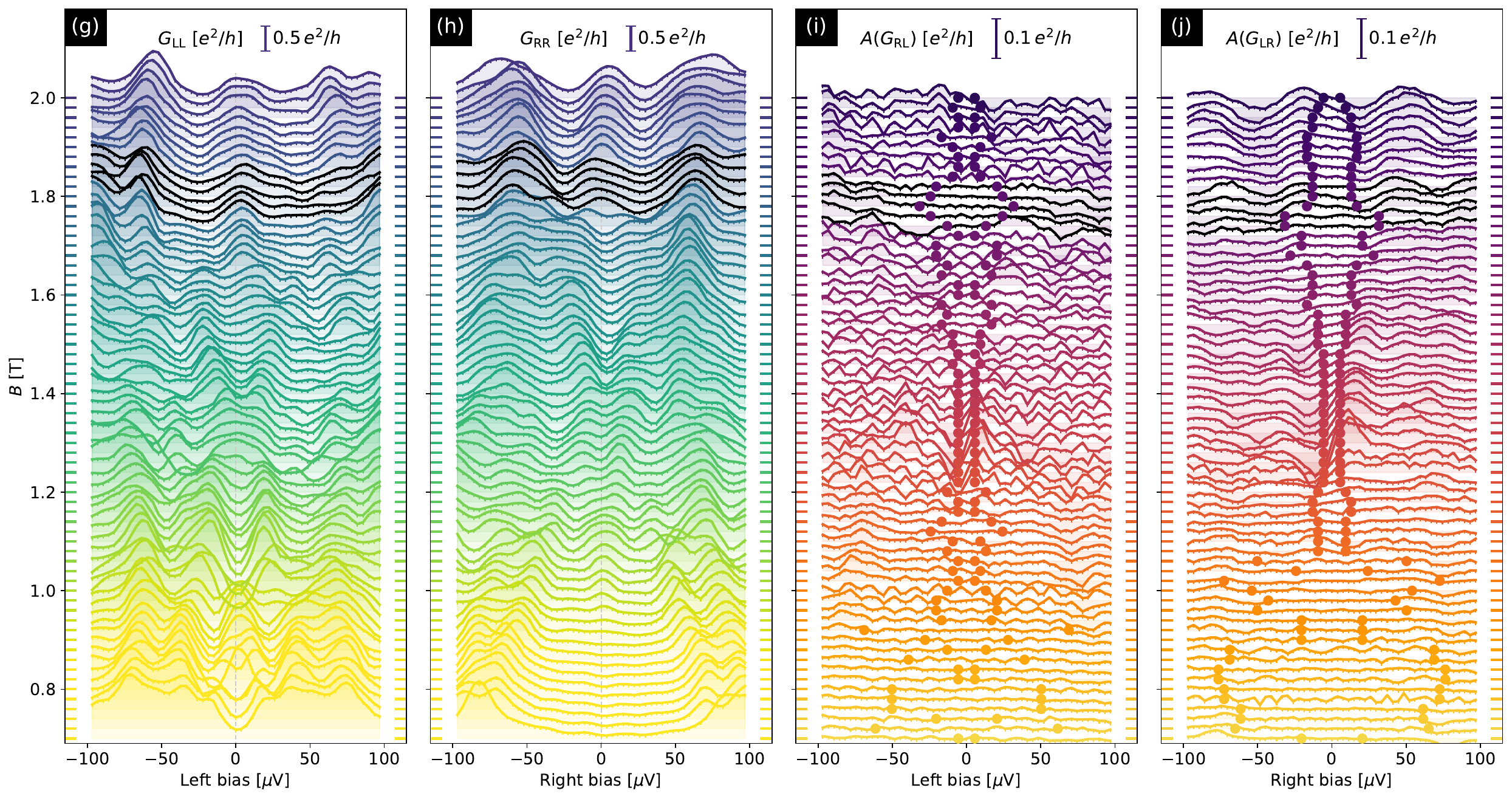}
\vskip -3mm
\caption{
Stage 2 data for device D.
There are two SOI$_2$s in the phase diagram.
(a)~The experimental phase diagram of device D that results from combining the clusters of stable ZBPs at both junctions with the map of the locus of zero/non-zero gap.
The boundary of the SOI$_2$s are interpreted as phase transition lines, consistent with visible gap closures along, respectively, 100\% and 93\% of them.
(b)~The experimental phase diagram, showing the trivial/topological phases, which the TGP identifies with the exterior/interior ($q = \pm 1$) of the SOI$_2$.
The color scale shows the size of the trivial (blue) or topological (red) gap.
The protocol assigns maximum topological gaps of \SI{17}{\micro\eV} and \SI{21}{\micro\eV} for the top and bottom clusters, respectively.
Measured local and anti-symmetrized non-local conductances along the horizontal line in panel b at $\Vp = -2.721\,$V: (c)~$\GLL$, (d)~$\GRR$, (e)~$A(\GRL)$, (f)~$A(\GLR)$.
The SOI$_2$ lies between the vertical lines.
Panels (g)-(j) are ``waterfall'' plots representing the same measured data.
All data shown in this figure were obtained for left/right $\Gag$ of approximately $0.8 e^2/h$.
The black curves in panels (e) and (f) and the dots in panels (i) and (j) indicate where the non-local signal drops below a threshold value, as described in the text.
}
\label{fig:deviceD_stage2}
\end{figure*}

\subsection{Experimental data from other devices}
\label{sec:other_devices}

Since disorder can destroy the topological phase, and different devices will have different disorder realizations, we can expect quantitative and qualitative differences between devices.
Indeed, we have measured devices in which we were not able to find a topological phase.
However, devices that have a narrow Al strip, zero-field induced gap to parent gap ratio in the required range, and weak disorder often pass the TGP while devices not meeting these requirements have never passed TGP, as expected from simulations.
For example, no devices with dielectric charge density above $3 \cdot \scu$, as extracted from a Hall bar on the same chip, have passed TGP.

In this section, we show data from devices B, C, and D, summarized in \Cref{fig:deviceB_stage2,fig:deviceC_stage2,fig:deviceD_stage2}, which also pass the TGP, thereby demonstrating that we can reproducibly fabricate devices passing the TGP.
These three devices are DLG devices.
Device B is built on the $\varepsilon$-stack and has $\Delta_\mathrm{Al} = 326 \pm \SI{29}{\micro\eV}$ and $\DeltaInd = 169 \pm \SI{11}{\micro\eV}$; hence, the ratio of the induced gap to the parent gap in device B is 0.52, which is slightly larger than in device A and very close to optimal.
It has the largest topological gap reported in this paper: $\DeltaMax = \SI{61}{\micro\eV}$.
Device C is built on the $\delta$-stack and has $\Delta_\mathrm{Al} = 292 \pm \SI{8}{\micro\eV}$ and $\DeltaInd = 104 \pm \SI{6}{\micro\eV}$; in device D, which is also built on the $\delta$-stack, the corresponding gaps are $\Delta_\mathrm{Al} = 293 \pm \SI{9}{\micro\eV}$ and $\DeltaInd = 117 \pm \SI{20}{\micro\eV}$.
The ratio of the induced gap to the parent gap in devices C and D, 0.35, and 0.4, respectively, is somewhat smaller than the nearly-optimal value of 0.52 that it takes in device B or even the value of 0.44 that it takes in device A.
The effective charged impurity densities at the interface with the dielectric are $\sc = 0.79 \cdot \scu$ in device B, $\sc = 1.1 \cdot \scu$ in device C, and $\sc = 1.0 \cdot \scu$ in device D, extracted by the procedure discussed in \Cref{sec:electrostatic_calibration}.%
\footnote{There is one subtlety here, which is that DLG devices have two different dielectric layers, one below the first gate layer and one between the gate layers.
The first dielectric layer is likely to control bulk properties of the wire while both dielectrics contribute to junction properties.
The quoted $\sc$ numbers are for Hall bars with both dielectric layers, but the difference with $\sc$ extracted from sibling chips is small.}
These values are smaller than in device A and satisfy the specification given in \Cref{sec:disorder_uniformity}.

In addition, we show data from devices E and F that do not pass the TGP.
They are DLG devices built on the $\delta'$ stack.
As noted previously, we do not expect all devices to pass the TGP, even if they were to have the same gap ratio and disorder levels as device A.
Hence, it is not surprising that some of our devices fail the TGP; indeed, it is required by consistency with our simulations.
Moreover, devices E and F have lower induced gap to parent gap ratios, which suppresses their expected probabilities of passing the TGP.
Device E has $\Delta_\mathrm{Al} = 415 \pm \SI{13}{\micro\eV}$ and $\DeltaInd = 90 \pm \SI{6}{\micro\eV}$ (induced gap to parent gap ratio of 0.22); in device F, the corresponding gaps are $\Delta_\mathrm{Al} = 338 \pm \SI{12}{\micro\eV}$ and $\DeltaInd = 92 \pm \SI{14}{\micro\eV}$ (induced gap to parent gap ratio of 0.27).
The effective charged impurity densities at the interface with the dielectric is $\sc = 3.1 \cdot \scu$ in device E and $\sc = 3.0 \cdot \scu$ in device F, extracted by the procedure discussed in \Cref{sec:electrostatic_calibration}.
These values are larger than in devices A-D, and, moreover, are large enough that do not satisfy the specification given in \Cref{sec:disorder_uniformity}.

Device E exhibits clusters of points in $(B,\Vp)$ space with stable ZBPs at both ends, thereby passing Stage~1 of the TGP.
However, non-local conductance measurements in the region of interest yield zero gap.
Device E is, thus, in a gapless phase and it fails Stage 2 of the TGP.
Device F fails even Stage 1 because it does not have clusters of $(B,\Vp)$ points with stable ZBPs at both ends.

Overall, the measurement data from devices A-F demonstrate the different qualitative phenomena observed in our devices.

\subsubsection{Device B passing TGP}

Since device B has a considerably larger zero-field induced gap than device A, a topological phase would have to occur at higher magnetic fields.
The TGP finds an experimental phase diagram that is consistent with this expectation.
There is a topological phase transition at $B=2\,$T, as shown in \Cref{fig:deviceB_stage2}(a,b).
The TGP assigns this device a maximum topological gap $\DeltaMax = \SI{61}{\micro\eV}$.
The median value of the topological gap across the orange region in \Cref{fig:deviceB_stage2}(a) is \SI{47}{\micro\eV}.
Device B has $|g_\mathrm{eff}| \approx 3.7$, which is smaller than that of the other devices reported in this paper.
This is consistent with device B's large $\DeltaInd / \Delta_\mathrm{Al}$ ratio, which implies that electrons in the lowest occupied sub-band have a higher amplitude to be in the superconductor, thereby inheriting both a larger induced gap and a smaller $|g_\mathrm{eff}|$.

\begin{figure*}
\includegraphics[width=18cm]{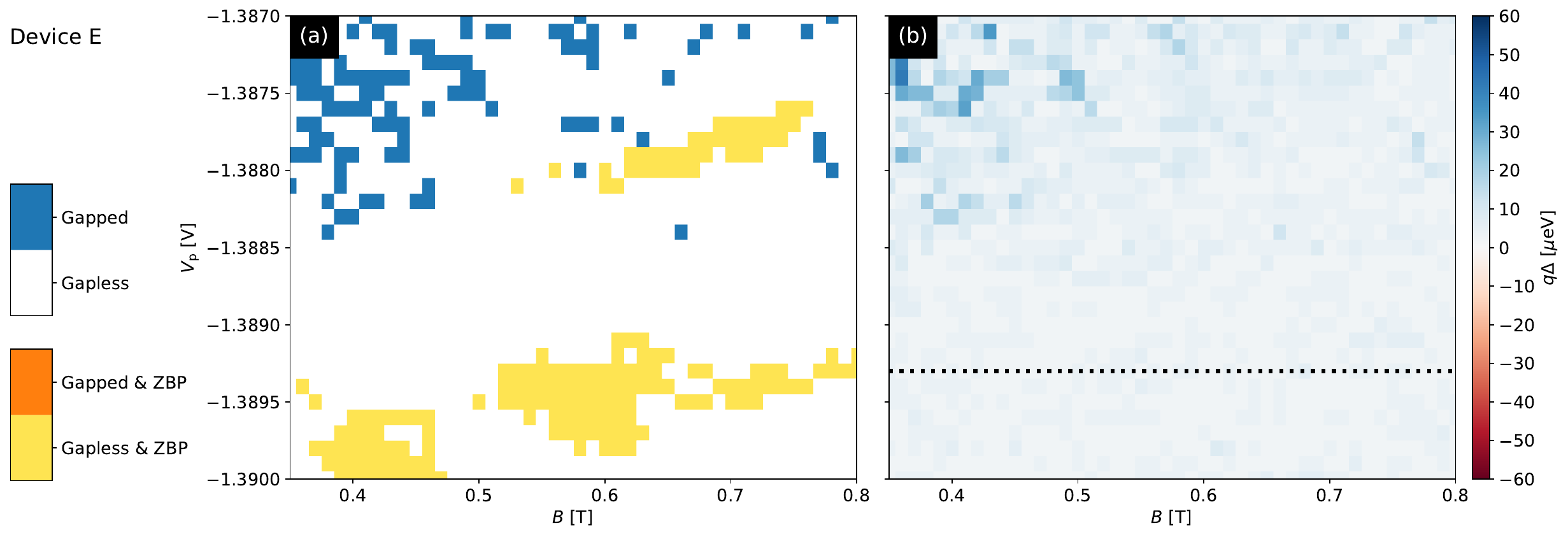}
\includegraphics[width=18cm]{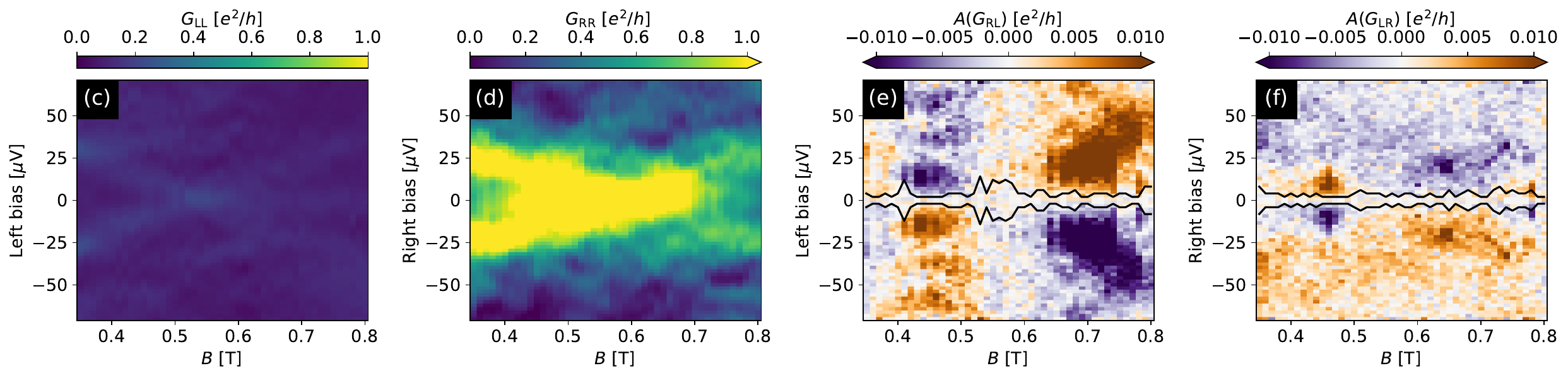}
\vskip -3mm
\caption{
Stage 2 data for device E.
(a)~The regions with stable ZBPs at both junctions.
The stability of ZBPs is determined by varying the cutter gate settings so that the above gap conductances are $0.6$ and $1.0 e^2/h$ on the left and $0.2$ and $0.4 e^2/h$ on the right.
(b)~The gap as function of $B$, $\Vp$.
It vanishes in the region of interest, so this device fails the TGP.
Measured local and anti-symmetrized non-local conductances along the horizontal line in panel b at $\Vp = -1.3893\,$V: (c)~$\GLL$, (d)~$\GRR$, (e)~$A(\GRL)$, (f)~$A(\GLR)$.
The local conductances in panels c and d show ZBPs, but there is no gap re-opening visible in the anti-symmetrized non-local conductances in panels (e) and (f).
The cut shown in (c)-(f) is at $\Gag$ of approximately $0.6 e^2/h$ on the left and $0.2 e^2/h$ on the right.
}
\label{fig:deviceE_stage2}
\end{figure*}

The extent of the SOI$_2$ is $\delta B = 0.2\,$T.
The measured extent in $\Vp$ is $\delta\Vp \approx 1\,$mV, however, Stage 2 did not go to $\Vp$ lower than $\Vp = -1.1580\,$V, where the SOI$_2$ still appears to be quite robust, so it is possible that this underestimates the size of the SOI$_2$.
It is also possible that the blue region centered around $B = 2.3\,$T and $\Vp = -1.573\,$V~--- which the TGP identifies as a gapped trivial region due to the absence of stable ZBPs~--- is actually topological but has MZMs that are poorly coupled to the leads.
There are some sign changes in the non-local conductance, such as the one that occurs at $B\approx 2\,$T in \Cref{fig:deviceB_stage2}(e,f).
Since $\GLR$ and $\GRL$ are suppressed at these sign changes, they lead to large values of the extracted gap, which can bias $\DeltaMax$ towards larger values.
However, even the median gap is \SI{47}{\micro\eV}, and there is a clearly identifiable gap edge at $\approx \SI{50}{\micro\eV}$ in $\GRL$ at $B = 2.1\,$T.
``Waterfall'' plots for local and non-local conductances at fixed plunger $\Vp = -1.15775\,$V are shown in \Cref{fig:deviceB_stage2}(g,j).
Additionally, in \Cref{fig:deviceB_stage2_waterfall_plunger} we show ``waterfall'' plots of conductances for fixed magnetic field $B = 2.12\,$T corresponding to the vertical line in \Cref{fig:deviceB_stage2}(b).

As noted above, device B has the largest $\DeltaInd$ value and the smallest $\sc$ value reported in this paper and, perhaps not surprisingly, the largest topological gap $\DeltaMax$ as well.
The topological gap is actually equal, within error bars, to the largest topological gap that we would expect for a \textit{perfectly clean}, infinitely-long DLG-$\varepsilon$ device.
This is not a contradiction.
Finite-size effects can increase the gap.
We measure a transport gap, which is the gap to extended states (for $L = \SI{3}{\micro\meter}$), and it can be larger than the gap in the spectrum if states at the gap edge have short localization lengths.

\begin{figure*}
\includegraphics[width=18cm]{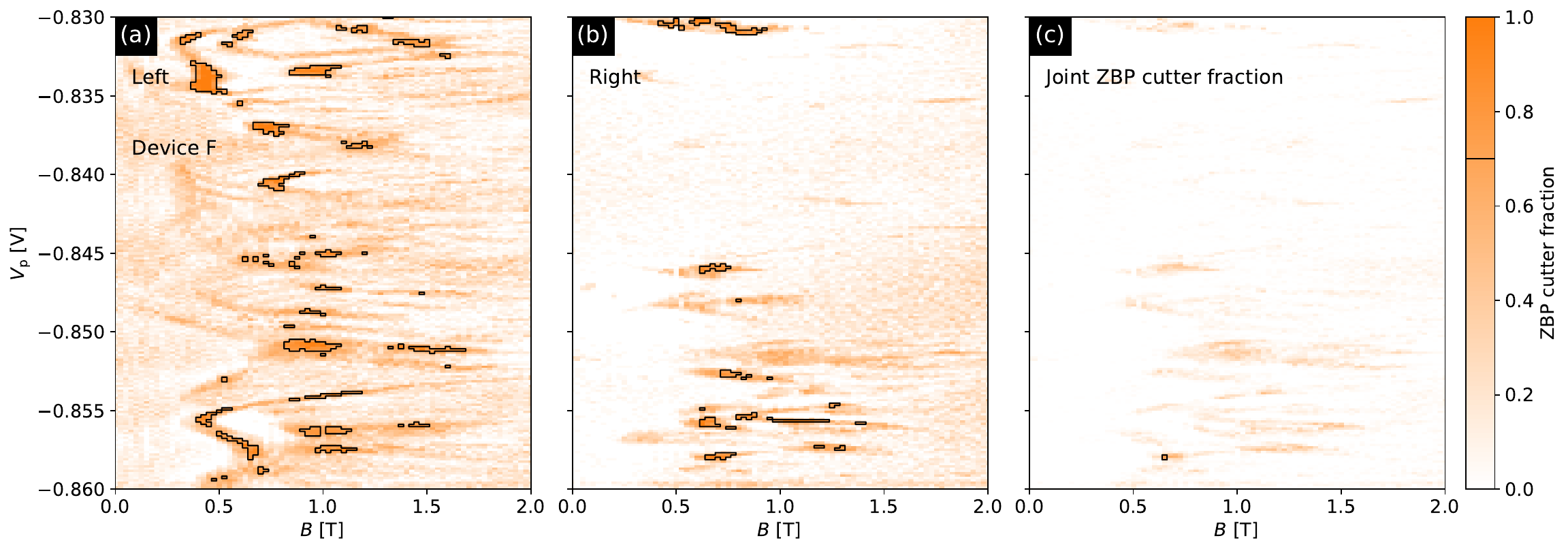}
\vskip -3mm
\caption{
Stage 1 data for device F.
The ZBP probabilities at the (a) left and (b) right junctions as a function $B$ and $\Vp$.
As may be seen in panel (c), there is no region that has stable ZBPs at both junctions.
}
\label{fig:deviceF_stage1}
\end{figure*}

\subsubsection{Devices C and D passing TGP}

The experimental phase diagram for device C is shown in \Cref{fig:deviceC_stage2}.
The TGP assigns this device a maximum topological gap $\DeltaMax = \SI{19}{\micro\eV}$, which is comparable to that of A1 and A3.
The lowest $B$ field at which the gap closes is $0.7\,$T, corresponding to $|g_\mathrm{eff}|\approx 4.4$, which is comparable to but smaller than that of device A.
The extent of the topological phase is $\delta\Vp \approx 1.5\,$mV and $\delta B = 0.2\,$T.
On the other hand, device C's ROI$_2$ is at significantly lower plunger gate voltage $\Vp \approx -2.3655\,$V than device A's.
We attribute this to differences in the dielectric that that are evident from Hall bar measurements.
In addition, the gap closing in device C is more clearly visible in $A(\GRL)$ than in $A(\GLR)$, and the ZBP has much higher amplitude in $\GLL$ than in $\GRR$.
The data is otherwise similar to that obtained from device A.

In \Cref{fig:deviceD_stage2}, we show data from device D.
There are two topological regions in the phase diagram.
The TGP assigns maximum topological gaps of \SI{17}{\micro\eV} for the top and bottom regions, respectively.
Even taken together, these regions occupy relatively small fraction of the phase diagram.
The top cluster is detected at $B = 1.1\,$T whereas the bottom one appears at $B \gtrsim 1.8\,$T.
The latter is a relatively high field, but it is consistent with the broad distribution of fields that we find in simulations.

Our results on device D show that the TGP can even identify small ROI$_2$s with small topological gaps.

\subsubsection{Devices E and F \textit{not} passing TGP}

In \Cref{fig:deviceE_stage2}, we show data from device E.
This device has large regions with stable ZBPs at each end and thus shows a relatively large ROI$_1$ in the Stage 1 of the TGP.
However, in Stage 2 the system appears gapless throughout the region of interest: The induced gap closes and there are no signatures of gap re-opening in the non-local measurements.
Since the system appears gapless in non-local conductances it is unclear if the correlated ZBPs correspond to a topological or trivial phase~--- either the topological gap is too small to be experimentally resolved or we observe a trivial state that couples to both sides because of the long coherence length that is expected in a system with very small gap.
The fact that the ZBP is significantly brighter in $\GRR$ than in $\GLL$ as well as the similarity of finite-bias features in local measurements hints towards the second scenario.
In both cases, such a phase would not be suitable for topological quantum computation.
This example demonstrates the importance of Stage 2 of the TGP where non-local conductance is measured in order to identify false positives from Stage 1.

Meanwhile, device F does not have any regions with stable ZBPs at both ends, as may be seen in \Cref{fig:deviceF_stage1}.
The absence of correlations may be due to an inhomogeneous slowly-varying potential along the nanowire.
The ZBPs that are seen at one end or the other are very similar to the ZBPs that are seen in devices passing the TGP.
However, the ZBP that show up in more than 70\% of cutter settings show no stable correlations between the two junctions.
They may originate from a trivial Andreev bound states (ABS) crossing zero energy or stable ZBPs, dubbed quasi-MZMs, that appear due to a slowly-varying potential near the junction.
Such states do not span over the whole length of the wire.
The absence of a large enough cluster of correlated ZBPs causes this device to fail Stage 1 of the TGP.

\section{Summary and discussion}
\label{sec:discussion}

In the previous sections, we have presented a summary of observed phenomena in gate-defined semiconductor nanowires coupled to a superconductor.
We have demonstrated that, when devices A-D are tuned to the single sub-band regime, they yield data passing the TGP at magnetic fields in the range $1$-$2.5\,$T.
We have observed topological gaps as high as $\DeltaMax \approx \SI{60}{\micro\eV}$.
We emphasize that our results are reproducible within the same cooldown and between different cooldowns, as we have shown for device A.
In short, our main empirical result is that multiple devices have passed the TGP.

These measurements represent strong evidence for the observation of a topological superconducting phase supporting MZMs.
The TGP has been tested with extensive simulations, reliably identifying topological regions of the phase diagram of simulated devices and correctly distinguishing trivial Andreev bound states from Majorana zero modes.
We found that, for a simulated device, there is $> 90\%$ probability that, when the TGP finds an ROI$_2$, there is a topological phase in this region of its phase diagram.

\begin{table*}
\begin{center}
\begin{tabularx}{0.85\textwidth}{|I|I|*{6}{C|}}
\cline{1-8}
\shortstack[l]{\noalign{\vskip 1.0ex} Device, \\ measurement} & 
\shortstack[l]{Design, \\ stack} & 
\shortstack[c]{\noalign{\vskip 0.7ex} $\DeltaMax$ \\ $[\si{\micro\eV}]$} &
\shortstack[c]{$|g_\mathrm{eff}|$ \\ $\phantom{1}$} &
\shortstack[c]{$\volSOI$ \\ $[\mathrm{mV}{\cdot}\mathrm{T}]$} & 
\shortstack[c]{\noalign{\vskip 0.7ex} $\volSOI$ \\ $[(\DeltaMax)^2]$} & 
\shortstack[c]{$\BavSOI$ \\ $[\mathrm{T}]$} &
\shortstack[c]{GB \\ $[\%]$} \\ [0.5ex]
\cline{1-8}
A, A1 & SLG-$\beta$ & 23 & 5.6 & 0.5 & 12 & 1.7 & 78 \\ 
\cline{1-8}
A, A2 & SLG-$\beta$ & 29 & 6.4 & 0.5 & 8.6 & 1.0 & 80 \\
\cline{1-8}
A, A3 & SLG-$\beta$ & 22 & 6.4 & 0.2 & 5.7 & 1.0 & 90 \\ 
\cline{1-8}
B & DLG-$\varepsilon$ & 61 & 3.7 & 0.2 & 0.3 & 2.1 & 92 \\ 
\cline{1-8}
C & DLG-$\delta$ & 19 & 4.4 & 0.1 & 2.0 & 1.0 & 100 \\
\cline{1-8}
D & DLG-$\delta$ & 17 & 4.5 & 0.1 & 2.5 & 1.8 & 100 \\ 
\cline{1-8}
D & DLG-$\delta$ & 21 & 4.5 & 0.1 & 1.4 & 1.1 & 93 \\ 
\cline{1-8}
\end{tabularx}
\end{center}
\vskip -3mm
\caption{
For devices A-D passing the TGP, we list the the measured maximum topological gap $\DeltaMax$; the effective $g$-factor $g_\mathrm{eff}$; the volume of the SOI$_2$ shown in \Cref{sec:experimental_data}, in units of mV$\cdot$T and $(\DeltaMax)^2$; the ``center-of-mass'' magnetic field $\BavSOI$ of the SOI$_2$; and the percentage of the boundary of the SOI$_2$ that is gapless (GB\%).
Device D has two SOI$_2$, and we list the values associated with both.
From simulations, we estimate the lever arm as $d\mu/d\Vp = 85\,\mathrm{meV/V}$ for device A, $78\,\mathrm{meV/V}$ for device B, and $79\,\mathrm{meV/V}$ for devices C and D, see \Cref{tab:effective_parameters}.
}
\label{tab:measured_topo_regions}
\end{table*}

We note that the observed topological gaps are in the range of 17-\SI{61}{\micro\eV} and occupy a correspondingly small size of the phase diagram.
Mesoscopic fluctuations are significant, so device-to-device variation and cooldown-to-cooldown variation for the same device cannot be neglected for current devices.

In order to guide incremental progress towards passing the TGP, we relied on our estimates of the material and disorder requirements that gate-defined nanowires must satisfy.
We developed designs and fabrication processes capable of meeting these requirements.

In the remainder of this section, we turn to the interpretation of our results.
The stability of the identified topological phase as a function of $B$ and $\Vp$ is an important consistency check for our results.
For the extracted effective $g_\mathrm{eff}$-factor in measurement A1, $\delta B \approx 500\,$mT corresponds to a Zeeman energy of \SI{80}{\micro\eV} and, for the calculated lever arm of $\approx 85\,$meV/V, $\delta\Vp \approx 1.5\,$mV corresponds to a chemical potential shift of \SI{128}{\micro\eV}, see \Cref{eq:H0}.
In other words, the phase space extent in field is $3.5$ times and the extent in gate voltage is $3.9$ times the maximum topological gap of \SI{23}{\micro\eV}.
In measurement A2, we found that the topological region in \Cref{fig:deviceA2_stage2} extends over a maximum $B$-field range $\delta B = 500\,$mT and a maximum $\delta\Vp = 2.5\,$mV, corresponding, respectively, to energy scales of \SI{80}{\micro\eV} and \SI{213}{\micro\eV}.
The other devices/measurements have similar stability in $B$ and $\Vp$.
Since the topological phase has an irregular shape, its area is smaller than the product of $B$ and $\Vp$; it is listed in \Cref{tab:measured_topo_regions}, where the areas of the other measured topological regions is also listed.
All of the devices have $\volSOI$ in the range 0.1-0.5 mV$\cdot$T, and they have $\volSOI$ values that are larger than $(\DeltaMax)^2$~--- in some cases, substantially larger~--- except device B.
Note, however, that device B may have a larger SOI$_2$ than what we can see in \Cref{fig:deviceB_stage2}, extending below $\Vp = -1.1580\,$V.
In other words, the observed candidate MZMs are stable with respect to parameter changes comparable to or larger than the maximum topological gap, which is an intrinsic energy scale of the problem that was determined from non-local conductance measurements.

The summary of the simulated data in the magnetic field range $B \le \SI{2.5}{\tesla}$, which is comparable to experiment, is shown in \Cref{tab:TGP_yield}.
Here we present a brief summary; the raw data set is available [\onlinecite{code_and_data}].
The TGP yield and the average volume of SOI$_2$ decreases with disorder strength (leaving aside the case of $\sc = 0.1 \cdot \scu$ which requires a different TGP calibration).
The observed phase space of the topological phase is consistent with the results of simulations presented in \Cref{tab:TGP_yield}, where we found that simulated SLG-$\beta$ and DLG-$\varepsilon$ devices had mean $\volSOIbar$ in the range 0.1-$0.2\,$mV$\cdot$T, but with a distribution that has long tails extending up to large volumes.
Similarly, our simulated SLG-$\beta$ devices have $\DeltaMax$ values in the range 23-\SI{35}{\micro\eV} while our simulated DLG-$\varepsilon$ devices have $\DeltaMax$ values in the range 25-\SI{29}{\micro\eV}, both with long tails at large gaps.
Device A would appear to be in the tail of the distribution, both with respect to $\DeltaMax$ and $\volSOI$ while device B appears to be in the tail of the distribution with respect to $\DeltaMax$.
In summary, the phase space for the topological phase is roughly as large as we would expect for a maximum topological gap of 20-\SI{60}{\micro\eV} and broadly consistent with our simulations.
We note that this is a larger gap than in early measurements of the $\nu = 5/2$ fractional quantum Hall state \cite{Willett87, Pan99b} and further note that the topological gap at $\nu = 5/2$ subsequently increased dramatically with material quality \cite{Kumar10, Zibrov17}.

The measured $\DeltaMax$, $\volSOI$, $\BavSOI$ values in \Cref{tab:measured_topo_regions} are consistent with the simulated values in \Cref{tab:TGP_yield}, which further validates the simulation model that was used to estimate the FDR.
The primary outliers are $\volSOI$ for measurements A1 and A2 and $\DeltaMax$ for device B; they are substantially larger than the mean values in simulations, but within the long tails of the non-Gaussian distributions found in simulations (see \Cref{sec:TGP_calibration_testing} for details).
This consistency between simulated and measured values validates the model used to test the TGP, as suggested in Ref.~\onlinecite{Akhmerov22}.

\begin{table}[t]
\begin{center}
\begin{tabularx}{\columnwidth}{|I|Y|C|U|C|V|C|}
\cline{1-6}
\shortstack[l]{\noalign{\vskip 1.0ex} Design, \\ stack} &
\shortstack[c]{$\sc$ \\ $[\scu]$} &
\shortstack[c]{Yield \\ $\le 2.5\,$T} &
\shortstack[c]{\noalign{\vskip 1.0ex} $\meanDeltaMax$ \\ $[\si{\micro\eV}]$} &
\shortstack[c]{$\volSOIbar$ \\ $[\mathrm{mV}\cdot\mathrm{T}]$} &
\shortstack[c]{$\BavSOIbar$ \\ $[\mathrm{T}]$}
\\ [0.5ex]
\cline{1-6}
\multirow{3}{*}{SLG-$\beta$}
& 1.0 & $47/50$ & 23 & 0.2 & 1.9 \\
\cline{2-6}
& 2.7 & $23/50$ & 34 & 0.1 & 1.5 \\
\cline{2-6}
& 4.0 & $24/49$ & 34 & 0.1 & 1.5 \\
\hline
\multirow{4}{*}{DLG-$\varepsilon$}
& 0.1 & $24/50$ & 35 & 0.2 & 2.2 \\ 
\cline{2-6}
& 1.0 & $28/50$ & 29 & 0.2 & 2.3 \\ 
\cline{2-6}
& 2.7 & $13/50$ & 30 & 0.2 & 2.2 \\ 
\cline{2-6}
& 4.0 & $16/50$ & 30 & 0.1 & 2.2 \\
\cline{1-6}
\end{tabularx}
\end{center}

\vskip -3mm
\caption{
Statistics of ROI$_2$ and SOI$_2$ properties for simulated transport data from SLG and DLG devices for the average disorder strengths given in the first column.
The third and fourth columns show the TGP yield as a fraction of devices that have at least one ROI$_2$.
Its denominator indicates the total number of simulated devices; the nomerator, the number of devices that pass the TGP.
The TGP yield decreases with increasing disorder.
All statistics in this table have been calculated with the magnetic field restricted to $B \le \SI{2.5}{\tesla}$.
The mean value of $\DeltaMax$ has weak dependence on the disorder strength, but the average phase space volume of an SOI$_2$, $\volSOIbar$, decreases with increasing disorder, as may be seen in the forth column.
For all disorder strengths simulated, $\volSOIbar \sim$ $O(0.1)\,\mathrm{mV}{\cdot}\mathrm{T}$.
The last column lists $\BavSOIbar$, the average $B$ field of an SOI$_2$, where the bar represents the statistical average over different disorder realizations.
These regions occur at significantly higher magnetic field in the $\varepsilon$-stack, as expected since $\DeltaInd$ is larger and $g^\star$ is smaller.
}
\label{tab:TGP_yield}
\end{table}

We now consider other potential explanations of our results.
According to our simulations, the probability that a device that passes the TGP does not have a topological phase overlapping the ROI$_2$ is less than 10\%, so any other explanation is extremely unlikely, though not impossible.
For an intuitive understanding of why other explanations are unlikely, let us discuss trivial ZBPs.
First, we observe that all of our devices have trivial ZBPs, even the ones that pass the TGP.
Often, they occur at only one junction, but they sometimes occur at both junctions and they can even be stable to changes in junction transparency and also to changes in $B$ and $\Vp$.
For instance, there are stable ZBPs at both junctions in measurement A1 for $1\,\mathrm{T} < B < 2\,$T and $\Vp \approx -1.173\,$V, as may be seen in \Cref{fig:deviceA1_stage2}.
In device E, there are stable ZBPs at both junctions for $0.5\,\mathrm{T} < B < 0.6\,$T and $\Vp\approx -1.13895\,$V, as may be seen in \Cref{fig:deviceD_stage2}.
These are all trivial and fall outside the topological region because the observed bulk transport gap is zero.

Now, let us consider the causes of trivial ZBPs.
One possible origin is a slowly-varying potential near the end of a device, which can be caused by certain tunnel barriers.
In sufficiently clean devices, this can lead to quasi-MZMs \textit{before} a bulk gap closing, where they are a precursor to the topological phase transition~\cite{Prada12, Kells12, Tewari14, Liu17, Pan21b}.
Quasi-MZMs do not appear \textit{after} a gap closing/re-opening.
If the bulk gap never re-opens, as in \Cref{fig:simulated_SLG_beta_R1_stage2}(g-j), then quasi-MZMs fail to become MZMs.
However, if the bulk gap re-opens, as in \Cref{fig:simulated_SLG_beta_R1_stage2}(c-f), then quasi-MZMs evolve into true MZMs.
By design, Stage 2 of the TGP weeds out stable ZBP clusters in which quasi-MZMs don't evolve into true MZMs.

Trivial ZBPs can also be induced by disorder in proximitized semiconductor nanowires with spin-orbit coupling~\cite{Bagrets12, Pikulin12, Pan22}.
However, this scenario does not entail a gap closing and re-opening in the non-local signal, from which we conclude that it does not apply to devices A, B, C, and D.
ZBPs can also be caused by trivial ABS, which can ``accidentally'' pass through zero energy.
It is very difficult to discern a trivial ABS from an MZM purely from the local conductance spectroscopy, which can be virtually identical.
However, as in the case of disorder-induced ZBPs, a trivial ABS need not be accompanied by a gap closing and re-opening in the non-local signal.

A gap closing/re-opening could be caused by the orbital effects of the magnetic field.
When half a flux quantum is threaded through the effective cross-sectional area of the device, the proximity effect is suppressed, and the gap closes.
It re-opens when the flux increases still further, closing a second time when $3/2$ flux quanta thread the active region.
For device A, we expect a gap closing due to orbital effects at $\gtrsim 2.5\,$T in higher sub-bands and even higher fields $> 3\,$T in the lowest sub-band.
The observed gap closing is at much lower fields.
Crucially, a gap closing due to orbital effects would be weakly-dependent on the gate voltage, as illustrated by the simulated phase diagrams of clean systems in \Cref{fig:simulated_infinite_clean_gap}, where the first closing due to orbital effects is a wide, nearly vertical white bar at $B \approx 2.5\,$T that intersects the topological lobes at $\Vp\approx -0.8\,$V and $\Vp\approx -1.0\,$V (SLG-$\beta$) or $\Vp\approx -0.95\,$V (DLG-$\varepsilon$).
We find similar behavior in the simulations that we used to test the TGP, which include disorder.
On the other hand, as may be seen in \Cref{fig:deviceA2_stage2}, the gap closing that is observed in measurement A2 is strongly dependent on the gate voltage: at $\Vp \approx -1.407\,$V, the gap closes at $B\approx 0.6\,$T, but at $\Vp \approx -1.398\,$V, it is still open at $B = 2\,$T.
Similarly, device C has a gap closing that occurs at $B \approx 0.7\,$T at $\Vp \approx -2.363$ but it is still open at $B = 1.3\,$T at $\Vp\approx -2.360$.
Devices B and D have ROI$_2$s that occur at higher fields, but still below $2.5\,$T, and the gap closings are at, respectively, $B\approx 1.6\,$T and $B \approx 1\,$T.
To determine the $\Vp$ dependence of the gap closing in device B, we performed a large scale Stage 1 scan, which shows clear dependence, with a trivial gap that that is still open at $B = 2\,$T for some values of $\Vp$.
Hence, an explanation relying on the orbital effect of the magnetic field is not consistent with either the $B$ field value or the $\Vp$-dependence of the observed gap closings/re-openings in devices A-D.

Of course, it is conceivable that the bulk gap closes and re-opens accidentally and that trivial ZBPs also accidentally occur for the same $B$ and $\Vp$.
But to pass the TGP, the accidental closing would have to occur over an entire curve in the $B$-$\Vp$ phase space and, moreover, the trivial ZBPs would have to persist over the enclosed region.
Of course, we cannot rule this out completely.
However, such coincidences do not require any physics that is not incorporated in our simulations.
Hence, they could occur in our simulations, and we can quantitatively bound their probability.
The false discovery rate (FDR) that we estimate from simulations is $\leqslant 8\%$ at a 95\% confidence level for all of the device designs, material stacks, and disorder levels that we have simulated, which implies that there is a low probability that data from devices A-D can be explained by trivial ZBPs that occur coincidentally with an accidental gap closing/re-opening instead of a topological phase.

\begin{figure}
\includegraphics{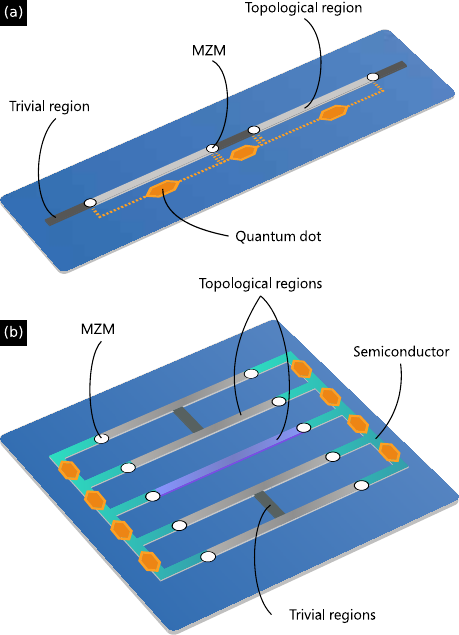}
\vskip -2mm
\caption{
(a)~The linear tetron, a minimal device for performing fusion.
The two outer regions must be tuned into the topological phase via the TGP, while the middle section must be in the trivial phase.
This results in a device with four MZMs.
(b)~Two two-sided tetrons, with which measurement-based braiding can be performed.
There are 5 topological sections.
The middle (purple) one is a coherent link that is used for connecting the left and right of the two tetrons.
}
\label{fig:qubits}
\end{figure}

\section{Looking ahead}
\label{sec:looking_ahead}

A reliable process for tuning devices into the topological phase is an essential step on the journey to topological quantum computation, which relies on the fusion and braiding of anyons.
Networks of such wires can be assembled into a many-qubit device, and this protocol can be used to tune each wire within a qubit into the topological phase.
The linear tetron qubit \cite{Fidkowski11b, Karzig17} is a minimal device for performing two non-commuting fusion operations;%
\footnote{See Refs.~\onlinecite{Plugge16, Plugge17, Vijay15, Vijay16a, Vijay16b} for related alternate qubit designs.} 
it is shown schematically in~\Cref{fig:qubits}(a).
There are two outer topological sections, separated by a trivial section.
Fusion outcomes for different pairs of MZMs are measured by coupling these MZMs to adjacent quantum dots.
By measuring different sequences of pairs of MZMs, we can directly measure the fusion rules, a topological invariant of the state.
The parameter space of such a device is simply too large to explore in the hope of finding a suitable operating point unless each of the outer sections of the nanowire is individually tuned into a topological phase, which can be achieved using the TGP.

Coherent manipulation of the encoded quantum state, for example through measurement-based implementations of braiding transformations~\cite{Bonderson08b, Bonderson08c}, requires an even more complex device, as shown in~\Cref{fig:qubits}(b).
This device consists of two superconducting islands, each comprised of two topological segments linked together by a short region of trivial superconductor, and thus contains two qubits.
We refer to this configuration as a two-sided tetron because MZMs are on each side of the device.
The additional topological wire between the qubit islands, shown in purple, is added to allow measurements between MZMs on opposite sides of the islands.
In this device, unitary Clifford gates on one of the two qubits can be performed by using the other as an auxiliary qubit and performing a sequence of single- and two-qubit measurements~\cite{Karzig17}.
Despite the even larger parameter space, these devices can be tuned using the TGP.
Continued improvement in simulation, growth, fabrication, and measurement capabilities will be required to achieve the topological gap required for such coherent operations.

\acknowledgments
We thank Greg Boebinger, Sankar Das Sarma, Michel Devoret, Klaus Ensslin, Leonid Glazman, Bert Halperin, Dale van Harlingen, Kam Moler, Krysta Svore, and Matthias
Troyer for extensive discussions.
We are grateful to Edward Lee and Todd Ingalls for their assistance with figures.
We thank Leo Kouwenhoven for helpful comments on a previous version of this manuscript.

\vspace{1mm}
\textbf{Correspondence and requests for materials} should be addressed to Chetan Nayak~(cnayak@microsoft.com).

\vspace{1mm}
$^\dagger$\small{Morteza Aghaee, Arun Akkala, Zulfi Alam, Rizwan Ali, Alejandro Alcaraz Ramirez, Mariusz Andrzejczuk, Andrey E. Antipov, Pavel Aseev, Mikhail Astafev, Bela Bauer, Jonathan Becker, Srini Boddapati, Frenk Boekhout, Jouri Bommer, Esben Bork Hansen, Tom Bosma, Leo Bourdet, Samuel Boutin, Philippe Caroff, Lucas Casparis, Maja Cassidy, Sohail Chatoor, Anna Wulf Christensen, Noah Clay, William S. Cole, Fabiano Corsetti, Ajuan Cui, Paschalis Dalampiras, Anand Dokania, Gijs de Lange, Michiel de Moor, Juan Carlos Estrada Salda\~{n}a, Saeed Fallahi, Zahra Heidarnia Fathabad, John Gamble, Geoff Gardner, Deshan Govender, Flavio Griggio, Ruben Grigoryan, Sergei Gronin, Jan Gukelberger, Sebastian Heedt, Jes\'us Herranz Zamorano, Samantha Ho, Ulrik Laurens Holgaard, William Hvidtfelt Padk{\ae}r Nielsen, Henrik Ingerslev, Linda Johansson, Jeffrey Jones, Ray Kallaher, Farhad Karimi, Torsten Karzig, Cameron King, Maren Elisabeth Kloster, Christina Knapp, Dariusz Kocon, Jonne Koski, Pasi Kostamo, Peter Krogstrup, Mahesh Kumar, Tom Laeven, Thorvald Larsen, Kongyi Li, Tyler Lindemann, Julie Love, Roman Lutchyn, Morten Hannibal Madsen, Michael Manfra, Signe Markussen, Esteban Martinez, Robert McNeil, Elvedin Memisevic, Trevor Morgan, Andrew Mullally, Chetan Nayak, Jens Nielsen, Bas Nijholt, Anne Nurmohamed, Eoin O'Farrell, Keita Otani, Sebastian Pauka, Karl Petersson, Luca Petit, Dima Pikulin, Frank Preiss, Marina Quintero-Perez, Mohana Rajpalke, Katrine Rasmussen, Davydas Razmadze, Outi Reentila, David Reilly, Richard Rouse, Ivan A. Sadovskyy, Lauri Sainiemi, Sydney Schreppler, Vadim Sidorkin, Amrita Singh, Shilpi Singh, Sarat Sinha, Patrick Sohr, Toma\v{s} Stankevi\v{c}, Lieuwe Stek, Henri Suominen, Judith Suter, Vicky Svidenko, Sam Teicher, Mine Temuerhan, Nivetha Thiyagarajah, Raj Tholapi, Mason Thomas, Emily Toomey, Shivendra Upadhyay, Ivan Urban, Saulius Vaitiek\.{e}nas, Kevin Van Hoogdalem, David Van Woerkom, Dmitrii V. Viazmitinov, Dominik Vogel, Steven Waddy, John Watson, Joseph Weston, Georg W. Winkler, Chung Kai Yang, Sean Yau, Daniel Yi, Emrah Yucelen, Alex Webster, Roland Zeisel, Ruichen Zhao.}

\vspace{5mm}

\appendix

\renewcommand{\thesection}{\Alph{section}}
\renewcommand{\thesubsection}{\Alph{section}.\arabic{subsection}}

\section{Effects of disorder}

\subsection{Projection to proximitized nanowire model}
\label{sec:projected_model}

The disorder-free limit of the single-sub-band regime
is well-represented by a Hamiltonian of the form of \Cref{eq:H0}:
\begin{equation}
\label{eq:projected_nanowire}
H = Z^\dagger
\bigl( H_\SM -
    \tilde{\Gamma}  O_\SC 
\bigr) Z
\end{equation}
In \Cref{eq:projected_nanowire}, the effective mass, $g$-factor, and spin-orbit coupling in $H_\SM$~--- obtained by projecting the full model for the device designs and material stacks described in \Cref{sec:layout,sec:material_stack} into the lowest sub-band~--- take the values given in \Cref{tab:effective_parameters}.
The parameters in \Cref{tab:effective_parameters} are weakly density-dependent within the range of $n\sim 0.02-0.04$nm$^{-1}$, which corresponds to the range of a Fermi wavelength $\kF^{-1} \simeq 40-80$\,nm.
Compared to \Cref{eq:H0}, there is a transformation $Z$, described below, which is due to the renormalization of the electrons in the semiconductor by their coupling $\Gamma$ to the Al, which has a superconducting gap $\DeltaAl = \SI{300}{\micro\eV}$ and a Zeeman potential $V_{x,\SC} = g_\SC \muB B/2$ with $g$-factor $g_\SC = 2$.%
\footnote{Note that this effective Hamiltonian is only valid provided $\DeltaAl$ is not too small.
In the more general case, integrating out the superconductor leads to an effective action that is non-local in time and has an energy-dependent renormalization.}
In the following we will neglect the orbital effect of the magnetic field on the superconductor, despite keeping the Zeeman splitting.
The justification for this is that the SC thickness is $d < 10\,$nm and we estimate the disordered coherence length in the Al, $\xiAl$, to be around $40\,$nm.
Orbital effects can be neglected when the depairing energy is small compared to the gap; this ratio goes as the square of the ratio of flux through the effective cross-section $d \xiAl$ to the flux quantum, yielding a condition $(B \xiAl d / \Phi_0)^2 \ll 1$~\cite{Maki64}.
For fields $B < 2.5\,$T, the left hand side is less than $1/4$, so we expect the corrections due to orbital effects to be correspondingly small.
The effective pairing depends on the Zeeman potential in the SC as $\tilde{\Gamma} = \Gamma \DeltaAl / \sqrt{\DeltaAl^2 - V_{x,\SC}^2}$.
Additionally, the Zeeman potential experienced by the SM now includes an induced contribution from the Al, $\tilde{V}_x = V_x + \Gamma V_{x,\SC} / \sqrt{\DeltaAl^2 - V_{x,\SC}^2}$.

In all three cases~--- SLG $\beta$-stack and DLG $\delta$- and $\varepsilon$-stacks~--- the transformation by $Z$ as well as the induced Zeeman energy from the Al suppress the effective $g$-factor below $g_\SM$.
Hence, as discussed in the previous section, the coupling between the semiconductor and the superconductor, $\Gamma$, should not be too large or else the $g$-factor will be strongly suppressed and the topological phase will be pushed to high $B$-fields where the Al approaches the Clogston limit.
On the other hand, $\Gamma$ cannot be too small, either, or else the induced gap $\DeltaInd \sim \Gamma \DeltaAl / (\Gamma + \DeltaAl)$ will be small.

We now turn to the renormalization factor $Z$.
The Green's function of a hybrid SM-SC system is given, in general, by
\begin{equation}\label{eq:Gamma0}
    G(\omega) 
    = \left[ \omega - H_\SM - \Sigma(\omega) \right]^{-1}
\end{equation}
with a self-energy $\Sigma(\omega)$ obtained by integrating out the SC degrees of freedom using the tunneling Hamiltonian model defined in Ref.~\onlinecite{Lutchyn12},
\begin{equation}
    \Sigma(\mathbf{r}, \mathbf{r}', \omega) 
    = - \int \! d\mathbf{x}_1 d \mathbf{x}_2 \,
    T^\dagger(\mathbf{r},\mathbf{x}_1) T(\mathbf{x}_2,\mathbf{r}') \,
    G_{\SC}(\mathbf{x}_1, \mathbf{x}_2, \omega).
\end{equation}
Here, the integrals are taken over the superconducting domain with $\mathbf{x} = (z,\mathbf{r})$ where $z$ and $\mathbf{r}$ are out-of-plane and in the plane of the interface, respectively; the tunneling matrix element reads
\begin{equation}
    T(\mathbf{x}_1, \mathbf{r}) 
    = t \delta(\mathbf{r}_1 - \mathbf{r}) 
    \delta (z_1) \partial_{z_1}   
\end{equation}
The delta function here corresponds to momentum conservation parallel to the interface whereas $\partial_{z_1}$ enhances transmission for electrons in the superconductor that are incident with momemtum normal to the interface.
After some algebra~\cite{Stanescu11}, one finds that $\Sigma(r,  r', \omega) \propto \delta(r-r') \int d \xi_k G_{\SC}(\omega, k)$ where $\delta(r-r')$ represents the rapid decay of the self-energy on the scale of the Fermi wavelength in the metal.

We now derive the expression for the self-energy in the presence of disorder in Al which is crucial for understanding the proximity effect~\cite{Stanescu11, Reeg18c, Kiendl19, Stanescu22, Thomas22}.
Indeed, disorder removes momentum conservation constraints and effectively increases the tunneling  rate $\Gamma$.
The disorder-averaged self-energy can be written as 
\begin{equation}
    \Sigma(\mathbf{r}, \mathbf{r}', \omega) 
    = \Gamma g(\omega) 
    \delta(\mathbf{r} - \mathbf{r}'),
\end{equation}
where the Usadel Green’s function $g(\omega)$ represents the diffusive limit of a spin-split s-wave superconductor.
Here we neglect the orbital contribution of the magnetic field, as justified above.
Analytic expressions for this Green's function in this limit have been recently reported in Refs.~\onlinecite{Faluke19, Khindanov21}.

Now, we would like to obtain a frequency-independent effective Hamiltonian with eigenvalues that approximate the poles of $G(\omega)$ \eqref{eq:Gamma0}, but, in order to do so, it is insufficient to take the static limit by replacing $\Sigma(\omega)$ by $\Sigma_0 \equiv \Sigma(0)$.
A better, widely used, approximation involves grouping the linear-in-frequency part of the self-energy, $\Sigma_1 \equiv \partial_\omega \Sigma(\omega) |_{\omega = 0}$, with the frequency of the Green's function to define an overall renormalization of the energy scale before taking the static limit.
Here we describe a generalization of that method.

Specializing the expressions in \cite{Faluke19, Khindanov21} to the static and linear-in-frequency parts, we obtain
\begin{equation}
\Sigma_0 = -\frac{\Gamma \left(-V_{x,\SC} \sigma_x \tau_z + \DeltaAl \sigma_y \tau_y \right)}{\sqrt{\DeltaAl^2 - V_{x,\SC}^2}},
\end{equation}
\begin{equation}
\Sigma_1 = -\frac{\Gamma \left( \DeltaAl^2 - V_{x,\SC} \DeltaAl \sigma_z \tau_x \right) }{\left( \DeltaAl^2 - V_{x,\SC}^2 \right)^{3/2}},
\end{equation}
and the Green's function in this linearized representation is
\begin{equation}
\tilde{G}(\omega) = \left[ \omega(1 - \Sigma_1) - H_\SM - \Sigma_0 \right]^{-1}.
\end{equation}
As a consistency check, we can consider the well-known $V_{x,\SC} = 0$ limit, where $\Sigma_0 = -\Gamma \sigma_y \tau_y$ and $\Sigma_1 = -\Gamma/\DeltaAl$ and recover
\begin{equation}
\tilde{G}_{V_{x,\SC}=0}(\omega) = \frac{\DeltaAl}{\DeltaAl + \Gamma} \left[ \omega - \frac{\DeltaAl}{\DeltaAl + \Gamma}(H_\SM - \Gamma \sigma_y \tau_y) \right]^{-1}
\end{equation}
with poles at the eigenvalues of the renormalized effective Hamiltonian
\begin{equation}
H_{\rm eff} = \frac{\DeltaAl}{\DeltaAl + \Gamma}(H_\SM - \Gamma \sigma_y \tau_y),
\end{equation}
recovering the previous results of Ref.~\onlinecite{Stanescu11}.
In the more general case where $V_{x,\SC} \neq 0$, $\Sigma_1$ is not proportional to the identity.
Hence, we have to be careful in factoring out the coefficient of $\omega$.
This is because we require a hermitian $H_{\rm eff}$, and in general the matrix $(1 - \Sigma_1)^{-1} (H_\SM + \Sigma_0)$ is not hermitian.
To resolve this, we take a symmetric product, defining $Z = (1 - \Sigma_1)^{-1/2}$ and obtain the result
\begin{equation}
    H_{\rm eff} 
    = Z^{\dagger} \left( H_\SM  + \Sigma_0 \right) Z.
\end{equation}

\begin{figure}
\includegraphics[width=8.6cm]{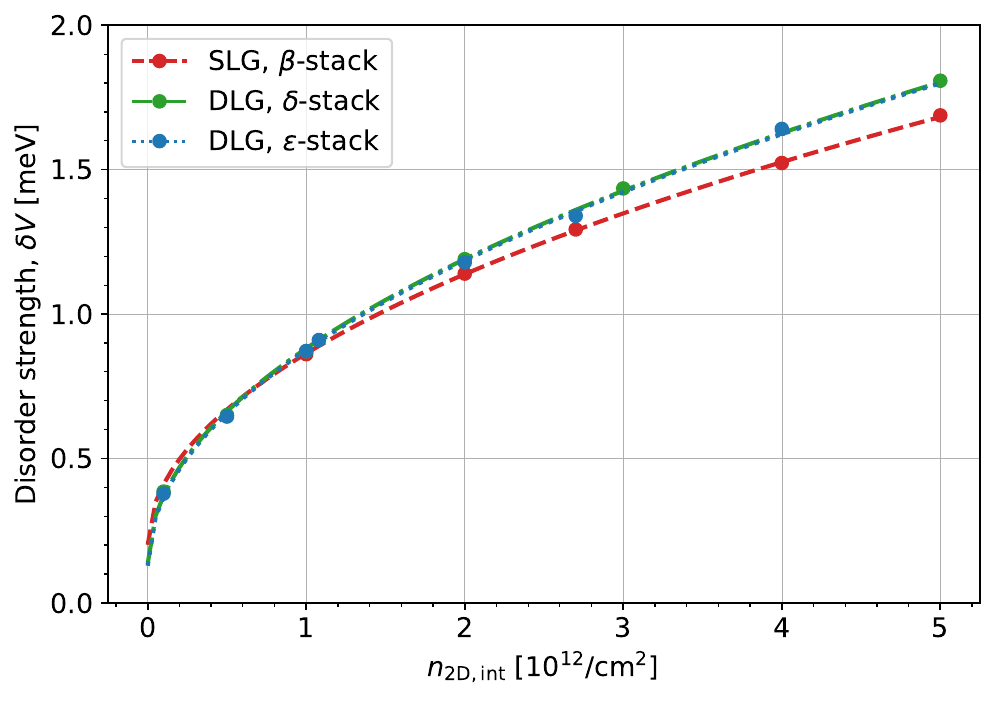}
\vskip -3mm
\caption{
The effective parameter $\delta V$ describing the disorder strength as a function of $\sc$, the density of charged defects at the semiconductor-dielectric interface, for the SLG-$\beta$, DLG-$\delta$, and DLG-$\varepsilon$ design and material stacks.
}
\label{fig:disorder_strength_vs_n2D}
\end{figure}

Finally, we now extend the model to include potential disorder in the semiconductor:
\begin{equation}
H = Z^\dagger \bigl( 
H_\SM - \tilde{\Gamma}  O_\SC + H_\mathrm{Dis} 
\bigr) Z,
\label{eq:nanowire_with_disorder}
\end{equation}
where the disorder potential, projected into the lowest subband, is
\begin{equation}
H_\mathrm{Dis} \!=\! \int\limits_0^L \!dx \, 
V(x)\,\psi_{\sigma}^\dag(x) \psi_{\sigma}(x).
\end{equation}
For the disorder mechanisms we have included, we find that the random potential can be approximately characterized by its sample-averaged autocorrelation:
\begin{equation}
\langle V(x) V(x') \rangle 
= \delta V^2 \exp\bigl(-|x-x'|/\kappa\bigr).
\label{eq:cumulant}
\end{equation}
Similar to our extraction of other single-sub-band parameters, we generate disorder potentials using the full 3D electrostatics model,%
\footnote{In this paper we have used the 3D Thomas-Fermi model for screening in device electrostatics simulations.
In the relevant parameter regime (i.e., in the lowest sub-band) we have compared the Thomas-Fermi (TF) approximation with Schr\"odinger–Poisson (SP) calculations and find values of $\delta V$ that agree within ~20\% accuracy.
SP calculations yield a slightly smaller $\delta V$ at low densities, and a slightly larger $\delta V$ at higher densities within the lowest sub-band.
We attribute this to the difference between the 3D density-of-states assumed by TF and the 1D density-of-states of the actual gate-defined nanowire.}
project the disorder potentials to the lowest subband, obtain the autocorrelation averaged over several disorder realizations, then fit the values $\delta V$, $\kappa$.
These projected disorder parameters, variance $\delta V < \SI{2}{\meV}$ and correlation length $\kappa = 80$, 115, and 120\,nm (for $\beta$-, $\delta$-, and $\varepsilon$-stacks, respectively), are obtained for the same design and materials stack as the uniform parameters given in \Cref{tab:effective_parameters}.
Here, the various sources of disorder and their strengths have been distilled to two numbers.
These parameters' relation to the underlying microscopic disorder depends on both the device geometry and the defect types and densities.
We show the relation between $\delta V$ and $\sc$ for SLG and DLG designs for the $\beta$, $\delta$, and $\varepsilon$ stacks in \Cref{fig:disorder_strength_vs_n2D}.

\subsection{Disorder-driven phase transition}
\label{sec:disorder_pd}

In general, disorder suppresses the topological phase~\cite{Motrunich01, Gruzberg05, Brouwer11a, Brouwer11b, Lobos12, Sau12, DeGottardi13, DeGottardi12, Adagideli14, Pekerten17, Ahn21}, causing important changes to the phase diagram of \Cref{fig:energy_spectrum_topophase}.
For weak disorder, this leads to a decrease of the topological gap and an increase of the disordered superconducting coherence length in the topological phase.
As the disorder strength is increased, these variations can create small non-topological regions in an otherwise topological wire, thus nucleating additional subgap states at the domain walls between topological and non-topological regions.
Eventually, these subgap states hybridize and lead to the breakdown of the topological phase through Griffiths' effects~\cite{Motrunich01}.
In the limit in which the clean topological gap $\DeltaT$ is small compared to the Fermi energy, the stability condition for the topological phase is $\ellLoc > \xiT$~\cite{Brouwer11b}, where $\ellLoc$ is the localization length in the normal state and $\xiT$ is the coherence length in the clean topological superconductor.
Equivalently, this can be rephrased in terms of energy scales as $\DeltaT \tau > \hbar/2$, where $\tau$ is the elastic scattering time.
In the effective model governed by $H = H_\SM + \DeltaInd O_\SC + H_\mathrm{Dis}$, the scattering rate due to the random potential $V(x)$ can be evaluated to lowest order in $\delta V$:
\begin{equation}
    \frac{\hbar}{\tau} 
    = \frac{\delta V^2}{2\EF}
    \frac{2\kF\kappa}{1+(2\kF\kappa)^2}.
    \label{eq:scattering_time}
\end{equation}
Here, $\EF$ and $\kF$ are the Fermi energy and momentum.
Thus, the transition at $\DeltaT \tau = \hbar/2$ occurs at a critical value $\delta V^2_c$:
\begin{equation}
    \delta V^2_c = 
    {4\EF}\DeltaT \bigl[{1+(2\kF\kappa)^2}\bigr]/2\kF\kappa.
    \label{eq:crit_disorder_avg}
\end{equation}
This expression for the critical disorder strength is valid in the regime of weak disorder and small gap, $\DeltaT \ll E_F$.
However, the nanowire in the single subband regime has a small Fermi energy, so the latter requirement is difficult to satisfy.
Moreover, disorder sources with correlation lengths $\kappa \gtrsim \kF^{-1}$ cause stronger scattering than short-range disorder $\kappa \kF \ll 1$, as may be seen from \Cref{eq:scattering_time}.
For realistic disorder levels, it is necessary to go beyond lowest-order in $\delta V$ \cite{Boutin22}.
Consequently, the precise location of the disorder-driven phase transition is more complicated than \Cref{eq:crit_disorder_avg} when the clean topological gap and disorder strength are comparable to the Fermi energy \cite{DeGottardi13, Boutin22}.
Finally, there are important differences between the thermodynamic limit and finite-sized systems, where the phase transition is rounded into a crossover.
In order to understand these additional complexities, we calculate the disordered coherence length numerically using the transfer matrix method, as we discuss in the next two subsections.

\subsection{Length scales and topological phases in finite systems}
\label{sec:length_scales}

Since we are concerned in this paper with topological phases in finite systems with disorder, we must pay attention to several important length scales.
We will denote the superconducting coherence length in the wire by $\xi(0)$.
It is the distance that a zero-energy unpaired electron can penetrate into the proximitized nanowire.
In the topological phase, $\xi(0)$ corresponds to the localization length of a Majorana zero mode.
The coherence length $\xi(0)$ diverges at the phase transition between the trivial and topological superconducting phases.
In the topological phase, when disorder is very weak, $\xi(0)$ approaches its clean value $\xiT$.
Increased disorder elongates $\xi(0)$ in the topological phase, while it shortens $\xi(0)$ in the trivial phase.

In a perfectly clean system, there would be no states below the clean topological gap $\DeltaT$, apart from the MZMs, and states above the gap would all be extended.
However, even weak disorder localizes all states except the zero-energy state at the critical point, as noted in \Cref{sec:proximitized_nanowire_model}.
It also causes localized states to appear below $\DeltaT$.
We will call the energy-dependent localization length, calculated for the Bogoliubov-de Gennes Hamiltonian~\eqref{eq:H0}, $\xi(E)$.
At low energy, $\xi(E) \to \xi(0)$.
When disorder is very weak, the density of states is very low for $E<\DeltaT$, and these states have localization lengths that grow smoothly as a function of $E$, increasing from $\xi(0)$ to $\xi(\DeltaT)$.
If $\xi(\DeltaT)>L$, then states above $\DeltaT$ will appear to be extended and there will be an apparent \textit{transport gap} $\Delta_\mathrm{tr}\lesssim\DeltaT$, which is the gap to ``extended states,'' namely the states whose localization lengths $\xi(E)$ are larger than the system size.
Turning now to the case of more general disorder strengths, we define $\xi(\Delta)$ as the maximum value of $\xi(E)$ for $E<\DeltaInd$.
Our devices are designed so that $\xi(\Delta) \gtrsim L$.
When this holds, we can measure the transport gap $\Delta_\mathrm{tr}$, which is the minimum $E$ for which $L/\xi(E)$ is small enough that states at energy $E$ are visible in bulk transport.
We explain these measurements in \Cref{sec:TGP} and \Cref{sec:TGP_measurements}.

The decay of subgap states at small energies $E\ll \DeltaT$ is controlled by $\xi(0)$ while transport by excited states is controlled by $\xi(\Delta)$.
Hence, if we observe a ZBP at one junction, the amplitude to observe it at the other junction decays as $e^{-L/\xi(0)}$, but if we observe an excited state of the bulk at one junction, the amplitude to observe the same state at the other junction decays at least as fast as $e^{-L/\xi(\Delta)}$.

Thus, depending on the disorder level and device length, there are three parameter regimes for the operation of a topological device.
The first regime is when $\xi(0) < \xi(\Delta) \ll L$.
In this limit, the device is longer than any of the finite length scales that characterize the topological phase.
Hence, there is no characteristic energy scale of the bulk topological phase that can be extracted from transport measurements.

In the second regime, $\xi(0)\ll L \lesssim \xi(\Delta)$.
In this regime, which we will call the asymptotic regime, the device is much longer than the localization length of a low-energy bound state.
When ZBPs are observed at both junctions, we know that there are two distinct bound states, one at each junction.
Since $L \lesssim \xi(\Delta)$, there are excited states that are effectively extended, i.e. have localization lengths that are comparable to or longer than the system size.
The transport gap is the energy gap to such states.

Finally, there is a third regime, in which $\xi(0)\lesssim L \lesssim \xi(\Delta)$.
This is a crossover regime in which the system is longer than the MZM localization length but it may not be so much longer that we are in the asymptotic limit.
In this case, there may be some bulk transport at zero energy even when $\Delta_\mathrm{tr}>0$ because the $e^{-L/\xi(0)}$ contribution is not negligible.
Consequently, the system will be intermediate between the asymptotic regime and the critical regime that we define in the next paragraph.
As $\xi(0)$ is reduced, the system will move more firmly into the asymptotic topological regime.
However, there isn't a particular value of the ratio $L/\xi(0)$ that separates the crossover regime from the asymptotic regime; the evolution from one to the other with increasing $L / \xi(0)$ is smooth.

None of the three regimes mentioned above~--- thermodynamic, asymptotic, and crossover~--- is possible near the critical point, where $\xi(0)=\infty$.
When $L \ll \xi(0)$, the system is in the critical regime, and we can't distinguish it from a critical system.
Even in a device of length $L=\SI{3}{\micro\meter}$, this is a fairly broad region.
As we shall see, for intermediate disorder strengths, over much of the parameter range in which the system would be in the topological phase for $L \to \infty$, the system is in the critical regime because the correlation length is long and $L \ll \xi(0)$.

\begin{figure*}
\includegraphics[width=15.9cm]{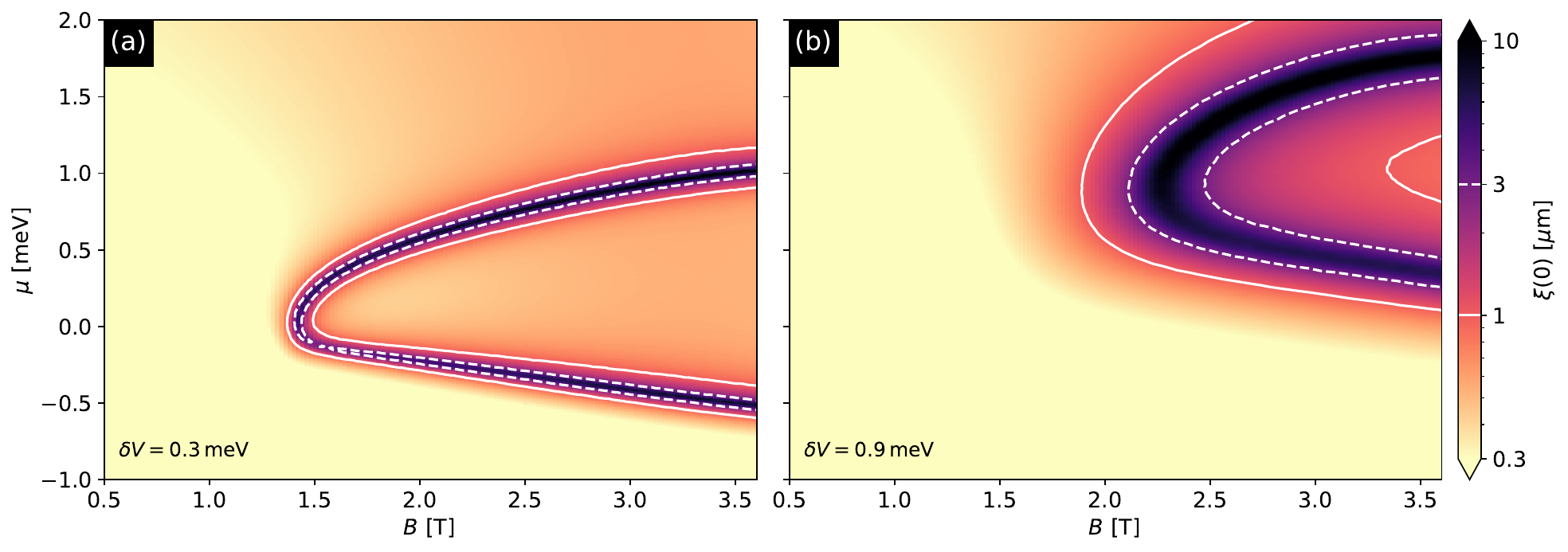}
\vskip -3mm
\caption{
The coherence length $\xi(0)$ evaluated numerically by the transfer matrix method for the model in \Cref{eq:nanowire_with_disorder} with DLG, $\varepsilon$-stack parameters from \Cref{tab:effective_parameters} and characteristic disorder strengths (a) $\delta V = 0.3$ meV and (b) $\delta V = 0.9$ meV.
The colorscale saturation highlights the divergence of $\xi(0)$, indicating the phase transition from the low-field trivial phase to the high-field topological phase.
The solid and dashed lines mark contours of constant $\xi(0) = 1$ and \SI{3}{\micro\meter} respectively.
For our \SI{3}{\micro\meter} simulations in \Cref{sec:pd_single_realization}, the region enclosed by dashed lines roughly represents the ``critical" regime, while the area between dashed and solid lines is in the ``crossover'' regime where $L > \xi(0)$, but we still expect finite-size effects and mesoscopic fluctuations.
}
\label{fig:xi0_epsilon_stack}
\end{figure*}

As an illustration of these various regimes, we calculate $\xi(0)$ for the model \Cref{eq:nanowire_with_disorder} using the transfer matrix method~\cite{Vishveshwara01}.
In \Cref{fig:xi0_epsilon_stack} we plot the result for DLG, $\varepsilon$-stack parameters from \Cref{tab:effective_parameters} and characteristic disorder strengths $\delta V = 0.3$ and $0.9\,$ meV.
As $\xi(0)$ is a function of the chemical potential and applied magnetic field, a finite wire can go from the critical to crossover to asymptotic regimes depending on these parameters.
When the system is in the thermodynamic, asymptotic, or crossover regimes, it is deep in a $\Delta_\mathrm{tr}>0$ phase.
The devices discussed in this paper are in the asymptotic or crossover regimes.
We discuss the implications for transport measurements when we described the topological gap protocol in \Cref{sec:TGP} and
\Cref{sec:TGP_measurements}.

\subsection{Finite-size behavior of topological invariants}
\label{sec:topological_invariants}

We now consider the finite-size behavior of the invariants distinguishing the topological and trivial phases.
We consider the ``scattering invariant''~\cite{Fulga11} and the Pfaffian invariant \cite{Kitaev01, Wimmer12}).
The scattering invariant is defined for an open system with a junction to a normal lead: $\mathrm{SI} = \mathrm{sgn} \det r \in [-1,1]$, where $r$ is the reflection matrix.
When $\mathrm{SI}=-1$, there is an MZM at the junction.
The scattering matrix $r$ depends on how open or closed the junction is.
In a finite-sized system, an MZM at a junction will hybridize with the MZM at the other end of the wire with strength $e^{-L/\xi(0)}$ (giving $\mathrm{SI}=+1$) unless it is coupled more strongly to the lead.
If the MZM couples poorly to the leads, then we will erroneously find $\mathrm{SI}=+1$.
If the MZM couples more strongly to the lead than to its partner at the other end, then we will have $\mathrm{SI}=-1$ in a finite-size system.
Note, however, that we will also find $\mathrm{SI}=-1$ if the second MZM is much closer than distance $L$ but has small hybridization with the one at the junction, namely the ``quasi-MZM'' scenario~\cite{Prada12, Kells12, Tewari14, Liu17, Pan21b}.
The Pfaffian invariant $\mathcal{Q}$ is defined for a closed system as the relative sign of the ground state parity between periodic (PBC) and anti-periodic (APBC) boundary conditions.
$\mathcal{Q} = \mathrm{sgn}[\mathrm{Pf}(A_\mathrm{\scriptscriptstyle PBC})] \, \mathrm{sgn}[\mathrm{Pf}(A_\mathrm{\scriptscriptstyle APBC})]$ when the Bogoliubov-de Gennes Hamiltonian is written in terms of real fermionic operators $\gamma_{2i-1} = c_i + c_i^\dagger$, $\gamma_{2i-1} = -i (c_i - c_i^\dagger)$ so that it takes the form $H = (i/2) \sum_{i,j} A_{ij} \gamma_i \gamma_j$.
In a finite system, we will have $\mathcal{Q} = -1$ when the hybridization of the two MZMs is larger via the periodic boundary condition that connects the two ends than the $e^{-L/\xi(0)}$ hybridization that occurs through the bulk of the wire.
In summary, both invariants rely on $e^{-L/\xi(0)}$ being smaller than the coupling to the lead or the boundary condition.
This is the limit in which the topological phase can be defined; it becomes more clearly distinct from the trivial phase in a continuous fashion as $e^{-L/\xi(0)} \to 0$.
In the next section, we will use the Pfaffian invariant to illustrate the combination of finite-size effects and disorder on the phase diagram.
In \Cref{sec:testing_TGP}, we will use the scattering invariant to test the accuracy of the TGP because the scattering invariant can be calculated for the same device geometry
and junction settings as used in transport simulations and measurements.

\section{Electrostatic calibration from Hall bars}
\label{sec:electrostatic_calibration}

Each of the topological gap devices described in the main text is accompanied by a Hall bar device subject to the same growth and fabrication processes; this Hall bar enables a characterization of the bulk material quality.
The full density dependence of the Hall mobility has been widely used previously to identify and quantify dominant scattering mechanisms in 2DEGs \cite{Laroche10, DasSarma14, Shojaei16, Tschirky17, Hatke17, Thomas18, Wickramasinghe18, Pauka20, Cimpoiasu20, Ahn21}.
In particular, across all samples, the low-density mobility rapidly increases with increasing density, consistent with the mobility being dominated there by scattering from the long-range Coulomb potential of remote impurities.

In our simulations, we calculate (i) the gate-voltage dependence of carrier density $n_\mathrm{e}(\Vg)$ for a 2DEG in our materials stack in a standard Schr\"odinger-Poisson framework; and (ii) the remote-impurity-limited mobility $\mu(\Vg)$ in the Boltzmann-Born formalism.
Both of these functions are parameterized by the density and location of impurities, and an effective composite dielectric permittivity.
These parameters are used to \textit{simultaneously} fit the model traces to experimental data.
As mentioned in the main text, our best fits to density and mobility measurements over a variety of samples suggested a simplification in which an effective 2D impurity density $\sc$ is placed in an ``impurity layer'' at the interface between the barrier and the gate dielectric.
Small changes in the position or width of the impurity layer can be compensated by tuning the impurity density, but large changes modify the overall density dependence of the mobility and result in poor fits.
Therefore, the values of charge impurity density that we quote here depend on the disorder model employed, with the goal of this model being to provide a \textit{consistent} description of the impact of fixed charges on \textit{both} the Hall mobility and TGP measurements.

\begin{figure*}
\includegraphics[width=16cm]{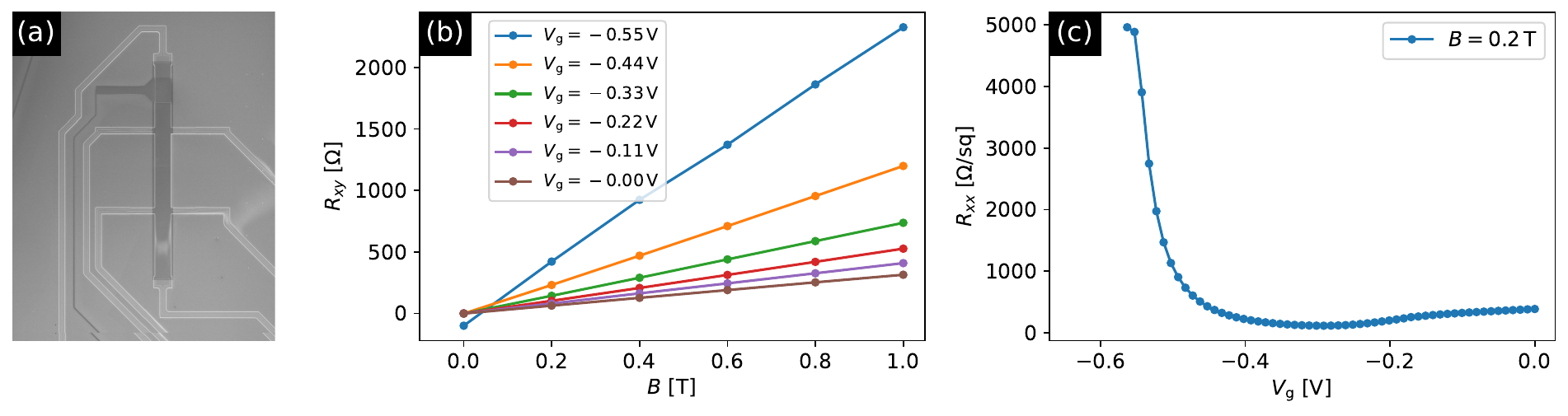}
\vskip -4mm
\caption{
Measurements of transverse and longitudinal resistance in the Hall bar device proximate to device A.
These raw data underpin the density and mobility data used in the impurity density extraction.
(a)~Image of the Hall bar device on the same chip as device A.
(b)~Transverse resistance ($R_{\textrm{xy}}$) as a function of the applied magnetic field at selected top gate $\Vg$ voltages.
(c)~Longitudinal resistivity as a function of applied top gate voltage $\Vg$ at $B = \SI{0.2}{\tesla}$ to avoid effects due to weak antilocalization.
Different colors correspond to sweeps down (blue) and up (orange) from $\left[V_{\textrm{th}}, V_{\textrm{sat}}\right]$, and down (green) and up (red) from $\left[V_{\textrm{th}} - \SI{0.5}{\volt}, V_{\textrm{sat}}\right]$ where $V_{\textrm{th}}$ is the threshold voltage and $V_{\textrm{sat}}$ is the saturation voltage of the Hall bar.
}
\label{fig:hall_bar_measurement}
\end{figure*}

\begin{figure}
\includegraphics[width=8.5cm]{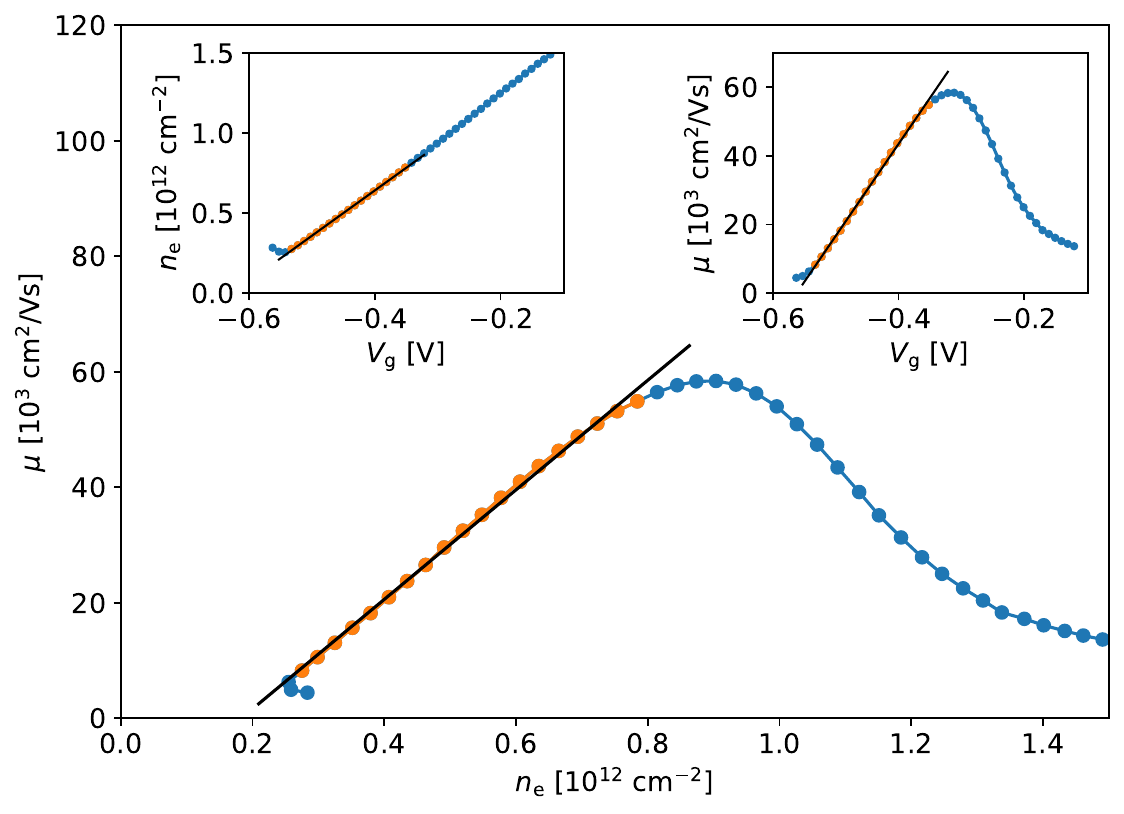}
\vskip -3mm
\caption{
Measured density and mobility vs gate voltage (as well as mobility vs density) from a Hall bar proximate to device A.
Orange points indicate the low-density data used in the fit.
Solid lines are the simulation result corresponding to the most-probable impurity density, $\sc = 2.7 \cdot \scu$, and dielectric permittivity.
}
\label{fig:hall_bar_mobility}
\end{figure}

Figure~\ref{fig:hall_bar_measurement}(a) shows an SEM image of a Hall bar used for 2DEG mobility measurements on the same chip as device A (with the proximity shown in the optical image in \Cref{fig:hall_bar_chip}).
Figure~\ref{fig:hall_bar_measurement}(b) shows the Hall resistance as a function of perpendicular field for a sequence of gate voltages, from which we extract the density as a function of gate voltage.
\Cref{fig:hall_bar_measurement}(c) shows the longitudinal conductance as a function of gate voltage.
By combining \Cref{fig:hall_bar_measurement}(b) and \ref{fig:hall_bar_measurement}(c), we obtain the mobility as a function of density.

Figure~\ref{fig:hall_bar_mobility} shows the mobility versus density for this Hall bar proximate to device~A [note that the low-density upturn in $n_\mathrm{e}(\Vg)$ and non-single-valued $\mu(\Vg)$ arise due to a measurement artifact and these points are excluded from the fitting].
The solid black lines represent our point estimate traces corresponding to the most-probable values of the effective impurity density, $\sc = 2.7 \cdot \scu$, and effective dielectric permittivity.

Using similar analyses, we extracted the corresponding effective charged impurity densities at the interface with the dielectric for devices C-F yielding $1.1 \cdot \scu$, $\scu$, $3.1 \cdot \scu$, and $3 \cdot \scu$, respectively.

\section{Localization length under aluminum}
\label{sec:localization_length}

In this appendix, we elaborate on the measurement of the localization length $\ellLoc$ in proximitized nanowires, which was briefly summarized in \Cref{sec:disorder_uniformity}.
We begin with the following observations.
A modest in-plane magnetic field ($B \sim 1\,$T) perpendicular to the wire will close the induced gap in all the segments of the semiconductor nanowire while the aluminum remains superconducting with a slightly suppressed parent gap $\DeltaAl\approx \SI{200}{\micro\eV}$.
When this occurs, $\GRL$ will be non-zero over the entire range of bias voltages from $\Vb = 0$ to $\DeltaAl$.
At small bias, the typical nonlocal conductance $\GRL$ depends on the length of the wire as $\GRL(\Vb) = A \exp(-2 L/ \ellLoc(\Vb))$ and similarly for $\GLR$.
To extract $\ellLoc$, we will measure the nonlocal conductances of segments of different length $L$.

\begin{figure}
\includegraphics[width=8.5cm]{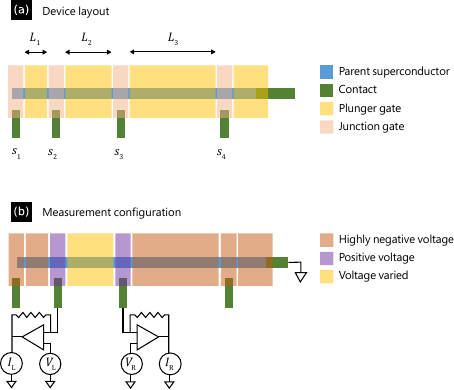}
\vskip -2mm
\caption{
(a)~Schematic of a nanowire device design with multiple gate defined sections of different lengths.
Sources $s_1$-$s_4$ contact the semiconductor quantum well.
(b)~The measurement configuration to measure conductance of a single section.
The junction gates for the section under measurement are set to a positive voltage to contact the semiconductor.
The plunger gate of this section will be varied during the measurement and all other gates are set to highly negative voltages to deplete all semiconductor states as described in the main text.
The non-local conductance is then measured with the aluminum nanowire grounded.
This measurement is then repeated for all sections of the nanowire.
}
\label{fig:1d_mobility_schematic}
\end{figure}

A schematic of the device used for this measurement is shown in \Cref{fig:1d_mobility_schematic}.
The device consists of a hybrid InAs/Al nanowire as described in the main text.
Sections of different length are defined by the plunger gates, having $L_{1-3}$ in the schematic.
Ohmic contacts are made to the semiconductor at several positions along the wire ($S_{1-4}$) and the coupling to the wire is controlled by junction gates.
By measuring $\GRL$ and $\GLR$ between sources 3 and 4, we obtain these conductances for a wire of length $L = \SI{1}{\micro\meter}$.
Between sources 2 and 3, a wire of length $L=\SI{2}{\micro\meter}$; between sources 2 and 4, a wire of length $L=\SI{3.5}{\micro\meter}$; sources 1 and 3, a wire of length $L=\SI{6.5}{\micro\meter}$; sources 1 and 4, a wire of length $L = \SI{8}{\micro\meter}$.
The wire width, charged disorder $\sc$, and junction transparencies are kept similar for different length wire segments.
The latter is accomplished by opening the junctions to reduce the dependence of $\GRL$ and $\GLR$ on the disorder configurations within each of the junctions.

We now describe the measurement of the semiconductor conductance for the section between $s_2$ and $s_3$.
The gate voltages are configured as illustrated in \Cref{fig:1d_mobility_schematic}(b).
The gates controlling the coupling between the contacts and the nanowire are set at a positive voltage, so that the semiconductor underneath these gates is in accumulation.
In this configuration each source has approximately the same contact resistance to the semiconductor under aluminum.

As noted above, a modest in-plane magnetic field is applied so that the semiconductor enters the normal state while the aluminum remains superconducting.
The non-local conductance is then measured between $s_2$ and $s_3$ while the aluminum is grounded.
All gates outside the section under measurement are set to highly negative voltages so that no semiconductor states are populated.
Transport is then allowed at energies below the aluminum gap between the two contacts, while carriers that tunnel through the barrier material into the aluminum are drained to ground and do not contribute to the measured conductance.
This behavior can be seen in \Cref{fig:hmp}(a), the conductance is approximately zero at energies above the aluminum gap.
At energies below the parent gap the conductance of the semiconductor is measured.

\section{TGP: Subtleties}
\label{sec:TGP_details}

\subsection{TGP measurements}
\label{sec:TGP_measurements}

First, to accelerate the search for ZBPs in Stage 1, rather then doing lengthy bias-voltage sweeps, we employ the third-harmonic ($3\omega$) technique described in Ref.~\onlinecite{Fornieri19}.
This gives a direct measurement of ${d^3}{I_\R}/d{V_\R^3}$, which is the curvature of the local conductance ${d}{I_\R}/d{V_\R}$.
When ${d^3}{I_\R}/d{V_\R^3}$ is negative and above the noise level, it indicates the presence of a ZBP.
Using this technique, we are able to scan over a large area in phase space by varying four parameters: $B, \Vp$, and the two cutter gate voltages $\Vlc$, $\Vrc$ that modulate the junction transparencies.
We further facilitate this by restricting the cutter gate voltages so that $\Gag$ at each junction is limited to the range $0.1$-$1\, e^2/h$ as a compromise between sufficient visibility and remaining in the tunneling regime.
In Stage 2, we will also sweep $\Vb$, but will restrict $(B, \Vp)$ to smaller ranges.

Second, $\GLR$ and $\GRL$ may receive contributions from line impedances in the measurement circuit, which we account for by taking the full impedance network into account \cite{Martinez21}.
In addition, we correct for finite frequency effects by calibrating the resistances and capacitances in the measurement circuit, as explained in \Cref{sec:measurement-circuit}.
Finally, we remove any remaining voltage divider corrections and improve SNR by focusing on the parts of the non-local conductances that are antisymmetric in bias voltage, $A(\GRL)$, $A(\GLR)$:
\begin{equation}
    A[(\GRL(\Vb)] \equiv \left[\GRL(\Vb) - \GRL(-\Vb)\right]/2
\end{equation}
and similarly for $\GLR$.
A discussion of relevant multi-terminal conductance symmetry relations may be found in Ref.~\onlinecite{Maiani22}.

Third, the transport gap extracted from $A(\GRL)$ will not, in general, be the same as that extracted from $A(\GLR)$.
The underlying transport gap is the same, but the two non-local conductances may not be the same due to the different ways in which local matrix elements enter $\GLR$ and $\GRL$.
[We will see an example of this in simulated data in \Cref{fig:simulated_DLG_epsilon_R1_stage2}(b,c) in \Cref{sec:testing_TGP}.]
This can obscure a gap narrowing or closing.
Hence, for any given $\Vp$ and $B$, we determine the induced gap as the lower of the gaps extracted from $A(\GRL)$ and $A(\GLR)$.
Consequently, the observation of non-vanishing $A(\GLR)$ \textit{or} $A(\GRL)$ at bias voltages approaching zero is a signature of a bulk gap closing.
Note that the energy spectrum can be gapless due to the presence of disorder-induced localized states at low energies.
The observed transport gap, extracted from $A(\GLR)$ and $A(\GRL)$, is the gap that we really care about since it is more predictive of qubit performance.

A fourth subtlety is that $A[\GLR({V_\R})], A[\GRL({V_\L})]$ can be either positive or negative, depending on whether transport at that bias voltage is primarily due to electrons or holes.
It can change sign as the matrix elements change as a function of $B$ or $\Vp$, passing through zero when this occurs~\cite{Wang21}.
Such a sign change can appear as a very sharp increase in the induced gap, centered about some $B$ or $\Vp$ value.
In determining the zero-field induced gap, we simply avoid these points.
The situation is slightly more complicated for the topological gap, since it varies between zero at the phase transition and a maximum value that determines the stability of the topological phase.
Hence, we will report both the median value of the gap over the SOI$_2$ and also its maximum value.
However, we will not extract the maximum from the single point at which it is largest.
It is not uncommon to have a very small region over which the extracted topological gap is very large because one of the aforementioned sign changes occurs, suppressing the signal in $\GLR$ and $\GRL$.
Hence, we define the ``maximum topological gap'' $\DeltaMax$ to be the upper quintile of measured gap values within SOI$_2$.

We add a note of caution here that the measured values of $\DeltaInd$ and $\DeltaMax$ are transport gaps, which are the lowest energy at which there is an excited state whose localization length is longer than the device length; they can be larger than the $\DeltaInd$ or $\DeltaMax$ expected for a clean system.

Fifth, we clarify the definition of the gapless boundary of an SOI$_2$.
The boundary of the region is defined to consist of all points inside the SOI$_2$ that neighbor the exterior of the SOI$_2$ on the side or diagonally.
Each such boundary point is then considered gapped if all of the neighboring points outside SOI$_2$ are gapped.
Otherwise the boundary point is considered gapless.
Using this definition, we extract the fraction of the boundary points that are gapless.

\subsection{TGP parameters}
\label{sec:TGP_parameters}

The TGP is parametrized by thresholds that were fixed by an initial set of calibration simulations described in \Cref{sec:thresholds} and then tested extensively by large-scale simulations for different disorder levels and device designs, as described in \Cref{sec:FDR}.
As a result, we have high confidence that ROI$_2$s overlap with the regions in parameter space where there is a topological phase, as we quantify in \Cref{sec:FDR}.

In principle, this protocol is designed to detect any topological phase with a sizable gap.
Finite experimental resolution and temperature, however, may obfuscate some of the topological signatures, giving rise to subtleties when interpreting the data that we discuss here.

A wire may have MZMs, but one or both of them may be slightly displaced from the end of the wire for some choices of junction transparency and, therefore, may not be visible.
(Indeed, we see in \Cref{fig:device_SLG}(c) that the local density of states can be peaked a few hundred nanometers away from the junction in a simulation of an ideal disorder-free device.)
For this reason, we do not insist that a ZBP be present for all junction configurations and, instead, consider a ZBP to be stable in Stage 1 if it is visible for at least 70\%,  of measured junction transparencies.
We define the ``cutter gate fraction'' as the fraction of junction transparencies (or, equivalently, cutter gate settings) for which a ZBP is present.
For instance, suppose we pick $5$ cutter gate voltages $V_\mathrm{rc}$ at the right junction such that $\Gag$ at the right junction takes the values 0.35, 0.49, 0.62, 0.76, $0.9e^2/h$ and similarly pick $5$ cutter gate voltages $V_\mathrm{lc}$ at the left junction so that $\Gag$ at the left junction ranges over the same five values.
The precise sampling over cutter pairs varies between the measurements presented in this paper, but there are always at least 20 cutter gate voltage pairs.
Then, a $(B,\Vp )$-point will be said to exhibit stable ZBPs at both junctions if there are ZBPs at both junctions for $> 70\%$ of cutter gate pairs $(V_\mathrm{lc}, V_\mathrm{rc})$.

For Stage 2, we set a threshold percentage $(\mathrm{ZBP}\%)_\mathrm{th}$ and define a stable ZBP as one that is visible for at least $(\mathrm{ZBP}\%)_\mathrm{th}$ of junction transparencies.
As we discuss further in \Cref{sec:thresholds}, we use calibration simulations to inform the choice $(\mathrm{ZBP}\%)_\mathrm{th}=60\%$ for the device parameters considered in this paper.
For the junction transparencies given as an example in the previous paragraph, we would perform Stage 2 measurements for 5 different cutter gate pairs $(V_\mathrm{lc},V_\mathrm{rc})$ such that $\Gag^R = \Gag^L = 0.35$, 0.49, 0.62, 0.76, $0.9e^2/h$.
A stable ZBP in Stage 2 would then need to be present for 3/5 cutter gate pairs.

A gap closing may not be visible even when it is present because $A(\GRL)$ and $A(\GLR)$ tend to be small at low bias voltage (due to their anti-symmetry in bias voltage) and will be suppressed even further by disorder and non-uniformity.
Another reason why an SOI$_2$ may be gapless along less than 100\% of its boundary is that neighboring gapped regions may have been misidentified as non-topological (e.g.~due to weak coupling of the MZMs to the leads or ZBP splitting due to a small topological gap).
In case of such misidentification, such regions should actually be included in the SOI$_2$, which would be larger and gapless along its entire boundary.
To account for both of these possibilities, we set a threshold percentage $(\mathrm{GB}\%)_\mathrm{th}$.
In order to qualify as an SOI$_2$, a cluster must be gapless along at least $(\mathrm{GB}\%)_\mathrm{th}$ of its boundary.
As we discuss in \Cref{sec:thresholds}, we use simulated transport data to inform the choice $(\mathrm{GB}\%)_\mathrm{th}=60\%$ for the device parameters discussed in this paper.

The non-local conductances $A(\GRL)$ and $A(\GLR)$ will never truly vanish at zero bias because there will at least be tunneling $e^{-L/\xi(0)}$ and thermally-activated $e^{-\Delta_\mathrm{tr}/T}$ contributions.%
\footnote{See also, Ref.~\onlinecite{Pan21a} for additional considerations that are relevant to the non-local conductance.}
Hence, we need to give an operational definition for $A(\GRL) \approx 0$ and $A(\GLR) \approx 0$.
To do this, we define a threshold value $G_\mathrm{th}$.
Then, if $A[\GLR({V_\R})]$, $A[\GRL({V_\L})] < G_\mathrm{th}$, we interpret this as $A[\GLR({V_\R})]$, $A[\GRL({V_\L})] \sim O(e^{-L/\xi(0)}, e^{-\Delta/T})$.
The extracted gap is obtained using the highest bias voltage $\Vb$ below which $A[\GLR(\Vb)] < G_\mathrm{th}$ and $A[\GRL(\Vb)] < G_\mathrm{th}$.
The choice of $G_\mathrm{th}$ should depend on $\Delta/T$ and $L/\xi(0)$ and also on the transparency of the junctions.
For the disorder strengths expected in our devices, we take $G_\mathrm{th}$ equal to $\exp(-3)\approx 0.05 $ times the maximal value $\max\{G_\NL\}$ of the non-local conductance at bias voltages greater than the induced gap (scanning over all $B$ for each $\Vp$ for a given cutter configuration).
As we discuss later in this section, this choice of $G_\mathrm{th}$ was set by applying the TGP to calibration data from simulated devices with $\sc = 2.4 \cdot \scu$.
For weaker disorder $\sc \ll \scu$, the optimal value of $G_\mathrm{th}$ should be smaller because $L/\xi(0) \gg 1$ and the transition becomes sharper in this limit.
If we don't take a smaller $G_\mathrm{th}$, the TGP will miss gap closings and will erroneously interpret $A(\GRL)$ and $A(\GLR)$ data as indicating that the gap remains open.
Defining $G_\mathrm{th}$ in terms of the high-bias conductance $\max\{G_\NL\}$ enables us to define it equally well for simulated data as for measured data (unlike, for instance, a $G_\mathrm{th}$ that depends on the noise level in a particular measurement setup).
When we plot either simulated or measured $A[\GLR(V)], A[\GRL(V)]$, we will use a black curve to indicate the bias voltage (as a function of $B$ field) below which each one is less than $G_\mathrm{th}$, see, e.g., \Cref{fig:simulated_DLG_epsilon_R1_stage2}(e,f).
We can restate the threshold defined in this paragraph as follows: there is a truly sharp distinction between the trivial and topological phases only in the infinite-size, zero-temperature, and infinitesimal transparency limits; hence the threshold gives a simulation-tested method for finding the rounded transition.

Finally, we note that the extracted transport gap and, therefore, the phase diagram depend on the junction transparencies (and, thereby, on the cutter gate voltages that control them).
The phase diagram must be stable to changes in the cutter gate voltages in the following sense: we require that a device passing the TGP must have a $\mathcal{C}^A_i \in T$ for a threshold percentage $(\mathcal{C}_i \%)_{\rm th}$ (see below) of cutter gate pairs $i$.
In other words, in order to pass the TGP, a fraction $(\mathcal{C}_i \%)_{\rm th}$ of cutter gate settings must have at least one $\mathcal{C}^A_i$ that is a subset of the ROI$_2$.
We will take $(\mathrm{C}_i\%)_{\rm th}=50\%$.
This combination of the stability thresholds for ZBPs, gap closing and reopening requirements and the overlap of the resulting SOI$_2$ is sufficient to virtually eliminate false positives when analyzing simulated data, as we shall see below in \Cref{sec:FDR}.

\section{Calibrating and testing the TGP with data from simulated devices}
\label{sec:TGP_calibration_testing}

In this section, we use simulations to quantify the reliability of the TGP.
Our transport simulations begin with three-dimensional models of the devices in \Cref{fig:device_SLG,fig:device_DLG} that include the electrostatic environment defined by the set of gate voltages.
We identify the $\Vp$ range for which the chemical potential is in the lowest sub-band and the cutter gate voltages for which the junction transparencies take the $5$ values $\Gag = 0.35$, 0.49, 0.62, 0.76, $0.9e^2/h$.
The resulting single sub-band parameters are given in \Cref{tab:effective_parameters} and \Cref{fig:disorder_strength_vs_n2D}.
For this gate voltage set, we perform transport simulations and calculate the scattering matrix of the system.
The local and non-local conductances $\GLL, \GRR, \GLR, \GRL$ are then obtained by convolution with the derivative of the Fermi function at temperature $T$.
We analyze this data according to the TGP according to the same procedure that we will use in \Cref{sec:experimental_data} to analyze experimental data.

\subsection{False discovery rate}
\label{sec:FDR}

The basic question that we wish to answer is: suppose the TGP returns an ROI$_2$ that passes; what is the probability that it does \textit{not} have any overlap with the topological phase? 
The goal of TGP calibration is to set thresholds that minimize this probability.
Once the TGP has been calibrated, we test it to assess whether this probability is low when the TGP is applied to a range of devices types and parameters: different junction designs, different material parameters such as spin-orbit coupling, different disorder strength.

To compute this probability in simulations, we compare ROI$_2$s with a topological index (the ``scattering invariant''~\cite{Fulga11}).
We classify ROI$_2$s as follows: if an ROI$_2$ has any overlap with a region of the simulated phase diagram with scattering invariant $-1$, then we will call it a true positive; otherwise, it is a false positive.
In \Cref{sec:TGP} (see \Cref{eq:FDR_preview}), we defined the classification of regions as TP and FP, and derived the FDR from these numbers in order to quantitatively measure the reliability of the TGP.

Note that this is a classification of regions, rather than a classification of devices.
Hence, if a device that does have a topological region were to pass the TGP but its ROI$_2$ were completely disjoint from the topological region, then this ROI$_2$ would be a false positive.
We do not attempt to count negative regions: an arbitrary region is very likely to be negative, so the number of negative regions is not a useful statistic.%
\footnote{On the other hand, a ``negative device,'' which is a device that fails the TGP, is a natural concept.
We return to it later when we define the TGP yield, which is the probability that a \textit{device} will pass the TGP~--- in other words, the complement of the probability of a negative device.}
Therefore, we do not compute the false positive rate (FPR), given by $\mathrm{FP}/(\mathrm{FP} + \mathrm{TN})$, where TN is the number of true negatives.
The FPR is less useful for the present discussion.

There is a further subtlety, which is that the scattering invariant depends on the junction transparencies.
Since our device has two junctions, the scattering invariant can be defined at either one: $\mathrm{SI}_i = \mathrm{sgn} \det r_i \in [-1,1]$, where $r_i$ is the reflection matrix and $i = \mathrm{L}$, $\mathrm{R}$.
When the junctions are completely closed, both $\mathrm{SI}_i$ are trivially equal to $+1$.
When the junctions are opened, $\mathrm{SI}_i = -1$ regions can appear, and they tend to grow as the junctions are opened further.
Thus, we must decide how to assign a topological index to a finite system.
We will define the topological region of the phase diagram as the union of the $\mathrm{SI}_i = -1$ regions over $i = \mathrm{L}$, $\mathrm{R}$ and the 5 different pairs of junction transparencies used in TGP Stage 2.
We will call this union the ``$\mathrm{SI} = -1$ region.''
Instead of taking the union of the $\mathrm{SI}_i = -1$ regions, we could have taken the intersections.
In our simulations, $\mathrm{SI}_\R = -1$ and $\mathrm{SI}_\L = -1$ regions overlap but not completely.
For some cutter gate settings, only $\mathrm{SI}_\R = -1$ while for others $\mathrm{SI}_\L = -1$ because a zero-energy state can couple poorly to the lead at one junction or the other for different cutter gate settings.
Our definition of the topological index is relatively insensitive to these details of the junctions.
If we had wanted to use the Pfaffian invariant \cite{Kitaev01}, we would have had to truncate the system to remove the junctions and then imposed periodic/anti-periodic boundary conditions.
This would no longer be the same device as we would be probing in transport.

A final technical detail: the conductance matrix is temperature dependent, and the output of the TGP has a resulting temperature dependence.
As we discuss in \Cref{sec:experimental_data}, the base temperature of our dilution refrigerators during these measurements is $20\,$mK, and the electron temperature is estimated to be $\lesssim 40\,$mK.
We use transport data at $30\,$mK for our calibration simulations.
We test the TGP using simulated data at $40\,$mK, and we used this simulated data in our estimates of the FDR and other statistical properties.

The main result of the subsections that follow is that we estimate that the FDR is $\leqslant 8\%$ at a 95\% confidence level for all device designs, material stacks, and disorder levels simulated.
Thus, when a device passes the TGP, there is $>92\%$ probability that it has a non-zero gap and $\mathrm{SI} = -1$.
The details are in \Cref{tab:TGP_FDR}.

\subsection{Setting the TGP thresholds}
\label{sec:thresholds}

As discussed in \Cref{sec:TGP_parameters}, there are three key thresholds which parametrize the TGP: $(\mathrm{ZBP}\%)_\mathrm{th}$, $(\mathrm{GB}\%)_\mathrm{th}$, and $G_\mathrm{th}$.
We choose these parameters so that the TGP has a low FDR.
If we were to make it very difficult to pass the TGP, then we would have very few false positives but also few true positives, and the FDR could be large.
If we make it too easy to pass the TGP, then will have many more true positives but also more false positives.
However, the right choices of $(\mathrm{ZBP}\%)_\mathrm{th}$, $(\mathrm{GB}\%)_\mathrm{th}$, and $G_\mathrm{th}$ lead to a TGP which is reliable because it has low FDR.

We performed an initial calibration of the TGP by analyzing simulated transport data  at $T = 30\,$mK from $28$ disorder realizations of an SLG-$\beta$ device with an average charged defect density of $\sc = 2.4 \cdot \scu$ and spin-orbit interaction $\alpha \simeq 13\,\mathrm{meV}{\cdot}\mathrm{nm}$.
We did not find any false positives: every ROI$_2$ has at least some subset with $\mathrm{SI} = -1$.
However, the number of true positives and, hence, the FDR varies with the threshold values.
We find that $(\mathrm{ZBP}\%)_\mathrm{th}=60\%$, $(\mathrm{GB}\%)_\mathrm{th} = 60\%$, and $G_\mathrm{th} = 0.05 \max\{G_\NL\}$ is close to optimal.
There are 53 ROI$_2$s spread across the $28$ devices (all true positives) for these threshold settings.
Hence, the FDR is $< 6.7\%$ at the $95\%$ confidence level.
Had $(\mathrm{ZBP}\%)_\mathrm{th}$ and  $(\mathrm{GB}\%)_\mathrm{th}$ been set lower, we would have found ROI$_2$s that did not contain a region with $\mathrm{SI} = -1$.
For higher $(\mathrm{ZBP}\%)_\mathrm{th}$ and  $(\mathrm{GB}\%)_\mathrm{th}$, we would have had fewer devices passing the TGP.
If we had set $G_\mathrm{th}$ too low, then too much of the phase diagram would have been classified as gapless, thereby concealing gap re-openings.
If we had set $G_\mathrm{th}$ too high, then too much of the phase diagram would have been classified as gapped, obscuring gap closings.
Note that the two thresholds, $(\mathrm{GB}\%)_\mathrm{th}$ and $G_\mathrm{th}$ are correlated.

We show two realizations of simulated SLG, $\beta$-stack devices in \Cref{sec:SLG_example}, one that passes the TGP and one that fails.

\subsection{Testing the TGP}
\label{sec:testing_TGP}

With the TGP thus calibrated and validated, we turn to simulations estimating the FDR.
Since we will be analyzing experimental data from devices with different designs, material stacks, and disorder levels, we apply the TGP to simulated SLG-$\beta$ and DLG-$\varepsilon$ devices with several different charge disorder levels.
For charge disorder given by $\sc \geqslant \scu$, we use the thresholds obtained in the calibration described above, whereas in the cleaner case of $0.1 \cdot \scu$ we lower $G_\mathrm{th}$ from $0.05$ to $0.01$ while leaving the other thresholds unchanged.
Although the \textit{optimal} values of the thresholds in the TGP depend on the temperature, design, material stack, and disorder level, our goal here is to show that, for the thresholds chosen, the TGP remains \textit{reliable} across a range of designs, material stacks, and disorder levels, and at a slightly higher temperature.
The motivation is that we would like to apply the TGP in cases in which neither the electron temperature nor the disorder level is known precisely.
Note, however, that if either the disorder level or temperature were very different, then we would probably need to adjust $(\mathrm{ZBP}\%)_\mathrm{th}$, $(\mathrm{GB}\%)_\mathrm{th}$, and $G_\mathrm{th}$.

To estimate the FDR, we performed large-scale simulations involving 349 different disorder realizations, distributed across SLG-$\beta$ and DLG-$\varepsilon$ devices with intermediate to strong disorder levels defined by charge impurity densities of 0.1, 1, 2.7 and 4 in units of $\scu$, as shown in \Cref{tab:TGP_FDR}.
The Hall bar measurements described in \Cref{sec:electrostatic_calibration} indicate that the devices experimentally measured in this paper have $\sc$ values in this range.
The defect density $\sc = 2.7 \cdot \scu$ is the largest of any of the measured devices reported in this paper that has passed the TGP, while $\sc = 4 \cdot \scu$ is larger than in of any of the devices reported in this paper, including those that failed the TGP.
Several of our measured devices have $\sc$  values at or below $\scu$.

We calculate the conductance matrix \Cref{eq:G_matrix} for each simulated disorder realization at $T = 40\,$mK and analyze this data according to the TGP, as we would with experimental data.
This analysis yields ROI$_2$s; many disorder realizations, especially with stronger disorder, have none, while some disorder realizations, typically with weaker disorder, have multiple ROI$_2$s.
We compare these ROI$_2$s to the scattering invariant and classify each ROI$_2$ as a TP or FP depending on whether any subregion of the ROI$_2$ has $\mathrm{SI} = -1$.
We find the statistics given in \Cref{tab:TGP_FDR}.
From the TP and FP values obtained from these simulations, we estimate the FDR by assuming a binomial distribution  and use the Clopper-Pearson confidence interval at a 95\% confidence level.

We do not have a single false positive in this data.
This does not mean that the TGP perfectly identifies the topological region.
Our results show that the TGP identifies an ROI$_2$ that has non-zero overlap with the $\mathrm{SI} = -1$ regions in the phase diagram.
However, part of the ROI$_2$s identified by the TGP do not have $\mathrm{SI} = -1$ and much of the $\mathrm{SI} = -1$ region lies outside the ROI$_2$, as we will see in the examples that we discuss in \Cref{sec:single_realization_example} and \Cref{sec:SLG_example}.
This is not surprising since, as we discussed in \Cref{sec:topological_invariants}, there is some inherent ambiguity in defining the topological phase in a finite system.
When we restrict the magnetic field to $B \leqslant 2.5\,$T, we find one false positive for an SLG-$\beta$ device with $\sc = 1.0 \cdot \scu$.
By restricting the magnetic field, the normalization of the conductance changes slightly, and a candidate SOI$_2$ for one cutter gate setting is classified as gapped with a very small gap; for a larger $B$ field range, it is classified as gapless and the regions fails the TGP.

Our analysis of the FDR indicates that TGP reliably identifies an $\mathrm{SI} = -1$ region for different disorder levels and for different device designs.
Indeed, as summarized in Table \ref{tab:TGP_FDR}, there are only small differences in the estimated FDR values for the above-mentioned parameters, and they are primarily due to the different numbers of ROI$_2$s for various disorder levels.
Thus, the TGP at these threshold values can be applied to a large class of topological gap devices with intermediate disorder strength.

Our simulations also give us information about how the disorder level and device design affect the probability that a device will pass the TGP.
We define the \textit{TGP yield} as this probability:
\begin{equation}
\textrm{TGP yield} \equiv P(\textrm{Device passes TGP}).
\end{equation}
As may be seen from \Cref{tab:TGP_yield}, the TGP yield depends strongly on $\sc$.
More disordered devices are less likely to pass the TGP because they are less likely to have a topological phase.

\begin{figure*}
\includegraphics[width=18cm]{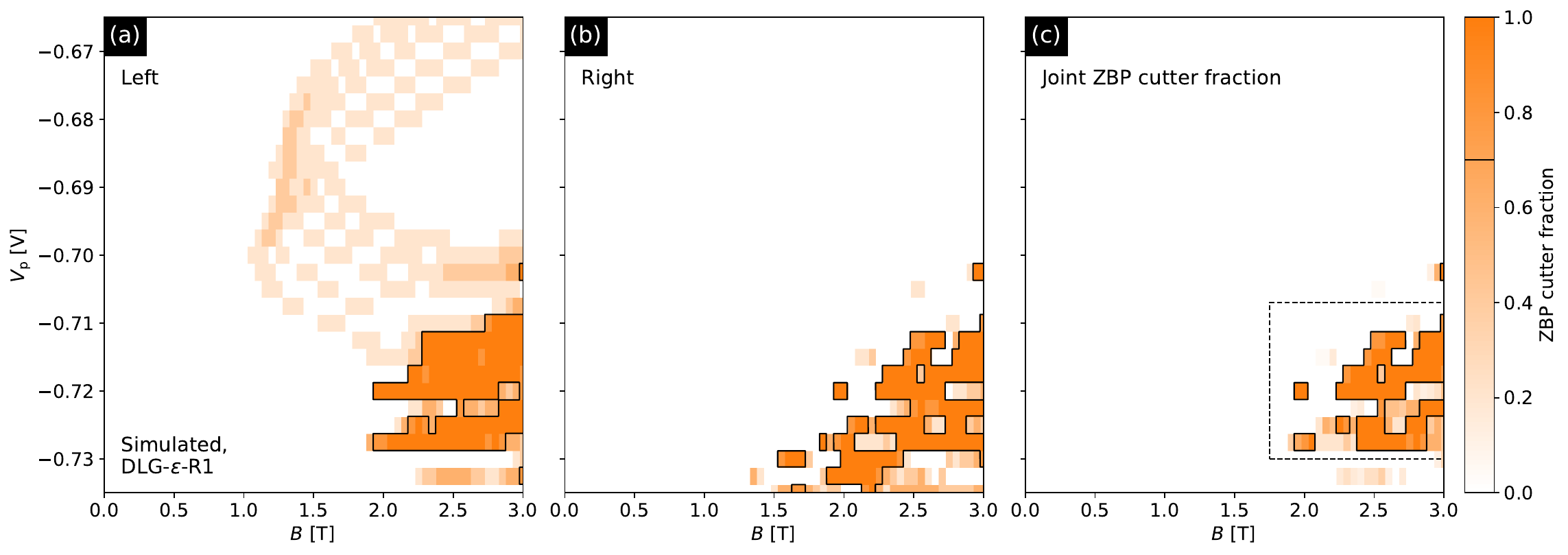}
\vskip -2mm
\caption{
The TGP applied to simulated transport data for DLG with $\sc = \scu$ and $40\,$mK temperature broadening, highlighting regions of phase space with stable ZBPs in (a)~$\GLL$ and (b)~$\GRR$.
(c)~Regions of phase space with stable ZBPs in both $\GLL$ and $\GRR$.
Simulated transport data such as this are used to test the TGP.
Here, we show one particular disorder realization for illustrative purposes.
Any possible visible resemblance to measured data is dependent on the disorder realization and does \textit{not} play a role in our analysis.
}
\label{fig:simulated_DLG_epsilon_R1_stage1}
\end{figure*}

We now consider the statistics of the SOI$_2$s that the TGP finds.
\Cref{tab:TGP_yield} shows these statistics when the magnetic field is restricted to $B\leqslant 2.5\,$T, while $B\leqslant 3\,$T in \Cref{tab:TGP_yield2}.
We find that $\DeltaMax$ has mean value $\meanDeltaMax$ that varies between 25 and \SI{35}{\micro\eV} for different disorder strengths and has a non-Gaussian distribution with long tails towards larger gap.
A smaller mean value is observed at the lowest disorder levels because there are significantly more SOI$_2$s at weak disorder, and many of them have small $\DeltaMax$.
The mean volume in the $B$-$\Vp$ parameter space of an SOI$_2$, denoted $\volSOIbar$, is of the order $0.1\,$mV$\cdot$T and decreases with increasing disorder strength, see \Cref{tab:TGP_yield}.
As with $\meanDeltaMax$, the distribution of $\volSOIbar$ is non-Gaussian with long tails.
Finally, the average $B$ field at which we observe SOI$_2$s ranges between 1.5\,T and 2.6\,T.
As expected, $\varepsilon$-stack devices have SOI$_2$s that occur at higher magnetic fields $> 2\,$T since they have larger $\DeltaInd$ and smaller $g^\star$.

\begin{table}
\begin{center}
\begin{tabularx}{\columnwidth}{|I|Y|C|U|C|V|C|}
\cline{1-6}
\shortstack[l]{\noalign{\vskip 1.0ex} Design, \\ stack} &
\shortstack[c]{$\sc$ \\ $[\scu]$} &
\shortstack[c]{Yield \\ $\le 3\,$T} &
\shortstack[c]{\noalign{\vskip 1.0ex} $\meanDeltaMax$ \\ $[\si{\micro\eV}]$} &
\shortstack[c]{$\volSOIbar$ \\ $[\mathrm{mV}\cdot\mathrm{T}]$} &
\shortstack[c]{$\BavSOIbar$  \\ $[\mathrm{T}]$}
\\ [0.5ex]
\cline{1-6}
\multirow{3}{*}{SLG-$\beta$}
& 1.0 &  $48/50$ & 23 & 0.2 & 2.1 \\
\cline{2-6}
& 2.7 &  $24/50$ & 34 & 0.1 & 1.5 \\
\cline{2-6}
& 4.0 &  $26/49$ & 36 & 0.1 & 1.5 \\
\hline
\multirow{4}{*}{DLG-$\varepsilon$}
& 0.1 & $48/50$ & 26  & 0.2 & 2.6 \\ 
\cline{2-6}
& 1.0 & $43/50$ & 29  & 0.2 & 2.6 \\ 
\cline{2-6}
& 2.7 & $33/50$ & 28 & 0.2 & 2.6 \\ 
\cline{2-6}
& 4.0 & $35/50$ & 28 & 0.2  & 2.6 \\
\cline{1-6}
\end{tabularx}
\end{center}
\vskip -3mm
\caption{
The analogous table to \Cref{tab:TGP_yield} with magnetic field restricted to the range $B\leqslant 3\,$T.
}
\label{tab:TGP_yield2}
\end{table}

\subsection{Example of the TGP applied to a single disorder realization}
\label{sec:single_realization_example}

To illustrate the TGP, we now focus on a particular disorder realization in a narrow \SI{3}{\micro\meter} long device based on the DLG, $\varepsilon$-stack design.
This is one of the devices that appears in \Cref{tab:TGP_FDR,tab:TGP_yield}.
We will call this simulated disorder realization DLG-$\varepsilon$-R1 for brevity.
We have applied the TGP to $T = 40\,$mK transport data for this device, which we discuss in detail below, explaining the different stages of the TGP through this example.
We also compare an SOI$_2$ identified by the TGP with the topological region determined by $\mathrm{SI} = -1$.
In \Cref{sec:SLG_example}, we discuss simulated data from two other devices that we call realizations SLG-$\beta$-R1 and SLG-$\beta$-R2.

\textit{Stage 1:} Stage 1 focuses on ZBPs in the local conductance.
From the local conductances $\GRR$ and $\GLL$, we can map out the regions in $(B,\Vp)$ space where there are stable ZBPs at the two junctions, where ``stable'' means that the ZBPs are present for a cutter gate fraction $> 70\%$, as described in \Cref{sec:TGP_parameters}.
For simulated disorder realization DLG-$\varepsilon$-R1, the locations of stable ZBPs at the left junction are shown in \Cref{fig:simulated_DLG_epsilon_R1_stage1}(a) and at the right junction in \Cref{fig:simulated_DLG_epsilon_R1_stage1}(b).
A topological phase should have stable ZBPs at both junctions at the same $B$ and $\Vp$, so \Cref{fig:simulated_DLG_epsilon_R1_stage1}(c) shows the phase space locations where there are stable ZBPs at both junctions.
This is the output of Stage 1 of the TGP.
The entire gate voltage range shown here, $-0.735\,\mathrm{V} \leqslant \Vp \leqslant -0.665\,$V, lies within the lowest sub-band.
There is a trivial zero-energy state at the left junction over a region in the $B-\Vp$ plane that traces out a parabolic shape starting around $(B,\Vp)=(1.2\,\mathrm{T},-0.69\,\mathrm{V})$.
However, this ZBP is unstable (i.e.~fine-tuned) with respect to cutter changes and, therefore, is filtered out by the TGP.
The ROI$_1$ identified by the TGP is a smaller region at higher $B$ and lower $\Vp$ where there are stable ZBPs at both junctions.

\begin{figure*}
\vskip -3mm
\includegraphics[width=18cm]{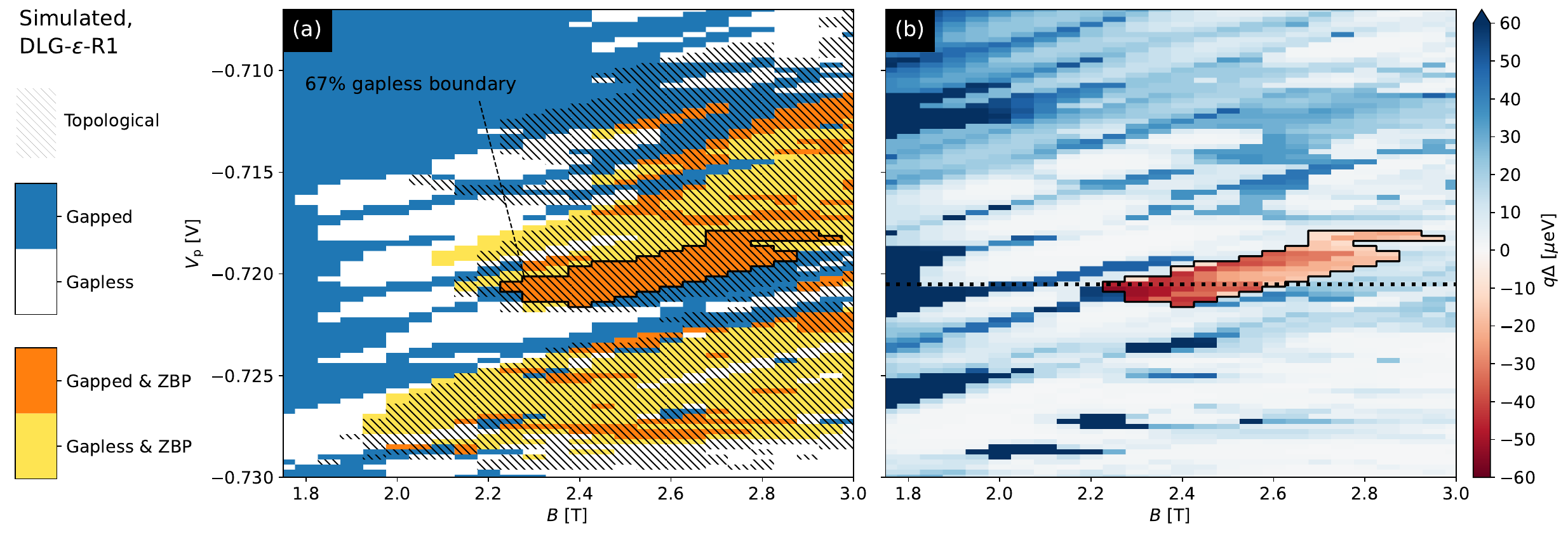}
\includegraphics[width=18cm]{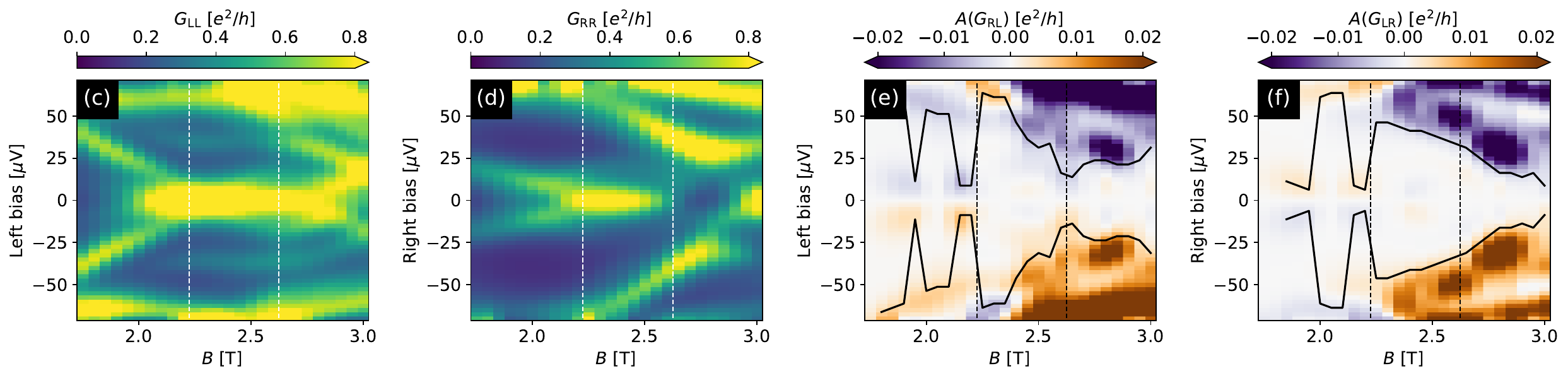}
\includegraphics[width=17.9cm]{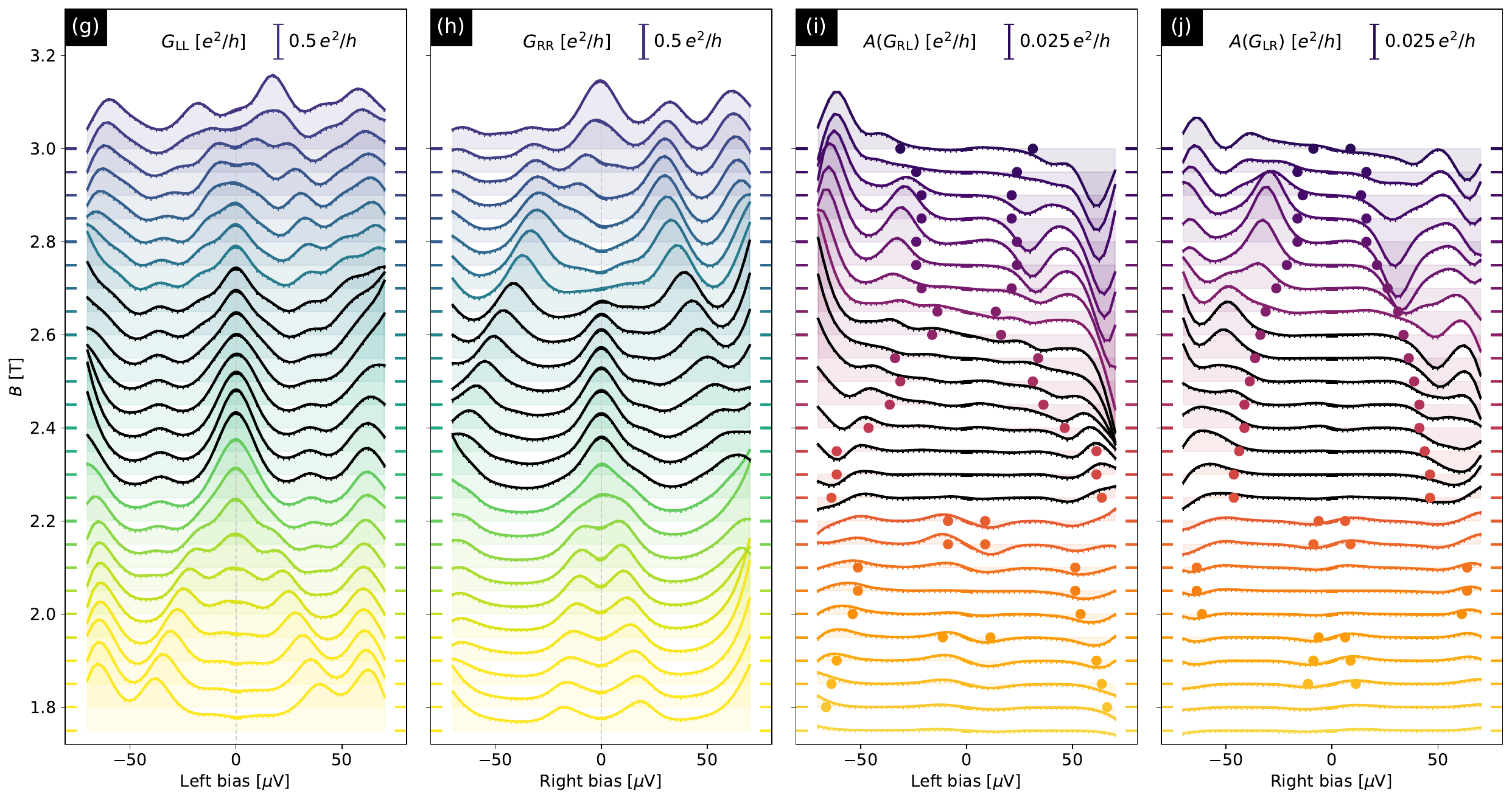}
\vskip -3mm
\caption{
(a)~The simulated phase diagram for DLG, $\varepsilon$-stack with $\sc = \scu$ at $T = 40\,$mK, combining the clusters of stable ZBPs at both junctions with the map of zero/non-zero gap.
We identify gapped/gapless regions, with/without stable ZBPs, according to the color legend on the left.
67\% of the boundary of the SOI$_2$ is gapless.
The hatched regions are where the topological invariant is negative.
(b)~The simulated phase diagram, showing trivial and topological phases, as identified by the TGP.
$q = \pm 1$ in the trivial/topological phase, so the color scale shows the size of the trivial (blue) or topological (red) gap.
The protocol assigns a maximum topological gap (defined as the top quintile of measured gaps within the SOI$_2$) of $\DeltaMax = \SI{41}{\micro\eV}$.
Simulated local and anti-symmetrized non-local conductances at $\Vp = -0.7205\,$V: (c)~$\GLL$, (d)~$\GRR$, (e)~$A(\GRL)$, (f)~$A(\GLR)$.
The field range between the vertical lines is in the SOI$_2$.
Panels (g)-(j) are ``waterfall'' plots representing the same simulated data.
The black curves in panels (e) and (f) and the dots in panels (i) and (j) are \textit{not} guides to the eye; they indicate where the non-local signal drops below a threshold value, as described in the text.
The details of these plots are disorder-dependent, and any visible resemblance to measured data does \textit{not} play a role in our analysis.
}
\label{fig:simulated_DLG_epsilon_R1_stage2}
\end{figure*}

\textit{Stage 2:}
In Stage 2, we focus on the neighborhood of the ROI$_1$ identified in Stage 1.
Stage 2 analyzes local and non-local transport data over a range of bias voltages.
In the simulated data, unlike in the experimental data discussed in \Cref{sec:experimental_data}, there is no drift of the plunger gate voltage $\Vp$, so the ROI$_1$ is automatically recovered from local transport data.
However, we do obtain the dependence of $\GRR$ and $\GLL$ on the bias voltage, as seen in \Cref{fig:simulated_DLG_epsilon_R1_stage2}(c,d).
The more significant new ingredient in Stage 2 is the bias-dependence of the non-local conductances $\GRL$ and $\GLR$, from which we determine the transport gap as a function of $B$ and $\Vp$.
For simulated disorder realization DLG-$\varepsilon$-R1, the non-local conductances $\GRL$, $\GLR$ and the derived gap (indicated by black curves) as a function of $B$ for $\Vp = -0.7205\,$V are shown in \Cref{fig:simulated_DLG_epsilon_R1_stage2}(e,f).
We give $5$ significant digits after the decimal point for $\Vp$ so that these values can serve as indices in the relevant data files; this may be convenient for readers wishing to look directly at the data underlying the figures in this paper.
Note that this cut passes through the topological phase transition and the topological phase; a gap closing and re-opening is seen in $\GRL$ and $\GLR$.
Combining the local and non-local information, we can classify any point in phase space as gapped without stable ZBPs, gapped with stable ZBPs, gapless without stable ZBPs, or gapless with stable ZBPs.
These are depicted in \Cref{fig:simulated_DLG_epsilon_R1_stage2}(a) as, respectively, blue, orange, white, or yellow.
If an orange region is surrounded by white or yellow along more than $60\%$ of its boundary, we identify it as an SOI$_2$, and give it a black boundary in \Cref{fig:simulated_DLG_epsilon_R1_stage2}(a).

In addition to stable ZBPs, we also observe some zero-energy states that have a non-topological origin, as noted in our Stage 1 analysis.
These trivial Andreev bound states (ABS) can be caused by resonances arising from the local electrostatic potential, local disorder, or a combination.
We give some examples of their spectroscopy in \Cref{sec:SLG_example}.
While the presence of these ABSs can be limited by careful design of the junctions, they can never be fully suppressed.
When running the TGP in experiments, we take care to tune away from pathological points in junction phase space as much as possible, while stepping between different junction configurations as we tune $\Gag$ to span from $0.1e^2/h$ to $e^2/h$.

As we can see in \Cref{fig:simulated_DLG_epsilon_R1_stage2}(a,b), the TGP finds a region of topological phase [shown in orange in panel (a) and red in panel (b)] around $B = 2.55\,$T, $\Vp = -0.72\,$V with a maximum topological gap $\DeltaMax = \SI{41}{\micro\eV}$, which is the upper quintile of the gaps extracted in the SOI$_2$.

\begin{figure*}
\includegraphics[width=18cm]{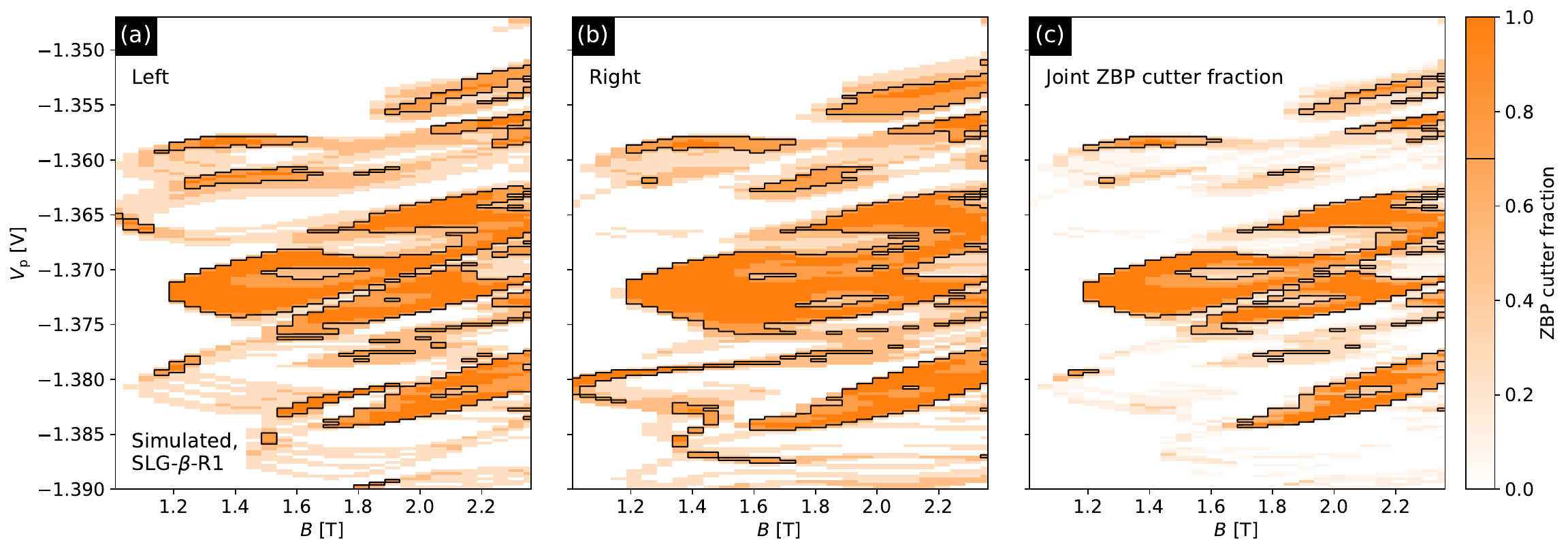}
\vskip -3mm
\caption{
The TGP applied to simulated transport data for a narrow \SI{3}{\micro\meter} long wire with $\sc = 2.4 \cdot \scu$, $\alpha \simeq 13 \mathrm{meV}\cdot\mathrm{nm}$ at $T = 30\,$mK, highlighting regions of phase space with stable ZBPs in (a)~$\GLL$ and (b)~$\GRR$.
(c)~Regions of phase space with stable ZBPs in both $\GLL$ and
$\GRR$.
Simulated transport data such as this are used to test the TGP.
Here, we show one particular disorder realization for illustrative purposes.
Any possible visible resemblance to measured data is dependent on the disorder realization and does \textit{not} play a role in our analysis.
}
\label{fig:simulated_SLG_beta_R1_stage1}
\end{figure*}

The topological gap increases from zero at the phase transition to $\DeltaMax$ in such a way that its median value over the red region within the black line in \Cref{fig:simulated_DLG_epsilon_R1_stage2}(b) is \SI{26}{\micro\eV}.
The TGP phase diagram is compared with the scattering invariant.
In \Cref{fig:simulated_DLG_epsilon_R1_stage2}(a), the region with $\mathrm{SI} = -1$ is hatched.
As may be seen in \Cref{fig:simulated_DLG_epsilon_R1_stage2}(a), the region identified by the TGP lies almost entirely within the region with negative topological index, i.e.~the TGP is fairly conservative and identifies a subset of the topological phase.
It is not a perfect match, of course, since the TGP is not directly calculating the scattering invariant and, moreover, the phase transition is rounded by finite temperature and finite-size corrections.
Note the similarity between the hatched regions with negative topological index in \Cref{fig:simulated_DLG_epsilon_R1_stage2}(a) and the bright orange regions.
The regions that pass the TGP are smaller \textit{pockets} within these splinters.

\Cref{fig:simulated_DLG_epsilon_R1_stage2}(b) is another version of the phase diagram for this disorder realization: it shows the transport gap extracted from $\GRL$ and $\GLR$ multiplied by $q = \pm 1$, depending on whether that point lies outside or inside the SOI$_2$.
Note that $q$ is \textit{not} the same as the topological invariant $\mathcal{Q}$ that is used analogously in the color scale in \Cref{fig:simulated_infinite_clean_gap} (where $\mathcal{Q}$ is the Pfaffian invariant); rather, $q$ is the proxy for $\mathcal{Q}$ that results from the TGP.
The color scale of \Cref{fig:simulated_DLG_epsilon_R1_stage2}(b) can be viewed intuitively as the magnitude of the bulk gap multiplied by a proxy for the sign of the topological invariant.
Darker red corresponds to larger topological gap.

As expected, there are some points along the boundary of the SOI$_2$ where the closing is not visible in either $\GRL$ or $\GLR$.
For this cutter gate pair in this particular disorder realization, 67\% of the boundary of the SOI$_2$ shows a gap closing in $\GRL$, $\GLR$, which is above $(\mathrm{GB}\%)_\mathrm{th}$.

A comparison between the orange region in \Cref{fig:simulated_DLG_epsilon_R1_stage2}(a) and the hatched region lends credence to the idea that the part of the boundary that is gapped is not actually a boundary at all, and the true boundary is at higher $B$ and lower $\Vp$.
The blue region below and to the right of the orange region is hatched, indicating that it has been misclassified as non-topological.
However, as may be seen in \Cref{fig:simulated_DLG_epsilon_R1_stage2}(b), this gapped (and ostensibly topological) region below and to the right of the SOI$_2$ is light blue, indicating that it has a small gap.
Hence, if it were topological, it would have a small topological gap and would not be useful for a topological qubit, in contrast to the darkest red regions within the SOI$_2$.
The same observation applies to all of the hatched $\mathrm{SI} = -1$ region that lies outside of the SOI$_2$: it may be topological, but it has a small transport gap.
Moreover, since they do not have stable ZBPs in $\GLL$ and/or $\GRR$, it is unlikely that it would be possible to couple to these ZBPs in a qubit.

In this section, we discussed a disorder realization for a DLG-$\varepsilon$ device that passed both Stages of the TGP and correctly found the topological phase.
In \Cref{sec:SLG_example}, we will show additional data from calibration simulations and discuss examples that fail the TGP.

\subsection{Simulated disorder realizations used for the calibration of TGP}
\label{sec:SLG_example}

\begin{figure*}
\vskip -3mm
\includegraphics[width=18cm]{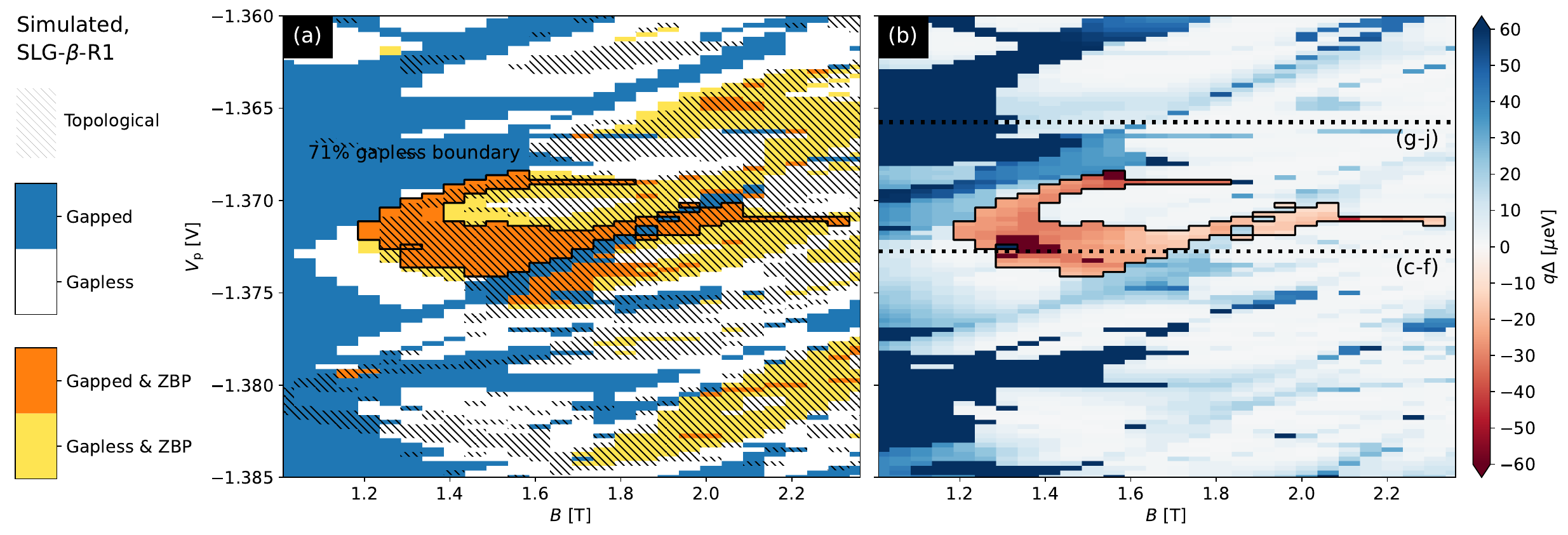}
\includegraphics[width=18cm]{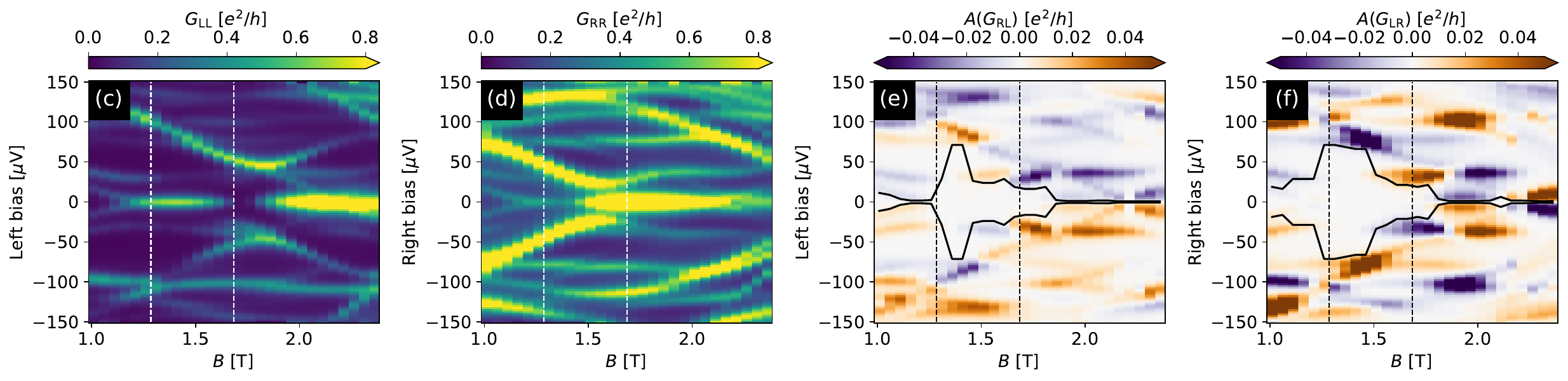}
\includegraphics[width=18cm]{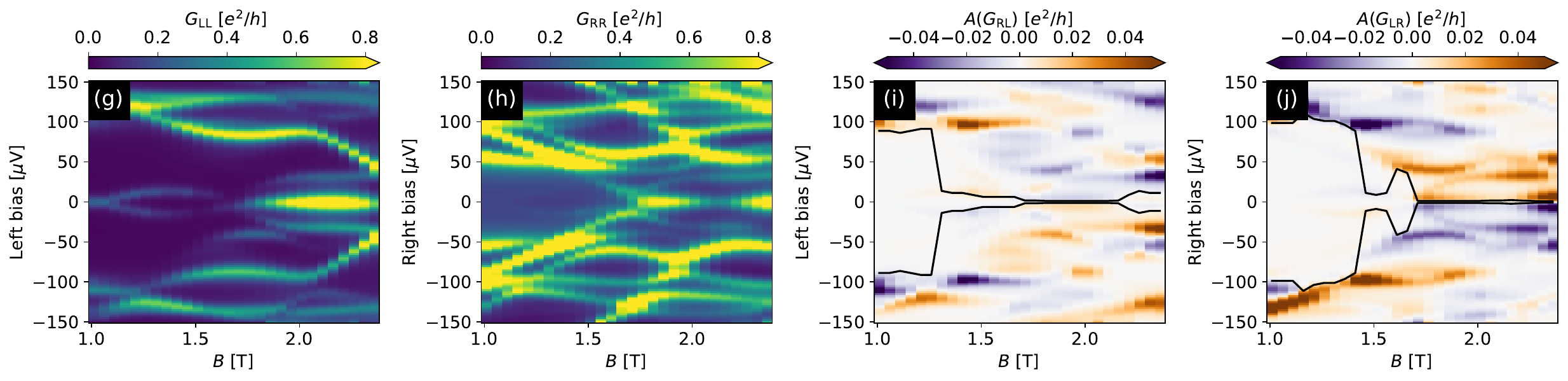}
\vskip -3mm
\caption{
(a)~The simulated phase diagram for SLG device (disorder realization R1) with $\sc = 2.4 \cdot \scu$, $\alpha \simeq 13 \mathrm{meV}\cdot\mathrm{nm}$ at $T = 30\,$mK, combining the clusters of stable ZBPs at both junctions with the map of zero/non-zero gap.
We identify gapped/gapless regions, with/without stable ZBPs, according to the color legend on the left.
71\% of the boundary of the SOI$_2$ is gapless.
The hatched regions are where the topological invariant is negative.
(b)~The simulated phase diagram, showing trivial and topological phases, as identified by the TGP.
$q = \pm 1$ in the trivial/topological phase, so the color scale shows the size of the trivial (blue) or topological (red) gap.
The protocol assigns a maximum topological gap (defined as the top quintile of measured gaps within the SOI$_2$) of $\DeltaMax = \SI{30}{\micro\eV}$.
The lower and upper horizontal dotted lines correspond to the cuts shown, respectively, in panels (c-f) and (g-j).
(c-f)~Simulated local and anti-symmetrized non-local conductances at $\Vp = -1.37275\,$V: (c)~$\GLL$, (d)~$\GRR$, (e)~$A(\GRL)$, (f)~$A(\GLR)$.
The field range between the vertical lines is in the SOI$_2$.
(g-j)~Conductances from the non-topological region ($\Vp = -1.36575\,$V, outside of SOI$_2$).
As may be seen by comparing panels (c,d) and (g,h), the local conductances are similar, but the non-local conductances in panels (i,j) lack a clear transport gap re-opening, distinguishing them from panels (e,f).
}
\label{fig:simulated_SLG_beta_R1_stage2}
\end{figure*}

In this Appendix, we discuss two examples of simulated data that was used to calibrate the TGP.
One example passes the TGP and one fails.
The parameters of the simulation are equivalent to those of SLG-$\beta$ devices with a larger spin-orbit coupling $\alpha \simeq 13\, \mathrm{meV}{\cdot}\mathrm{nm}$.

The first realization, called SLG-$\beta$-R1, represents the \SI{3}{\micro\meter}-long device with $\sc = 2.4 \cdot \scu$.
Here we assume $T = 30\,$mK.
Stage 1 data is shown in \Cref{fig:simulated_SLG_beta_R1_stage1}.
There are stable ZBP clusters in $\GRR$ and $\GLL$ as well as accidental ZBPs present for some cutter settings which correspond to disorder-induced subgap states at the junction.
Due to the larger level of disorder in SLG-$\beta$-R1 than in DLG-$\varepsilon$-R1, there are more such accidental sub-gap states than in \Cref{fig:simulated_DLG_epsilon_R1_stage1}.
The joint ZBP map is shown in \Cref{fig:simulated_SLG_beta_R1_stage1}(c).

The Stage 2 analysis of simulated device SLG-$\beta$-R1 is presented in \Cref{fig:simulated_SLG_beta_R1_stage2}.
Simulated device SLG-$\beta$-R1 has an SOI$_2$ around $(B, \Vp) \approx (1.5\,\mathrm{T}, -1.37275\,\mathrm{V})$.
71\% of the boundary of of this SOI$_2$ is gapless;
the maximum and median topological gaps are \SI{30}{\micro\eV} and \SI{25}{\micro\eV}, respectively, for this cutter gate setting.
Local and non-local conductances for $\Vp = -1.37275\,$V are shown in \Cref{fig:simulated_SLG_beta_R1_stage2}(c-f).
There are stable ZBPs at both junctions over a range of $B$ fields of extent $\approx 1.3$-$1.7\,$T, and there is a clear gap closing and re-opening as a function of $B$.
Also, note the similarity between the hatched regions with negative SI in \Cref{fig:simulated_SLG_beta_R1_stage2}(a) and the bright orange regions in the simulated
Stage 1 data in \Cref{fig:simulated_SLG_beta_R1_stage1}.
 
For the sake of comparison, we also present trivial ZBPs seen at a different value of $\Vp = -1.36575\,$V in \Cref{fig:simulated_SLG_beta_R1_stage2}(g-j) [see upper dotted line in \Cref{fig:simulated_SLG_beta_R1_stage2}(b)].
However, the non-local conductance data in \Cref{fig:simulated_SLG_beta_R1_stage2}(i,j) clearly show that the system is gapless over the corresponding range of $B$ field values.

\begin{figure*}
\includegraphics[width=18cm]{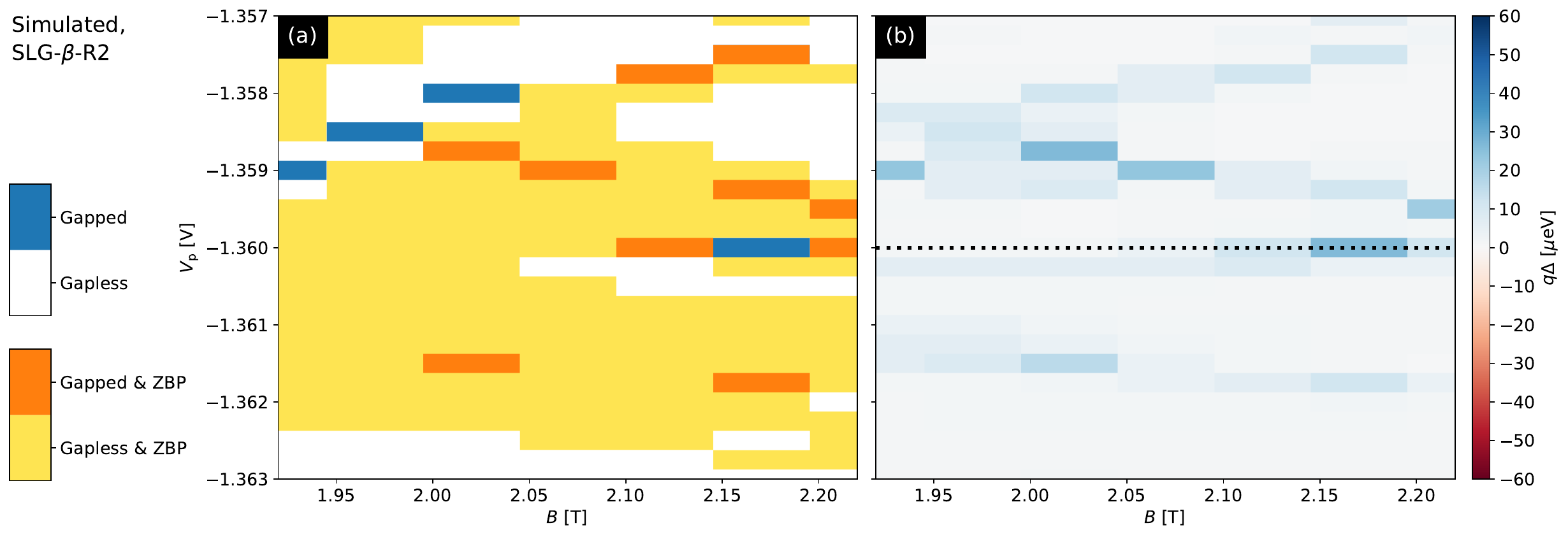}
\includegraphics[width=18cm]{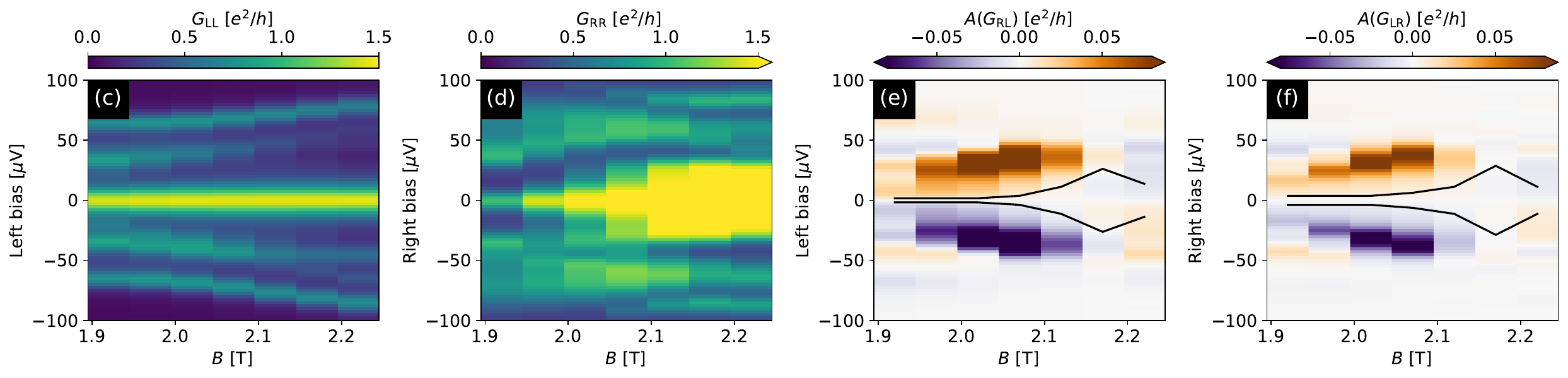}
\vskip -3mm
\caption{
Stage 2 analysis for realization SLG-$\beta$-R2 at $T = 30\,$mK.
(a)~The regions with stable ZBPs at both junctions.
(b)~The gap as function of $B$ and $\Vp$.
It vanishes in the region of interest, so this device fails the TGP.
(c)-(f)~Local and anti-symmetrized non-local conductances at $\Vp = -1.36\,$V.
The local conductances in panels (c) and (d) show ZBPs, but there is no gap re-opening visible in the anti-symmetrized non-local conductances in panels (e) and (f).
}
\label{fig:simulated_SLG_beta_R2_stage2}
\end{figure*}

In \Cref{fig:simulated_SLG_beta_R2_stage2}, we present Stage 2 of TGP for another disorder realization, which we call SLG-$\beta$-R2.
This is one of the realizations that passed Stage 1 but failed Stage 2.
There are stable ZBPs at both junctions around $\Vp = -1.36\,$V, as may be seen in \Cref{fig:simulated_SLG_beta_R2_stage2}(c,d).
However, the non-local conductances in \Cref{fig:simulated_SLG_beta_R2_stage2}(e,f) yield zero gap.
In fact, the, system is gapless over most of the scanned region, so this simulated transport data does not pass Stage 2 of the TGP.
This example once again reinforces the fact that local measurements alone cannot reliably detect a topological phase.

\section{Comparison of SOI$_2$ for different cutter pairs}
\label{sec:comparison_cutters}

\begin{figure*}
\includegraphics[width=18cm]{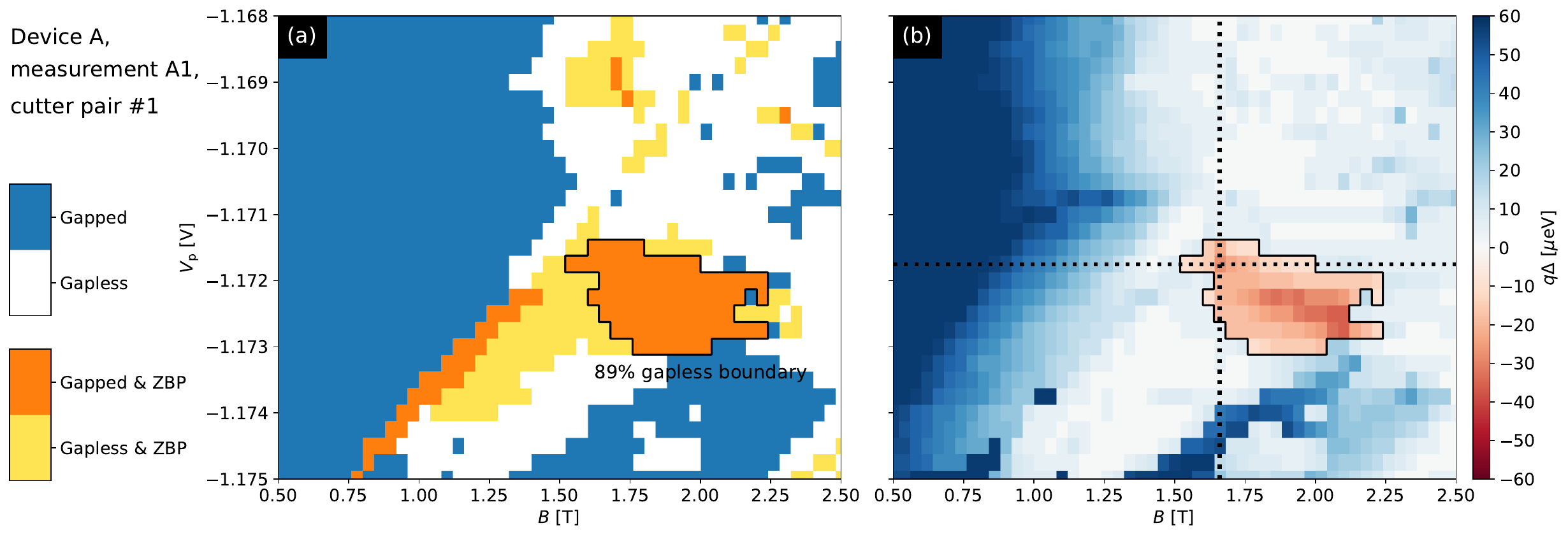}
\includegraphics[width=18cm]{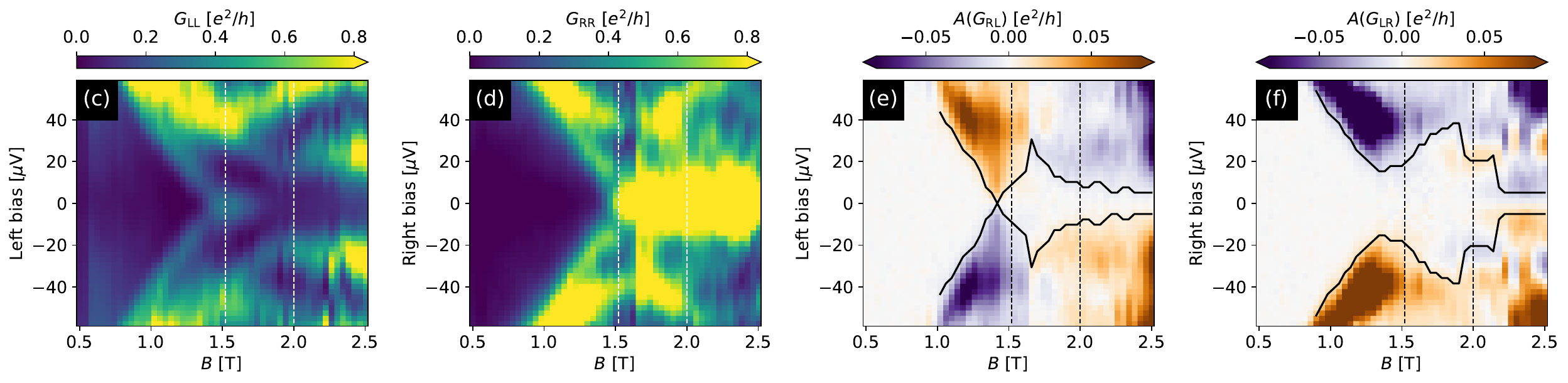}
\includegraphics[width=17.9cm]{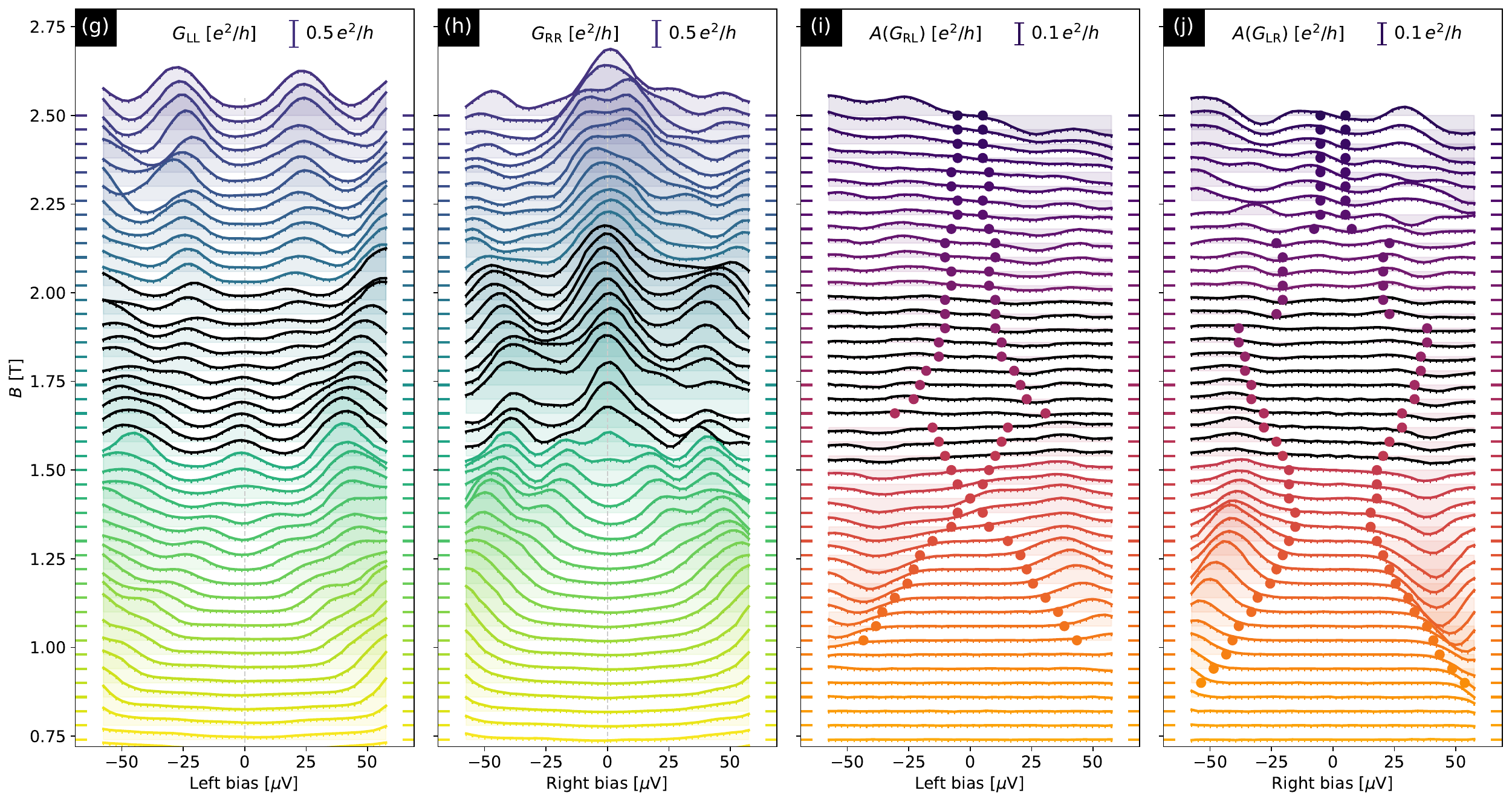}
\vskip -3mm
\caption{
TGP stage 2 analysis of the device A (measurement A1).
Figure shows the same as in \Cref{fig:deviceA1_stage2} but for cutter pair \#1 corresponding to $\Gag \approx 0.5 e^2/h$ for both sides.
(a)~The boundary of the SOI$_2$ is interpreted as a phase transition line, consistent with a visible gap closure along 89\% of it.
(b)~The protocol assigns a maximum topological gap $\DeltaMax = \SI{26}{\micro\eV}$.
}
\label{fig:deviceA1_stage2_cutter1}
\end{figure*}

\begin{figure*}
\includegraphics[width=18cm]{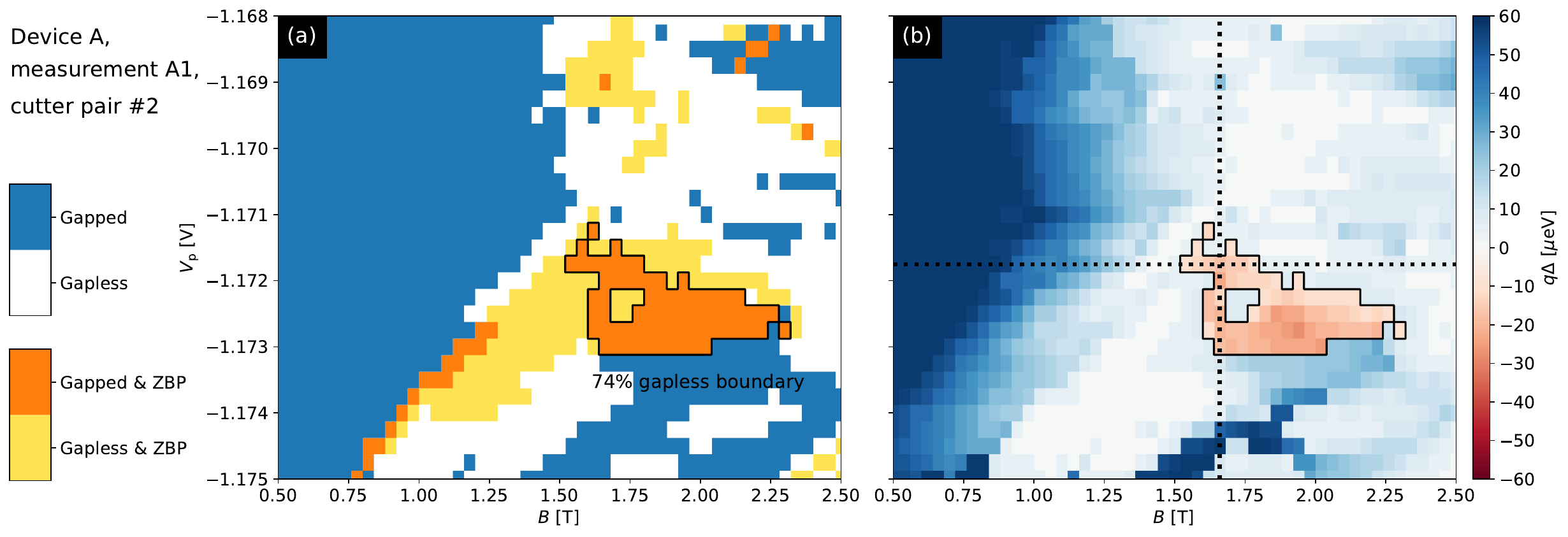}
\includegraphics[width=18cm]{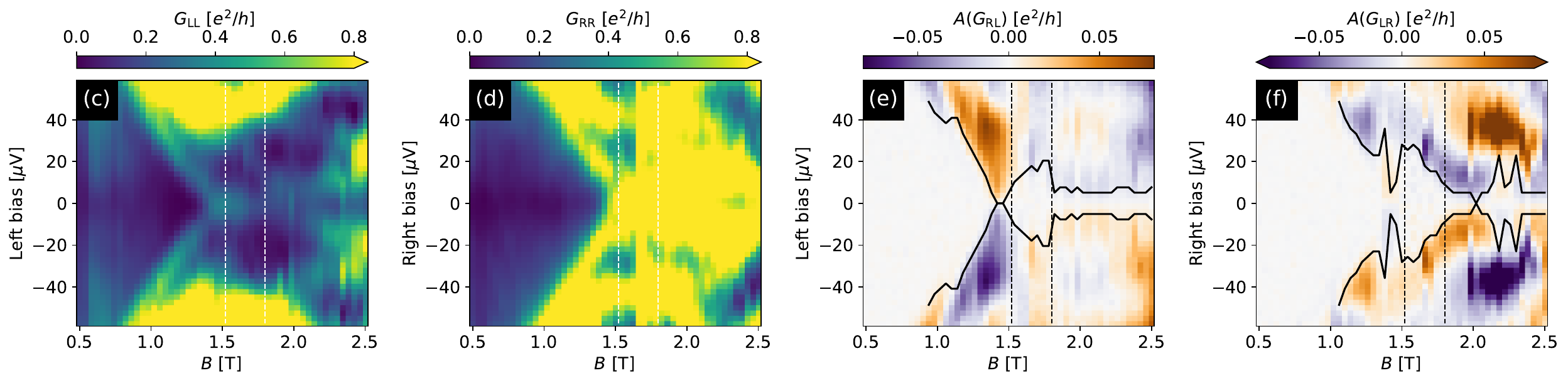}
\includegraphics[width=17.9cm]{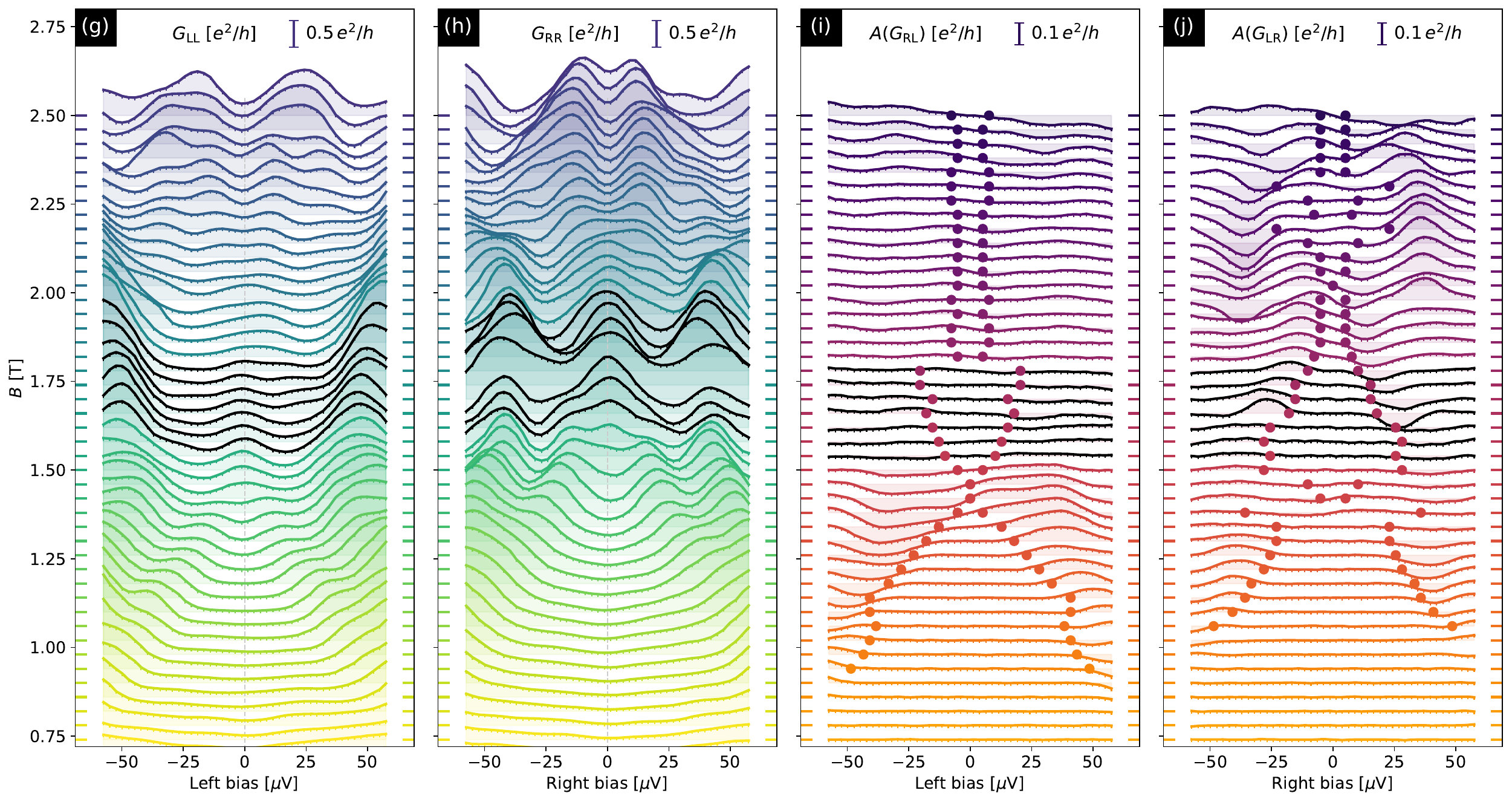}
\vskip -3mm
\caption{
TGP stage 2 analysis of the device A (measurement A1).
Figure shows the same as in \Cref{fig:deviceA1_stage2} but for cutter pair \#2 corresponding to $\Gag \approx 0.7 e^2/h$ for both sides.
(a)~The boundary of the SOI$_2$ is interpreted as a phase transition line, consistent with a visible gap closure along 74\% of it.
(b)~The protocol assigns a maximum topological gap $\DeltaMax = \SI{21}{\micro\eV}$.
}
\label{fig:deviceA1_stage2_cutter2}
\end{figure*}

In this section we compare the TGP stage 2 outcome for different cutter pairs.
\Cref{fig:deviceA1_stage2} demonstrates the results for cutter pair \#0  (with $\Gag \approx 0.3 e^2/h$ for both sides) showing 78\% gapless boundary and $\DeltaMax = \SI{23}{\micro\eV}$.
Results for cutter pairs \#1 ($\Gag \approx 0.5 e^2/h$) and \#2 ($\Gag \approx 0.7 e^2/h$) are demonstrated in \Cref{fig:deviceA1_stage2_cutter1,fig:deviceA1_stage2_cutter2} showing 89\% gapless boundary and $\DeltaMax = \SI{26}{\micro\eV}$ for cutter pairs \#1 and 74\% gapless boundary and $\DeltaMax = \SI{21}{\micro\eV}$ for cutter pairs \#2.
This comparison shows that SOI$_2$s corresponding to the different cutter pairs are similar.
The vertical and horizontal dotted lines in panel (b) are the same in \Cref{fig:deviceA1_stage2,fig:deviceA1_stage2_cutter1,fig:deviceA1_stage2_cutter2}; the intersection of this two lines is always inside the SOI$_1$ which clearly demonstrates that they overlap.

\section{Three-terminal conductance measurements with several hundred Hz excitations}
\label{sec:measurement-circuit}

The conductance matrix is measured with a standard lock-in technique, applying simultaneous voltage excitations with amplitudes $\Vinput_\L(f_\L)$ and $\Vinput_\R(f_\R)$ to the inputs of the left and right terminals, respectively.
They are at frequencies $f_\L$ and $f_\R$, respectively, and currents $\Iinput_\L(f_\L)$ and $\Iinput_\L(f_\R)$ are measured at the left terminal while $\Iinput_\R(f_\L)$ and $\Iinput_\R(f_\R)$ are measured at the right terminal.
However, due to the finite impedance network between the signal input and the measured sample, the voltages applied and the currents measured deviate from those at the sample inputs.
At lock-in frequencies of several hundred Hz, the capacitances of the cryostat lines and filters (on the order of several nF) combined with the line and filter resistances have a non-negligible contribution to the measured currents.
Furthermore, a finite resistance at the drain contact imposes an effective voltage modulation of the opposing contact as a voltage divider effect.
As a result, the sample experiences voltage modulations $\Vsample_\L(f_\L)$ and $\Vsample_\L(f_\R)$ at the left and $\Vsample_\R(f_\L)$ and $\Vsample_\R(f_\R)$ at the right input.
The currents at the the left terminal are $\Isample_\L(f_\L)$ and $\Isample_\L(f_\R)$ while the currents at the right terminal are $\Isample_\R(f_\L)$ and $\Isample_\R(f_\R)$.
Noting that $\Isample_{i}(f) = G_{ij} \Vsample_{j}(f)$ where $i, j = \mathrm{L}, \mathrm{R}$, the conductances are obtained from
\begin{equation}
\begin{pmatrix} 
G_\LL & G_\LR \\ G_\RL & G_\RR
\end{pmatrix} 
= 
\begin{pmatrix} 
\Isample_\L(f_\L) & \Isample_\L(f_\R) \\ \Isample_\R(f_\L) & \Isample_\R(f_\R)
\end{pmatrix} 
\begin{pmatrix} 
\Vsample_\L(f_\L) & \Vsample_\L(f_\R) \\ 
\Vsample_\R(f_\L) & \Vsample_\R(f_\R)
\end{pmatrix}^{-1}.
\label{eq:ConductanceEvaluation}
\end{equation}
Note that we recover \Cref{eq:G_matrix} if we take $\Vsample_\L(f_\R) = \Vsample_\R(f_\L) = 0$.
To proceed, we need a map
\begin{equation}
\Mm \equiv \begin{pmatrix} \Mm_{VV} & \Mm_{VI} \\ \Mm_{IV} & \Mm_{II} \end{pmatrix}
\end{equation}
between the voltages and currents at the sample, $\Vv = [\Vsample_\L(f), \Vsample_\R(f)]^\mathrm{T}$ and $\Iv = [\Isample_\L(f), \Isample_\R(f)]^\mathrm{T}$, and the voltages applied and the currents measured, $\Vv_0 = [\Vinput_{\L, 0}(f), \Vinput_{\R, 0}(f)]^\mathrm{T}$ and $\Iv_0 = [\Iinput_{\L, 0}(f), \Iinput_{\R, 0}(f)]^\mathrm{T}$:
\begin{equation}
\begin{pmatrix} \Vv \\ \Iv \end{pmatrix} = 
\begin{pmatrix} \Mm_{VV} & \Mm_{VI} \\ \Mm_{IV} & \Mm_{II} \end{pmatrix}
\begin{pmatrix} \Vv_0 \\ \Iv_0 \end{pmatrix}.
\end{equation}

\begin{figure*}
\includegraphics[width=15.0cm]{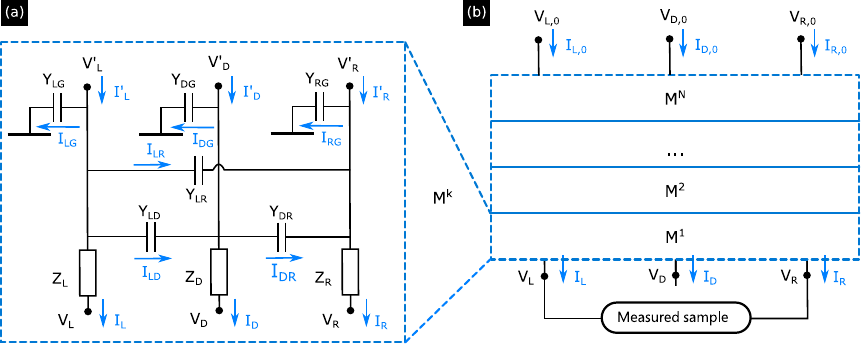}
\vskip -2mm
\caption{
(a)~The stage of the impedance network that leads from the input ports to the measured sample.
(b)~The full network consists of multiple stages in series.
}
\label{fig:fridge_circuit_model}
\end{figure*}

The impedance network between the input ports and the measured sample consists of a sequence of stages, each of which has the general form illustrated in \Cref{fig:fridge_circuit_model}(a): every line has a capacitance to ground and to every other line in parallel and a resistance in series to the next stage.
The drain of the measured device is also connected to ground at the input and is thus included in the circuit here and will be reduced from the final mapping later.
By denoting the terminals as $k = \{\mathrm{L}, \mathrm{R}, \mathrm{D}\}$ and $l = \{\mathrm{L}, \mathrm{R}, \mathrm{D}, \mathrm{G}\}$ (where G denotes ground), the output voltages $dV_k$ and currents $dI_{k}$ are determined by the input voltages $dV'_k$ and currents $dI'_{k}$, as well as cross-currents $dI_{kl}$ as
\begin{multline}
dI_k = dI_k' - \sum_{l \neq k} dI_{kl} 
= dI_k' - \sum_{l \neq k} Y_{kl} (dV'_k - dV'_l) \\
= -\biggl(\sum_{l \neq k} Y_{kl}\biggr) dV'_k + \sum_{l \neq k} Y_{kl}dV'_l + dI_k',
\label{eq:Current}
\end{multline}
\begin{multline}
dV_k = dV'_k - Z_k dI_k\\
= \biggl(1 + Z_k \biggl(\sum_{l \neq k} Y_{kl}\biggr)\biggr) dV'_k - Z_k \sum_{l \neq k} Y_{kl} dV'_l - Z_k dI'_k,
\label{eq:Voltage}
\end{multline}
where $Z_k$ are the impedances (resistances for sufficiently low frequencies) of the lines, $Y_{kl} = 2\pi i f C_{kl}$ are the parallel admittances via capacitors $C_{kl}$ between $k$ and $l$, and $V_\G = 0$.
Equations~\eqref{eq:Current} and \eqref{eq:Voltage} can be expressed in matrix form as
\begin{widetext}
\begin{equation} 
\begin{pmatrix}
V_\L \\
V_\R \\
V_\D \\
I_\L \\
I_\R \\
I_\D
\end{pmatrix}
=
\begin{pmatrix}
1 + Z_\L \sum_{k = \R, \D, \G} Y_{\L k}& -Z_\L Y_\LR & -Z_\L Y_{\L\D} & -Z_\L & 0 & 0 \\
-Z_\R Y_\LR & 1 + Z_\R \sum_{k = \L, \D, \G} Y_{Rk} & -Z_\R Y_{\R\D} & 0 & -Z_\R & 0  \\
-Z_\D Y_{\L\D} & -Z_\D Y_{\R\D} & 1 + Z_\D \sum_{k = \L, \R, \G} Y_{\D k} & 0 & 0 & -Z_\D  \\
-\sum_{k = \R, \D, \G} Y_{\L k} & Y_\LR & Y_{\L\D} & 1 & 0 & 0 \\
Y_\LR & -\sum_{k = \L, \D, \G} Y_{\R k} & Y_{\R\D} & 0 & 1 & 0 \\
Y_{\L\D} & Y_{\R\D} & -\sum_{k = \L, \R, \G} Y_{\D k} & 0 & 0 & 1 \\
\end{pmatrix}
\begin{pmatrix}
V'_\L \\
V'_\R \\
V'_\D \\
I'_\L \\
I'_\R \\
I'_\D
\end{pmatrix}
\label{eq:IVrelation}
\end{equation}
\end{widetext}
This has the general form
\begin{equation}\begin{pmatrix} \Vv \\ \Iv \end{pmatrix} = \begin{pmatrix} \Mm^1_{VV} & \Mm^1_{VI} \\ \Mm^1_{IV} & \Mm^1_{II}\end{pmatrix} \begin{pmatrix} \Vv' \\ \Iv' \end{pmatrix} \equiv \Mm^1 \begin{pmatrix} \Vv' \\ \Iv' \end{pmatrix}.
\end{equation}
With multiple stages in a sequence, the mapping from the input to the measured device
\begin{equation}
\begin{pmatrix} \Vv \\ \Iv \end{pmatrix} = \Mm^\textrm{tot} \begin{pmatrix} \Vv^0 \\ \Iv^0 \end{pmatrix}
\end{equation}
is given by matrix multiplication $\Mm^\textrm{tot} = \prod_k \Mm^k$ where the matrices $\Mm^k$ are given by \Cref{eq:IVrelation}.
The $6 \times 6$ map $\Mm^\textrm{tot}$ is further reduced to a more useful $4 \times 4$ form by the following two steps.

\textit{Step 1.} Imposing current conservation ($dI_\D = - dI_\L - dI_\R$) at the sample, which then reduces the map to a $5 \times 5$ matrix with $k \in \{\Vsample_\L, \Vsample_\R, \Vsample_\D, \Isample_\L, \Isample_\R\}$ and $l \in \{\Vinput_\L, \Vinput_\R, \Vinput_\D, \Iinput_\L, \Iinput_\R\}$ which now takes the form 
\begin{multline}
M^\textrm{red}_{k l} = M^\textrm{tot}_{k l} - \left(M^\textrm{tot}_{ \Isample_\L \Iinput_\D} + M^\textrm{tot}_{ \Isample_\R  \Iinput_\D} + M^\textrm{tot}_{\Isample_\D \Iinput_\D}\right)^{-1} \,\times\\ \left(M^\textrm{tot}_{ \Isample_\L l} + M^\textrm{tot}_{ \Isample_\R l} + M^\textrm{tot}_{ \Isample_\D l}\right) M^\textrm{tot}_{k  \Iinput_\D}.
\label{eq:reduction1}
\end{multline}

\textit{Step 2.}
Defining $\Vsample_\L$ and $\Vsample_\R$ as potentials with respect to $\Vsample_\D$ at the sample, as well as $\Vinput_\D = 0$, which leads to the final $4 \times 4$ matrix $\Mm$ with
$k \in \{\Vsample_\L, \Vsample_\R, \Isample_\L, \Isample_\R\}$ and $l \in \{\Vinput_\L, \Vinput_\R, \Iinput_\L, \Iinput_\R\}$
that takes the form
\begin{equation}\begin{split} 
&M_{V_{L(R)} l} = M^\textrm{red}_{V_{L(R)}l} - M^\textrm{red}_{V_Dl} \\
&M_{I_{L(R)} l} = M^\textrm{red}_{I_{L(R)} l}.
\label{eq:reduction2}
\end{split}
\end{equation}

The effect on input voltages and measured currents of additional filtration from the measurement instruments should also be incorporated.

For our setup, the mapping $\Mm$ is comprised of four stages $\Mm^k$ where the resistances are determined by the installed filters and, for the final stage, are the sample lead resistances which are extracted independently from the linear $I$-$V$ response in DC measurements.
To calibrate the capacitances in the setup, we measure $\Iinput_\L$ and $\Iinput_\R$ as a function of the excitation frequency $f_\L$ of a voltage modulation first applied to the left port, then as a function of frequency $f_\R$ of the modulation applied to the right port when the device is fully pinched off (i.e.\ the zero conductance limit).
We then use capacitances between the lines and the ground as fit parameters relating the measured $\Iinput_\L$ to $f_\L$ and $\Iinput_\R$ to $f_\R$.
Finally, the mutual capacitance between $L$ and $R$ is obtained as a fit parameter relating the measured $\Iinput_\L$ and $\Iinput_\R$ to $f_\R$ and $f_\L$, respectively.
In the experiments described in the main text, the extracted capacitances of the fridge lines and filters, together with known cut-off frequencies of the voltage sources and current preamplifier outputs, are then used to calculate the sample conductances based on the measured $\Iinput_\L$ and $\Iinput_\R$.
Full details on this calibration routine, including all extracted parameters that were used to process measured data, can be found in the accompanying data repository~\cite{code_and_data}.

Note that a further effect of finite resistances on the stages of the measurement circuit is to rescale the applied DC bias voltages.
As described in Ref.~\onlinecite{Martinez21}, the DC voltages at the sample $\mathbf{V} = (V_\L, V_\R)^\mathrm{T}$ are related to the applied DC voltages $\mathbf{V^0} = (V^0_\L, V^0_\R)^\mathrm{T}$ and DC currents $\mathbf{I^0} = (I^0_\L, I^0_\R)^\mathrm{T}$ through the relationship:
\begin{align}
    \label{eq:v_tilde}
    \mathbf{V} &= \mathbf{V^0} - R \ \mathbf{I^0},
\end{align}
where $R$ is a matrix containing the total resistances on the left, right and drain lines:
\begin{align}
    R = \left(
    \begin{matrix}
        R_\L + R_\D & R_\D\\
        R_\D & R_\R + R_\D
    \end{matrix}
    \right).
\end{align}
For all the experimental datasets presented in this manuscript, the average magnitude of these corrections is 1-6\% of the applied bias voltages.
For this reason, we do not apply these corrections and present the data as a function of the applied bias voltages, and not of the bias voltages at the sample.

For the purposes of induced gap estimation as in \Cref{sec:subgap}, full characterization of the measurement circuit is not required, since the relevant features are the presence of conductance peaks and the DC voltage biases at the sample.
In this case, we apply a simpler correction procedure for finite frequency effects consisting of rotation of the acquired signal in the complex plane (typically a few degrees), projection along the axis that maximizes signal to noise ratio, and subtraction of a small residual conductance offset (typically on the order of $10^{-3} e^2/h$).

\section{Subgap density of states at zero-field}
\label{sec:subgap}

\begin{figure*}
\includegraphics[width=14cm]{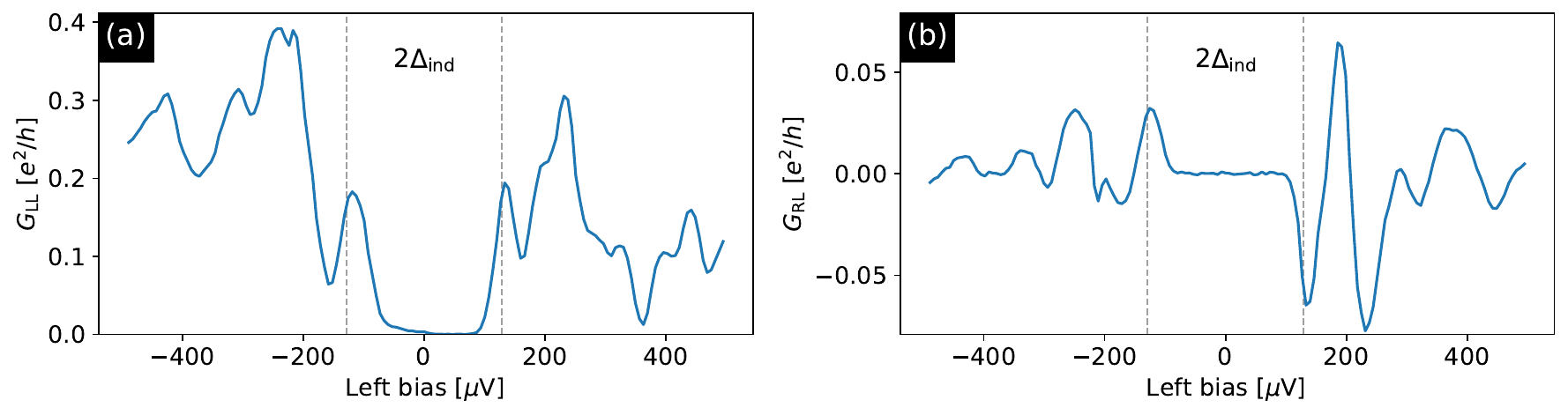}
\vskip -3mm
\caption{
Example cut of local (a) and non-local (b) conductance vs bias voltage for device A at $B = 0$.
The zero-field bulk induced gap as extracted from this dataset is $\DeltaInd \sim \SI{125}{\micro\eV}$, as estimated from the locations of the lowest-lying peak in the non-local data.
This number is consistent with $\DeltaInd = 129 \pm \SI{12}{\micro\eV}$ quoted in the main text (shown by vertical dashed lines), obtained by aggregating over several such bias traces from device A.
}
\label{fig:deviceA_characterization}
\end{figure*}

We now discuss the subgap density of states measured in Device A at zero magnetic field.
The presence of a finite sub-gap conductance can be attributed to coupling to the lead, temperature broadening, and/or the interfacial disorder, as discussed in Ref.~\onlinecite{Chang14}.
Indeed, the first-generation of proximitized nanowires~\cite{Lutchyn18} revealed a ``soft'' induced gap which was primarily associated with disorder at the superconductor-semiconductor interface.
The local and non-local conductances as a function of bias voltage for device A are shown in \Cref{fig:deviceA_characterization}.
Note that in order to have large enough signal in the non-local conductance, one needs to keep the junctions sufficiently open.
Datasets for both devices show well-defined coherence peaks at the edge of the induced gap and a strong suppression of the local conductance below the gap.
The sub-gap non-local conductances are below the noise floor of the measurement and, thus, consistent with zero.
The magnitude of the induced gap as determined from the coherence peak location in the local conductance agrees with the location of the lowest peak in the non-local conductance, indicating that, for these bias cuts, no sub-gap states were observed in the junction.
In general, sub-gap states may appear in the local spectroscopy, which we interpret as localized bound states at the junction, likely originating from the defects residing close to junction.

In order to understand the residual subgap density of states observed in local spectroscopy, we can compare the measured sub-gap suppression to the theory model described in \cite{Beenakker92}.
The sub-gap suppression is consistently below the single-channel limit, indicating that we have a ``hard'' proximitized gap with multiple conduction channels at the junction.
Indeed, the above-gap-conductance at the cutter and plunger values was approximately $ 0.18\, e^2/h$ whereas the zero-bias conductance was comparable to the noise floor, which is estimated to be $0.001 \, e^2/h$.
This is consistent with having two or more conduction channels in the lead.
 
Finally, we note that a better way to quantify the level of disorder in our devices is by measuring the localization length under the superconductor at finite B-field, outlined in \Cref{sec:localization_length}.
This measurement is more sensitive to disorder and is easier to interpret than the zero-field local conductance measurement.
Nevertheless, the residual subgap conductance data is consistent with having ultra-clean proximitized nanowires as discussed above.

\bibliography{topogap}

\end{document}